\documentclass[letterpaper]{aa}
\usepackage{graphicx,url}
\usepackage{natbib}
\bibpunct{(}{)}{;}{a}{}{,} % to follow the A&A style

\begin{document}

%\thesaurus{3 (02.01.1; 02.18.5; 11.01.2; 11.10.1; 11.17.4: 3C\,273 ) }

\newcommand{\de}{{\rm d}}
\newcommand{\vect}[2]{\left(\begin{array}{rr}#1\\#2\end{array}\right)}
\newcommand{\eg}{e.\,g.}
\newcommand{\ie}{\mbox{\emph{i.\,e., }}}
\newcommand{\cf}{\emph{cf.}\ }
\newcommand{\vv}{\emph{vv.}}
\def \kpc{\mbox{${\rm\ts kpc}$}}
\def \mJy{\mbox{${\rm\ts mJy}$}}
\def \microJy{\mbox{${\rm\ts\mu Jy}$}}
\def\micron{\hbox{\,$\mu$m}}
\def \nm{\mbox{${\rm\ts\,nm}$}}
\def \cm{\mbox{${\rm\ts cm}$}}
\def \privcom{{\it priv.comm.\/}}
\def\Bmin{\mbox{$B_\mathrm{min}$}}
\def\gmax{\mbox{$\gamma_\mathrm{max}$}}
\def\alo{\mbox{$\alpha_\mathrm{low}$}}
\def\ahi{\mbox{$\alpha_\mathrm{high}$}}
\def\nub{\mbox{$\nu_\mathrm{b}$}}
\def\nuc{\mbox{$\nu_\mathrm{c}$}}
\def\dL{\mbox{$d_\mathrm{L}$}}
\def\aopt{\mbox{$\alpha^\mathrm{UV}_\mathrm{opt}$}}
\def\aBRI{\mbox{$\alpha_{BRI}$}}
\def\aIRo{\mbox{$\alpha^\mathrm{opt}_\mathrm{IR}$}}
\def\arIR{\mbox{$\alpha^\mathrm{IR}_\mathrm{radio}$}}
%alpha radio-optical
\def\aro{\mbox{$\alpha_\mathrm{RO}$}}
\def\aKO{\mbox{$\alpha_{KO}$}}

\title{The radio--ultraviolet spectral energy distribution of the jet
in 3C\,273\thanks{Based on observations made
with the NASA/ESA Hubble Space Telescope, obtained at the Space
Telescope Science Institute, which is operated by the Association of
Universities for Research in Astronomy, Inc.\ under NASA contract
No.~NAS5-26555.  These observations are associated with proposals
\#5980 and \#7848.  Also based on observations obtained at the NRAO's
VLA.  The National Radio Astronomy Observatory is a facility of the
National Science Foundation operated under cooperative agreement by
Associated Universities, Inc.}}

\author{Sebastian~Jester \inst{1}\thanks{\emph{Present address:}
   Fermilab MS 127, PO Box 500, Batavia IL 60510, USA}
   \and Hermann-Josef~R\"oser \inst{1}
   \and Klaus~Meisenheimer \inst{1}
   \and Rick~Perley \inst{2}
}

\offprints{S. Jester, \email{jester@fnal.gov}}
\date{Received January 6, 2004; accepted October 21, 2004}
\institute{
   Max-Planck-Institut f\"ur Astronomie, K\"onigstuhl~17,
   69117~Heidelberg, Germany
   \and National Radio Astronomy Observatory, P.\,O.~Box 0, Socorro, NM~87801, USA
}

\authorrunning{Jester et al.}
\titlerunning{The radio-ultraviolet spectra of the jet in 3C\,273}

\abstract{We present deep VLA and HST observations of the large-scale
jet in 3C\,273 matched to 0\farcs3 resolution.  The observed spectra
show a significant flattening in the infrared-ultraviolet wavelength
range.  The jet's emission cannot therefore be assumed to arise from a
single electron population and requires the presence of an additional
emission component.  The observed smooth variations of the spectral
indices along the jet imply that the physical conditions vary
correspondingly smoothly.  We determine the maximum particle energy
for the optical jet using synchrotron spectral fits.  The slow decline
of the maximum energy along the jet implies particle reacceleration
acting along the entire jet.  In addition to the already established
global anti-correlation between maximum particle energy and surface
brightness, we find a weak positive correlation between small-scale
variations in maximum particle energy and surface brightness.  The
origin of these conflicting global and local correlations is unclear,
but they provide tight constraints for reacceleration models.

}

\maketitle
\section{Introduction}\label{s:intro}
Jets are collimated outflows, thought to be launched from an accretion
disk around a central compact object, which can be a supermassive
black hole, a stellar-mass black hole or neutron star, or a young
stellar object (YSO). They transport mass, energy, both linear and angular
momentum as well as electromagnetic fields outward from the central
object.  A detailed understanding of the formation of these jets,
their connection to the accretion disk from which they are launched,
and the physics governing their internal structure and observable
properties has not yet been achieved. Here we consider the synchrotron
emission from the kiloparsec-scale jet of the quasar \object{3C\,273},
which is one of the brightest and largest and therefore an instructive
sample case.

Jets became part of the standard model of extragalactic radio sources
\citep{BBR84} to link the lobes, which emit the bulk of the
synchrotron radio luminosity, with the active galactic nucleus (AGN)
of the host galaxy. In this model, jets merely transport energy to
feed the lobes. In the most powerful sources, a double shock structure
terminates the jets, consisting of an outer bow shock and contact
discontinuity separating the jet material from the external medium and
an internal shock (Mach disk) at which the relativistic flow is
decelerated and bulk kinetic energy is channeled into highly
relativistic particles through a shock acceleration mechanism. These
particles emit the synchrotron radiation observed from the lobes. The
radio hot spot is usually assumed to coincide with the Mach disk. The
{\it optical} synchrotron emission observed from some hot spots can
also be explained by first-order Fermi acceleration at a
jet-terminating shock \citep{MH86,HM87,magnumopus89,hs_II}.

In this model, the electrons responsible for radio synchrotron
emission from the jets themselves are accelerated near the black hole
and then simply advected with the jet flow. However, observations of
high-energy (optical and X-ray) synchrotron radiation from 3C\,273 and
other jets force us to revise this picture.  Electrons with the highly
relativistic kinetic energies required for synchrotron emission in the
infrared and optical have a very short radiative lifetime. This
lifetime is much less than the light-travel time along the jet in
3C\,273 \citep[already noted by ][]{GS64} and other jets. Observations
of optical synchrotron emission from such jets \citep{RM91,M87,SU02}
as well as from the ``filament'' near Pictor~A's hot spot
\citep{RM87,PRM97} therefore suggest that in addition to a localized,
``shock-like'' acceleration process operating in hot spots, there is
an extended, ``jet-like'' mechanism at work in radio sources in
general and 3C\,273's jet in particular \citep{hs_II}.  The extended
mechanism may also be at work in the lobes of radio galaxies, where
the observed maximum particle energies are above the values implied by
the losses within the hot spots \citep{Mei96} and by the dynamical
ages of the lobes \citep{BR00}.  We therefore make a clear distinction
between the emission from the hot spot itself, the lobes and the body
of the jet. Although all emit synchrotron radiation, the physical
processes accelerating the particles emitting in these regions may be
quite different.

The lifetime problem has been exacerbated by observations at even
higher frequencies: Einstein and ROSAT observations, for example,
showed X-ray emission from the jets in M87 \citep{SGF82,BSH91,NMRF97}
and 3C\,273 \citep{HS87,Roe00}. More recently, observations with the
new X-ray observatory Chandra showed extended X-ray emission from many
more jets, like PKS~0637$-752$ \citep{Sch00} and Pictor\,A
\citep{Wil01} as well as other jets and hot spots.  Chandra also
supplied the first high-resolution X-ray images of the jets in 3C\,273
and M87 \citep{Mareta01,Sam01,Mar02}. The X-rays from these objects
seem to be of non-thermal origin \citep[for an overview, see][]{HK02}:
they could at least partially be due to synchrotron emission
\citep{Roe00,Mareta01,Mar02}. Alternatively, inverse-Compton
scattering could be responsible for the X-rays. The photon seed field
can be provided by the synchrotron source itself if it is sufficiently
compact, for example in the hot spots of Cygnus A
\citep{HCP94,WilsonCygA}. If the bulk flow of a jet is still highly
relativistic on large scales, the boosted energy density of the cosmic
microwave background radiation field can lead to the observed X-ray
fluxes \citep{Cel00,Tav00}. In all cases, those electrons producing
the radio-optical synchrotron emission suffer additional losses from
the inverse-Compton scattering, decreasing their cooling timescale
even below the synchrotron cooling scale.

Thus, the fundamental question posed by the observation of optical
extragalactic jets is the following: how can we explain high-frequency
synchrotron and inverse-Compton emission far from obvious acceleration
sites in extragalactic jets? While information on the source's
magnetic field structure may be obtained from the polarization
structure, the diagnostic tool for the radiating particles is a study
of the synchrotron continuum over as broad a range of frequencies as
possible, \emph{i.\,e.,} from radio to UV wavelengths, and with
sufficient resolution to discern morphological details. The shape of
the synchrotron spectrum gives direct insight into the shape of the
electron energy distribution, thus also constraining the emission by
the inverse-Compton process at other wavelengths.  Here, we consider
the shape of the synchrotron spectrum of the jet in 3C\,273.

\subsection{The  jet in 3C\,273}

This optical jet was first detected on ground-based images. Like M87, its
optical brightness and length are so unusually large that it was
detected even before radio jets were known. It appears to consist of a
series of bright knots with fainter emission connecting them (see
Fig.\,\ref{f:obs.nir.IRmap}).  \citet{GS64} described the jet's
optical spectrum as ``weak, bluish continuum'', suspecting that this
was synchrotron radiation. This was confirmed by \citet{RM91} through
optical polarimetry.

3C\,273's radio jet extends continuously from the quasar out to a
terminal hot spot at 21\farcs4 from the core, while optical emission
has been observed only from 12\arcsec\ outward.\footnote{For the
conversion of angular to physical scales, we assume a flat cosmology
with $\Omega_{\mathrm{m}}=0.3$ and $H_{0} = h_{70}\times
70\,\mathrm{km\,s^{-1}Mpc^{-1}}$, leading to a scale of $2.7
h_{70}^{-1} \mathrm{kpc}$ per second of arc at 3C\,273's redshift of
0.158.}  We concentrate on this ``outer'' part of the jet here, and
will report observations of optical emission from the inner jet with
the VLT in a future publication \citep[see
also][]{Marea03}. \citet{Baheta95} presented the first HST imaging of
this jet, noting the jet resembles a helical structure.

Prior to the present work, synchrotron spectra have been derived for
the hot spot and the brightest knots using ground-based imaging in the
radio \citep{jetII}, near-infrared $K^{\prime}$-band \citep{NMR97} and
optical $I, R, B$-bands \citep{RM91} at a common resolution of
1\farcs3 \citep{MNR96,Roe00}. This radio-to-optical continuum can be
explained by a single power-law electron population resulting in a
constant radio spectral index\footnote{We define the spectral index
$\alpha$ such that $f_\nu \propto \nu^{\alpha}$.} of $-0.7$, but with
a high-energy cutoff frequency decreasing from $10^{17}\,$Hz to
$10^{15}$\,Hz outward along the jet.

Here, we present VLA and HST NICMOS observations (\S\ref{s:obs}) in
addition to the WFPC2 data already published in \citet{Jes01}.
Together, these constitute a unique data set in terms of resolution
and wavelength coverage for any extragalactic jet --- only M87 is
similarly well-studied \citep[][and references
therein]{MRS96,SBM96,HB97,Pereta99,Per01}. Using these observations at
wavelengths 3.6\cm, 2.0\cm, 1.3\cm, 1.6\micron, 620\nm\ and 300\nm, we
derive spatially resolved (at 0\farcs3) synchrotron spectra for the
jet (\S\ref{s:results}). By fitting model spectra according to
\citet{HM87}, we derive the maximum particle energy everywhere in the
jet in order to identify regions in which particles are either
predominantly accelerated, or predominantly lose energy
(\S\ref{s:ana}).  The model spectra reveal excess near-ultraviolet
emission above a synchrotron cutoff spectrum accounting for the
emission from radio through optical, which implies that a
two-component model is necessary to describe the emission. The
radio--optical--X-ray spectral energy distributions (SEDs) suggest a
common origin for the UV excess and the X-rays from the jet
\citep[\S\ref{s:disc.pop}; see also ][]{Jes02}. By considering just
the optical spectral index, we concluded in \citet{Jes01} that
particles must be reaccelerated along the entire jet. Here we confirm
this conclusion by using the full spectral information from radio to
near-ultraviolet (\S\ref{s:disc.acc}). We show that the observed
changes of cutoff energy and surface brightness along the jet can be
jointly explained as effects of changes in the magnetic field and the
Doppler beaming parameter along the jet (\S\ref{s:disc.corr}).  We
conclude in \S\ref{s:summary}.

\section{Observations and data reduction}\label{s:obs}

\subsection{Radio observations}\label{s:obs.radio}

The jet has been observed at all wavelength bands available at the
NRAO's Very Large Array (VLA), \ie at 90\,cm, 20\,cm, 6\,cm, 3.6\,cm,
2\,cm, 1.30\,cm and 0.7\,cm. Observations were carried out between
July 1995 and November 1997, to obtain data with all array
configurations (thus covering the largest range of spatial
frequencies). Total integration times are of order a few times
10,000\,s in each band. At 3.6~cm, the resolution set by the maximum
VLA baseline of just over 32\,km is 0\farcs24, with higher resolution
at shorter wavelengths. However, at 0.7\,cm, fewer than half antennas
were equipped with receivers at that time, and the brightness of the
jet is so low relative to the noise that only the hot spot is detected
even at a fairly low resolution of 0\farcs35. 

The VLA data were edited, calibrated and CLEANed according to standard
procedures. Table \ref{t:obs.vla.dynran} quotes the achieved dynamic
ranges.  The present analysis considers the data at 3.6\,cm, 2\,cm and
1.3\cm. This allows to fix the common resolution for the entire study
at 0\farcs3, slightly inferior to the resolution of the data at
3.6\,cm.  The remainder of the VLA data set will be discussed in a
future publication, which will also contain details about the data
processing.

\begin{table}
 \caption{\label{t:obs.vla.dynran}Dynamic ranges for
the VLA images}
\begin{minipage}[t]{\hsize}
\renewcommand{\footnoterule}{}
\centering
\newcounter{mpc}
\stepcounter{mpc}
\begin{tabular}{lcccc}
\hline\hline {\bf VLA}&{\bf $\lambda$}&{\bf Peak
flux}&{\bf RMS noise}&{\bf Dynamic}\\
{\bf band} & \cm & \mJy & \mJy & {\bf range}\\
\hline
 X & 3.6 & 33.0 & $4.5\times 10^{-4}$ & 75,000\\
 U & 2.0 & 28.3 & $2.6\times 10^{-4}$ & 110,000\\
 K & 1.3 & 23.4 & $4.0\times 10^{-4}$ & 59,000\\
 Q\footnote{image not used for spectra} & 0.7 & 20.9 & $2.5\times 10^{-3}$ & 9,000\\
\hline\hline
\end{tabular}
\end{minipage}
\end{table}

\subsection{Optical and near-ultraviolet
observations}\label{s:obs.optuv}

Optical ($\lambda 620\nm$) and near-ultraviolet ($\lambda 300\nm$)
images were obtained under HST proposal \#5980, using WFPC2 and
filters F622W (total exposure time 10,000\,s) and F300W (exposure time
35,500\,s).  The data reduction and jet images are described in
\citet{Jes01}.

\subsection{Near-infrared observations}\label{s:obs.nir}

\subsubsection{Data}

Observations were carried out under HST proposal \#7848 using NICMOS
camera 2 (NIC2) on board the HST, which has $256\times256$ pixels of
nominal scale 0\farcs076.  Filter F160W was used, with a central
wavelength of $1.6\micron$, yielding a diffraction limit of 0\farcs17.
The total exposure time on the jet was 34560\,s distributed over 30
individual exposures with integer-pixel offsets. Each exposure was
read out non-destructively every 256\,s.  Here we give only an outline
of the data reduction; for details, see \citet{JesterDiss}.

Nearly all of the frames reduced using the CALNICA pipeline provided
by STScI show an offset in the background level between the detector
quadrants as well as an imprint of the flat-field pattern.  The
quadrant offsets are ascribed to spatial and temporal variations of
the detector bias level which have been termed ``shading'' and
``pedestal'' \citep{nic-anomaly}. The recommended use of
temperature-dependent dark files \citep{nic-dark} did not improve the
quality of the reduced images, nor did any of the otherwise available
correction tools. We therefore employed a custom reduction routine
which initially estimates the sky and dark current by filtering the
jet signal from all individual readouts. Any remaining
quadrant-to-quadrant variation after subtracting the sky and dark
current is ascribed to an additive component.  These residual offsets
are removed by subtracting the modal value from each
quadrant. Finally, cosmic-ray and bad pixels are rejected using a
pixelwise median filter.

The resulting images are not perfectly flat individually, suggesting
that there may be a residual problem with the flat-field. However, no
attempt is made to correct this because there is no information on
what the correct flatfield might be. Residual background structures
(including possible large-angle scattering wings from the quasar core)
are removed by modeling the background around the jet using
second-order polynomials along detector rows, whose coefficients are
smoothed in the perpendicular directions \citep[identical to the
method used for the WFPC2 images, see][]{Jes01}.  The photometric
calibration is performed using the appropriate conversion factor from
the \verb+synphot+ package provided by STScI.  In the conversion, we
do not correct for variations in the spectral index but always assume
a flat spectrum in $f_\nu$; this correction would be at most 2\%.

\subsubsection{Map of near-infrared brightness}\label{s:obs.nir.IRmap}
\begin{figure*}
% turn sidecaption off for astro-ph 
% \sidecaption
% \includegraphics[width=12cm,clip]{H_spike.eps}
\includegraphics[width=\hsize,clip]{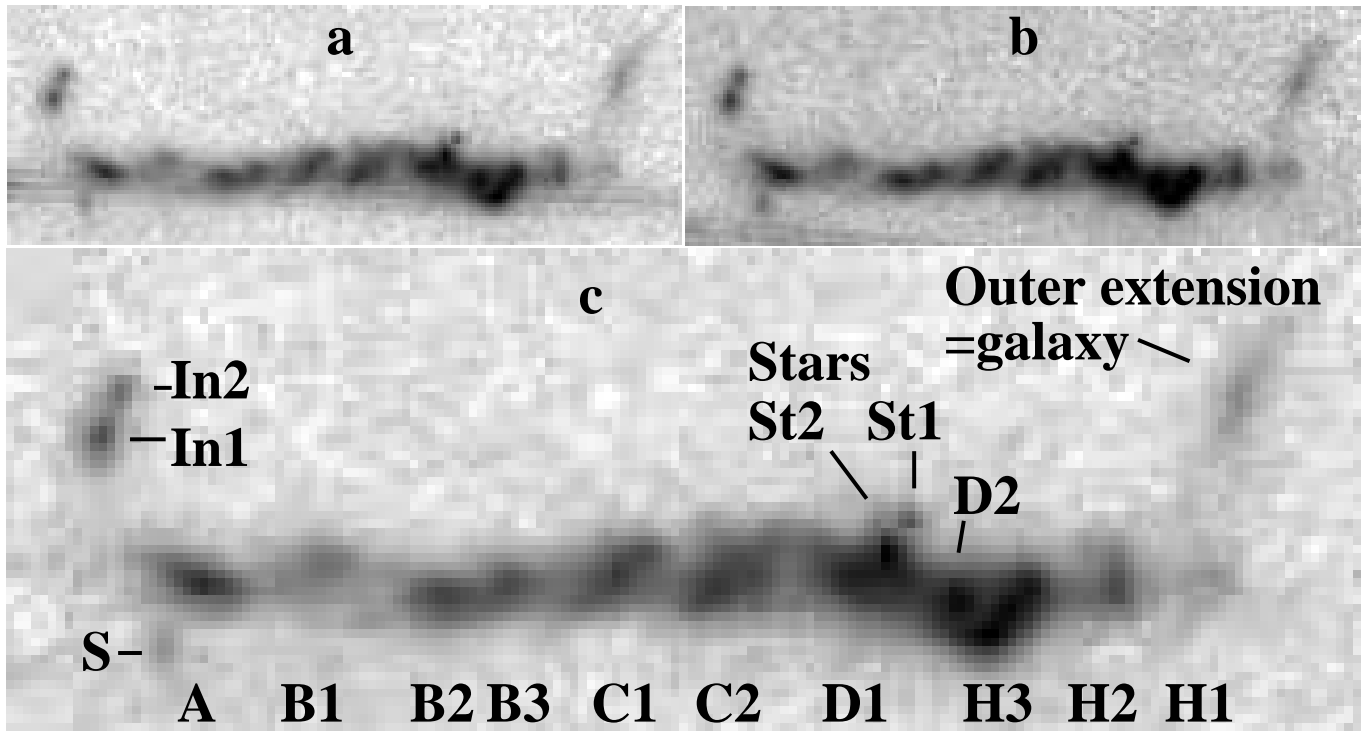}
\caption{Map of near-infrared brightness ($\lambda 1.6\mu$m), rotated
with position angle $222\degr$ along the horizontal.  a: Sum of images
with total exposure time 23\,040\,s, showing the diffraction spike
passing through part of the jet. b: Sum of remaining images with total
exposure time 11\,520\,s, diffraction spike clear of the jet.  c: as
a, after subtraction of diffraction spike modeled on
b.\label{f:obs.nir.IRmap}}
\end{figure*}
Figure\,\ref{f:obs.nir.IRmap} shows the reduced and summed images. The
overall appearance of the jet's morphology at 1.6\,$\mu$m is very
similar to that at visible wavelengths \citep[compare the WFPC2 images
in ][ and Fig\,\ref{f:res.ima.images} below]{Jes01}. The only
significant difference (apart from the second star St2, see below) is
that the tip of the jet H1 (apparently downstream of the hot spot H2)
is visible on the infrared image, but not at higher frequencies. The
jet appears equally well collimated in the near-infrared as in the
optical. All optical extensions are detected, the southern extension S
here being much weaker than knot A, while they are comparable in the
UV.  In1 is clearly brighter than In2, while again their near-UV
brightness is comparable, indicating a marked color difference between
the inner extension's two knots.

There is signal from one of the quasar's diffraction spikes superposed
on the jet emission collected in the first 2/3 of the total exposure
time (Fig.\,\ref{f:obs.nir.IRmap}). In addition, an IR-bright object
St2 is located within the jet, close to the faint star St1 just north
of the jet which is also detected on the optical image. We modelled
the spike by summing the 10 exposures in which the spike is clear of
the jet (in these, the telescope has been rotated by 4\degr\ compared
to the previous 20; Fig.\,\ref{f:obs.nir.IRmap} b).  A scaled version
of the model is then subtracted from each individual image. The result
of subtracting the spike model from the ``contaminated'' sum frame is
shown in Fig.\,\ref{f:obs.nir.IRmap} c.

To assess whether the IR-bright object St2 is part of the jet or an
unrelated foreground object, we first checked whether its brightness
profile is consistent with that of a point source. The widths of
Gaussians fitted to 1D cuts along rows and columns of the sum images
are consistent with the known width of the PSF.  Secondly, there are
no correlations with total or polarised brightness features of the
radio jet.  We are therefore confident that this object is a star
which by chance appears superimposed on the jet image.  We modelled
this star on the sum images as a Gaussian sitting on top of a sloping
plane which accounts for the underlying jet flux.  An appropriately
scaled version of the star model is subtracted from all 30 individual
frames \citep[for details, see][Section 2.2.5]{JesterDiss}.  The flux
removed in this manner is $2.0\,\microJy$. The jet flux in an
equivalent aperture (ignoring any possible proper motion) on the F622W
image of $0.2\,\microJy$ is an upper limit to this object's flux in
this band, giving it a $R-H \ga 3$.

\section{Photometry}\label{s:photo}

In order to determine the synchrotron spectrum over the entire jet, we
perform beam-matching aperture photometry at a grid of positions
covering the jet.  The photometry was done by our own \verb+MPIAPHOT+
software \citep[\emph{cf.} ][]{RM91}. This uses a weighted summation
scheme, equivalent to a convolution, to match the different point
spread functions to a common beam size. At the same time, it allows an
arbitrary placement of apertures with respect to the pixel grid of
individual images without loss of precision.  All images are matched
to a final resolution of 0\farcs3 FWHM, slightly inferior to that of
the data set with the lowest resolution (the 3.6\,cm radio data imaged
with 0\farcs25 resolution; \cf \S\ref{s:obs.radio}).  Following
considerations given in \citet[][ Appendix A]{Jes01}, this requires a
relative image alignment accurate to 0\farcs03 in order to limit
photometry errors from misalignment to 5\% for point sources.  The
flanks of point sources have the steepest possible intensity
gradients, the photometry error for extended sources will usually be
an order of magnitude smaller. This accuracy is required to avoid the
introduction of spurious spectral index features.

We achieve the desired alignment accuracy by using a grid of
photometry positions, defined as offsets on the sky relative to the
quasar core, which is assumed to coincide at all wavelengths.  This
grid is transformed into detector coordinates on each individual data
frame, accounting for telescope offsets and geometric
distortion. There is one frame each for the three VLA wavelengths, 30
for HST-NICMOS, four and 14, respectively, for HST-WFPC2 at 620\nm and
300\nm.  Because of 3C\,273's location near the celestial equator, and
because all offsets between individual HST frames are small (below
30\arcsec), detector and celestial coordinate offsets are related by a
simple linear transformation.

The grid transformation is straightforward for the VLA images, which
contain both the quasar and the jet and do not suffer from saturation
effects. For the HST images, we first obtain the precise location of
the quasar core from one short exposure (a few seconds) obtained
during each HST ``visit''. The telescope offsets between these short
exposures and the deep science exposures of the jet are obtained from
the engineering (``jitter'') files provided with the data.  The
aperture pixel position is calculated from the quasar's pixel
position, the desired offset on the sky and the telescope offset. We
account for geometric distortion by using the wavelength-dependent
cubic distortion correction for WFPC2 as determined by
\citet{trauger-geom}, and the quadratic NICMOS coefficients given in
\citet{NIC-distortion}. This procedure also takes care of the slightly
differing image scales along the NIC2 detector's x- and
y-directions. We stress here that the unsaturated quasar image has
been crucial in achieving the necessary alignment accuracy.

We use a rectangular grid of aperture positions. The grid extends
along position angle 222\fdg2, starting at a radial distance of
$r=12\farcs0$ from the quasar and extending to
$r=23\farcs0$. Perpendicular to the radius vector, the grid extends to
$\pm 1\farcs0$. Individual grid points are spaced 0\farcs1 apart,
yielding a good sampling of the 0\farcs3 effective resolution, so that
there are 111 radial grid points and 21 points perpendicular to
the radius vector, \ie 2331 in total.  All points are transferred to
the individual data frames, and the flux per 0\farcs3 aperture
centered at each grid position is determined.

The photon shot noise is below 0.5\% per beam in all bands. The HST
images have a flat-field error of 1\% (WFPC2) and 3\% (NICMOS) added
in quadrature. The uncertainty in the background estimation is
estimated from the scatter in blank sky regions as 0.01\,$\mu $Jy per
beam for WFPC2, and 0.03\,$\mu $Jy for NICMOS, forming an error floor
significant only for the faintest parts of the jet. An error source
unique to the interferometric radio data is the error from the
deconvolution, i.e., errors in the sense that the inferred brightness
distribution does not correspond to the true distribution on sky, in
particular for the fainter parts of the jet. This error is very hard
to quantify and we use a 3\% error to account for this. All these
error sources limit the accuracy of relative photometry within one
waveband. In addition, all wavebands will suffer an error from the
absolute photometric calibration, typically 2\%.
 
\section{Results}\label{s:results}
\subsection{Jet images at 0\farcs3}\label{s:res.images}

\begin{figure*}
\parbox[c]{0.1\hsize}{\hspace*{0.1\hsize}}\parbox[c]{.9\hsize}{\includegraphics[width=.9\hsize,clip]{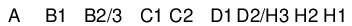}}
\parbox[c]{0.1\hsize}{radius/\arcsec}\parbox[c]{.9\hsize}{\includegraphics[width=.9\hsize,clip]{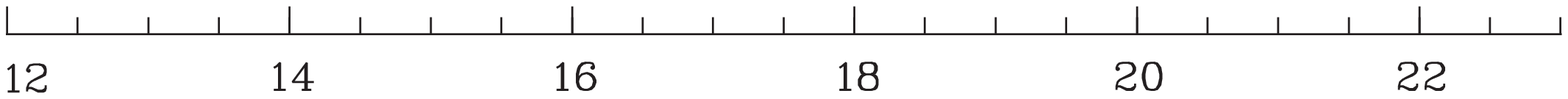}}
\parbox[c]{0.1\hsize}{$U$\\300\,nm\\0.9\microJy}\parbox[c]{.9\hsize}{\includegraphics[width=.9\hsize,clip]{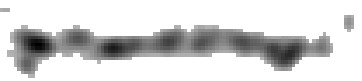}}
\parbox[c]{0.1\hsize}{$R$\\620\,nm\\1.25\microJy}\parbox[c]{.9\hsize}{\includegraphics[width=.9\hsize,clip]{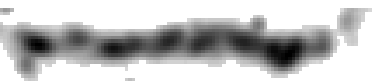}}
\parbox[c]{0.1\hsize}{$H$\\1.6\,$\mu$m\\5.75\microJy}\parbox[c]{.9\hsize}{\includegraphics[width=.9\hsize,clip]{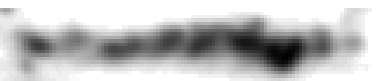}}
\parbox[c]{0.1\hsize}{K-band\\1.3\,cm\\95\mJy}\parbox[c]{.9\hsize}{\includegraphics[width=.9\hsize,clip]{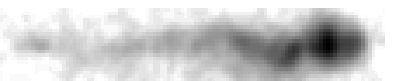}}
\parbox[c]{0.1\hsize}{U-band\\2.0\,cm\\150\mJy}\parbox[c]{.9\hsize}{\includegraphics[width=.9\hsize,clip]{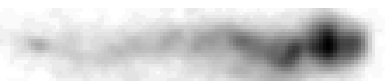}}
\parbox[c]{0.1\hsize}{X-band\\3.6\,cm\\265\mJy}\parbox[c]{.9\hsize}{\includegraphics[width=.9\hsize,clip]{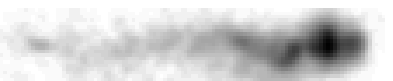}}
\parbox[c]{0.1\hsize}{radius/\arcsec}\parbox[c]{.9\hsize}{\includegraphics[width=.9\hsize,clip]{skala.eps}}
\parbox[c]{0.1\hsize}{\hspace*{0.1\hsize}}\parbox[c]{.9\hsize}{\includegraphics[angle=0,width=.9\hsize,clip]{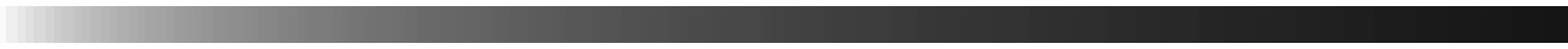}}
\caption[Photometry of the jet in 3C\,273 at 0\farcs3 effective beam
size]{\label{f:res.ima.images}Photometry of the jet in 3C\,273 at
0\farcs3 effective beam size. Clipped to show only measurements with
aperture signal-to-noise ratio $>5$. Grey levels runs from 0 to the
peak flux/beam with a pseudo-logarithmic stretch as indicated by the
greyscale bar. Jet features are labelled as in
Fig.\,\ref{f:obs.nir.IRmap}.  The offset of 0\farcs2 between radio and
optical hot spot position can be made out clearly.}
\end{figure*}
In order to compare the images at different wavelengths, the
photometry results are reassembled into the images shown in
Fig.\,\ref{f:res.ima.images}.  We compare the jet's morphological
features at different wavelengths before considering the spectra.

\subsubsection{Jet morphology from radio to UV}\label{s:res.im.morph}

\begin{figure}
\resizebox{!}{\hsize}{\includegraphics{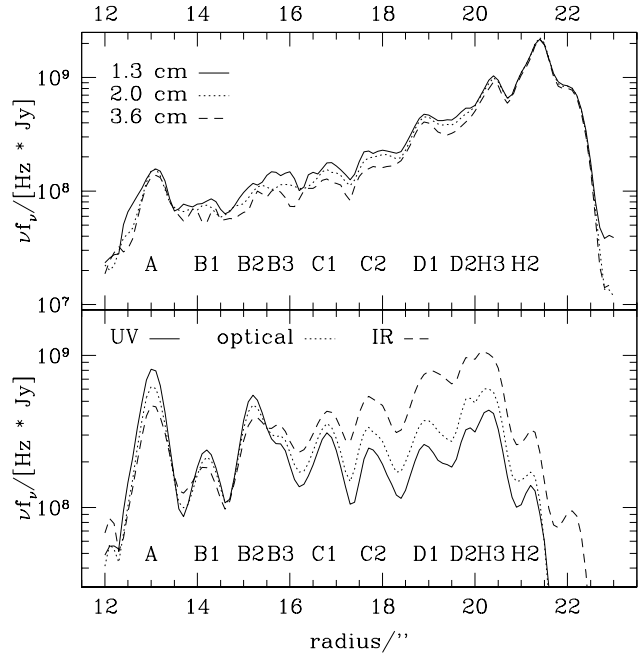}}
\caption{\label{f:res.im.morph.ridge}Plot of surface brightness per
beam along the jet's ridge line, \ie showing the brightest point per
column from Fig.\,\ref{f:res.ima.images}.}
\end{figure}
\begin{figure*}
\resizebox{\hsize}{!}{\includegraphics[angle=270]{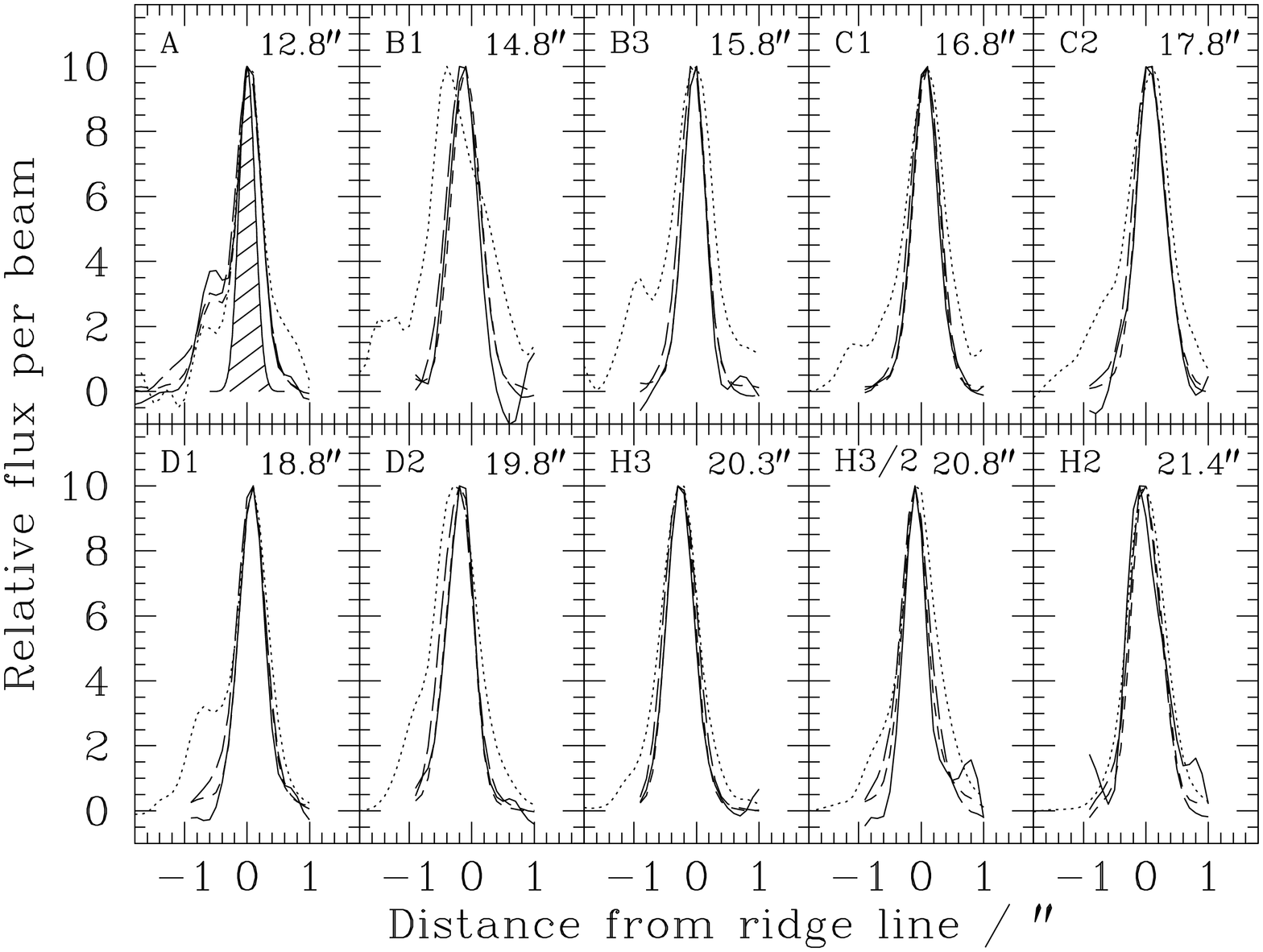}}
\caption{\label{f:res.im.morph.cuts} Normalised cuts through the jet
profile at various distances from the quasar at 300\,nm (solid line),
620\,nm (short dash), 1.6\micron\ (long dash), and 2.0\,cm (dotted);
resolution is 0\farcs3. Negative offsets are to the south of the
jet. The shaded Gaussian in the first panel shows the resolution of
0\farcs3 FWHM.  The ``southern extension'' is visible in the
short-wavelength profiles at $r=12\farcs8$. However, the apparently
corresponding feature in the radio profile is in fact
unrelated. Instead, it is due to the radio cocoon around the jet:
\protect\citet[][\cf their Fig. 3]{jetIV} noted that the radio
emission is more extended, in particular to the south, than the
optical emission. There is a similar tendency for the near-infrared
profile to be more extended than the optical/UV. Other than that, the
profiles show a similar jet width at all wavelengths. The cut through
B1 clearly shows the transverse offset in optical and radio emission
between the jet's two strands.  In D2, the radio peak appears offset
slightly to the south from the optical peak, but the interpretation of
this shift is uncertain.}
\end{figure*}
To facilitate a comparison of the jet morphology at different
wavelengths beyond a direct inspection of the panels in
Fig.\,\ref{f:res.ima.images}, we show the flux profile along the jet
in Fig.\,\ref{f:res.im.morph.ridge} and normalised transverse profiles
in \ref{f:res.im.morph.cuts}.  There is a close correspondence of
morphological features at all wavelengths, \ie a coincidence of local
brightness maxima and the occurrence of ``knots'' across the entire
observed wavelength range \citep[\emph{cf.} ][ who describe the same
fact by noting that the jet's features have similar angular sizes at
all wavelengths]{Baheta95}.  The sole exception within the jet is
region B1, in which the jet's apparently double-stranded nature is
most conspicuous.  At HST frequencies, B1's northern strand is
considerably brighter than the southern, while on the radio images,
the situation is exactly the opposite. This is also the only location
in the jet which might be classified as edge-brightened (\cf
\S\ref{s:res.im.vol}).  Note that the two bright knots of the jet
preceding B1 are the sources of the brightest X-ray emission, perhaps
with a small offset between the locations of the optical and X-ray
peaks \citep{Mareta01}.

Apart from this discrepancy, only the relative brightness of the knots
changes with wavelength.  Relative to knot~A at the onset of the
optical jet, the radio peak brightness increases by a factor of about
5--10 for H3, and another factor of two for the radio brightness peak
H2, which has historically been called the radio ``hot spot''.
However, in the near-infrared at 1.6\,\micron, the brightness peaks at
H3, and H2 are already fainter than most of the remainder of the jet.
In the near-UV at 300\nm, H2 is the faintest feature, while A is the
brightest.  As already noted above, the tip of the jet H1 is detected
up to the near-infrared, but not at shorter wavelengths (any emission
so far detected at 600\nm beyond H2 is related to the nearby galaxy,
not to H1).  Thus, the brightness profile tends to invert from radio
to near-ultraviolet. This trend continues up to X-rays: A dominates
the jet's X-ray luminosity \citep{Mareta01}, while H2 dominates the
radio luminosity.  This change in brightness profile with wavelength
is equivalent to a change in the spectrum along the jet, which will be
considered below (\S\ref{s:res.alpha}).

The transverse cuts (Fig.\,\ref{f:res.im.morph.cuts}) confirm the
findings of \citet[][ \emph{cf.} their Fig. 3]{jetIV}: the width of
the radio and optical jet is very similar, but there is extended radio
emission without an optical counterpart to the south of the jet.
Confirming the findings of \citet{NMR97}, there is a tendency for the
near-infrared emission to be more extended to the south than the
optical and ultraviolet emission.  This strengthens their conclusion
that the optical jet traces the emission of the jet channel as such.
The more extended and fainter low-frequency emission corresponds to a
surrounding cocoon of material, interpreted as material which is
back-flowing after having passed through the hot spot \citep{jetIV}.

\begin{figure}
% too big for astro-ph
% \resizebox{\hsize}{!}{\includegraphics{hotspot_r_k-7_newhorz.eps}}
\resizebox{\hsize}{!}{\includegraphics{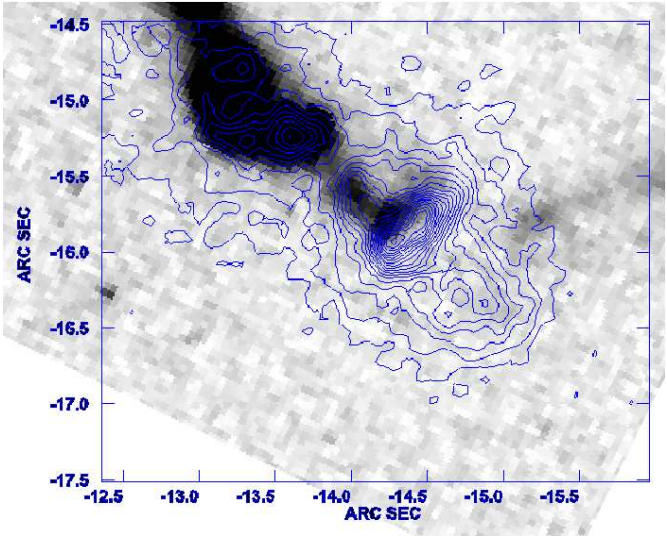}}
\caption{\label{f:res.im.morph.HS}Comparison of hotspot position at
optical and radio wavelengths. Greyscale shows the 620\,nm image at
original resolution, contours show the 1.3\,cm radio map at 0\farcs125
resolution. Coordinates are relative to the quasar core, using which
both images were aligned. The radio hot spot is clearly offset by
0\farcs2 from the optical hot spot, while the radio contours closely
agree with the optical image in the preceding region (D2/H3).}
\end{figure}
Figures\,\ref{f:res.ima.images} and \ref{f:res.im.morph.ridge} show a
radial offset of about 0\farcs2 between the position of H2 at radio
and optical wavelengths.  This offset is confirmed by the overlay of
radio and optical data at the respective instrumental resolution in
Fig.\,\ref{f:res.im.morph.HS}. The uncertainty of the offset
determination is likely dominated by the systematic pointing
uncertainty of the HST data with respect to the VLA data, which we
kept to better than 0\farcs03 (\cf \S\ref{s:photo}).  We discuss the
interpretation of this offset in \S\ref{s:disc.HS}.

\subsubsection{Jet volume}\label{s:res.im.vol}
It is necessary to know the jet volume to obtain an estimate for the
magnetic field by the minimum-energy argument (\S\ref{s:ana.equi}
below).  We just noted that the optical jet delineates the jet
channel as such, while the radio emission contains contributions from
the surrounding material. The geometry of the jet channel is therefore
best constrained by considering the optical morphology. We assume the
jet has isotropic emissivity which is constant along individual lines
of sight through the jet and neglect relativistic beaming
effects. (Even in the presence of beaming, the conclusions are
unaltered as long as the beaming does not vary significantly along any
given line of sight.)

The jet is center-brightened at all wavelengths on images resolving
its width (the only exception being B1, as noted above in
\S\ref{s:res.im.morph}).  If the emission region was confined to a
cylindrical shell at the jet surface, the resulting brightness
distribution would be edge-brightened, both for uniform emissivity
resulting from a tangled magnetic field geometry, and for an ordered
helical field \citep{MeiHabil,Laing81}. To lowest order, the jet is
therefore considered as a cylinder completely filled with emitting
plasma.  The small-scale structure seen on the optical images and the
0\farcs2 optical spectral index map \citep{Jes01} suggests that the
true internal structure of the jet is more complicated -- so
complicated that a more accurate model than the simple one assumed
here requires a detailed understanding of the internal structure,
composition and flow parameters governing the fluid dynamics of the
jet. However, any model with more free parameters than a filled
cylinder is not constrained by the available data. We therefore assume
that the jet is a cylinder extending along position angle 222\fdg2
whenever a value of the jet volume is required, and next determine the
appropriate value for the radius of this cylinder.

\subsubsection{Width of the jet}\label{s:res.im.width}

\begin{figure}
\resizebox{\hsize}{!}{\includegraphics{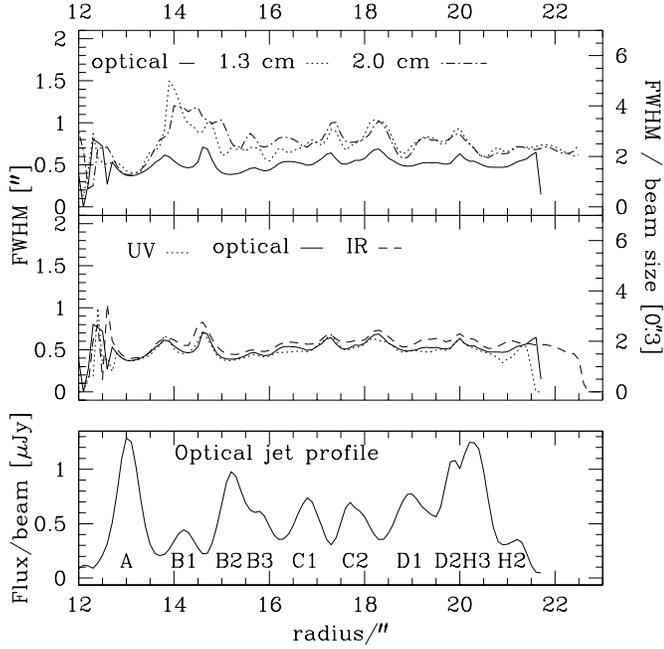}}
\caption{\label{f:res.im.width.fwhm}Comparison of jet full-width at
half the maximum intensity at different wavelengths.  The width is
determined column by column on the images in
Fig.\,\ref{f:res.ima.images} as the full width at half the maximum
intensity along the column. The bottom panel shows the optical
($\lambda$620\nm) flux profile (maximum intensity) for reference. The
middle panel shows the width at $\lambda$300\,nm (UV), $\lambda
620$\nm (optical) and $\lambda 1.6\,\mu$m (IR). The upper panel shows
again the optical width for reference and the width at the radio
wavelengths of $\lambda 1.3\,$cm and $\lambda 2.0\,$cm.  The
right-hand-side axis for the upper two panels expresses the observed
width in units of the effective resolution of 0\farcs3.  This plot
shows that although the jet is clearly wider in the radio than at HST
frequencies, the increasing isophotal width in the radio is mainly
caused by the increase in brightness, not by an actual widening, which
would be seen as an increase in the jet's FWHM.}
\end{figure}

Comparing the radio and the optical images
(Fig.\,\ref{f:res.ima.images}), it appears that the radio emission is
widening significantly towards the hot spot, while the optical
emission is of smaller and constant width.  However, the transverse
jet profiles (Fig.\,\ref{f:res.im.morph.cuts}) show that the radio and
optical width are, in fact, comparable throughout. The apparent
widening of the radio isophotes is predominantly due to the increasing
brightness, and hence and signal-to-noise ratio, as a larger part of
the point spread function's (PSF) wings is visible above the
background noise.  Finite-resolution isophotes should therefore not be
used to judge the widths of jets.

With sufficient signal-to-noise, the appropriate comparison would be
using a deconvolution. Here we instead give the run of the jet FWHM in
Fig.\,\ref{f:res.im.width.fwhm}.  As expected from
Fig.\,\ref{f:res.im.morph.cuts}, the radio jet does not widen in the
way suggested by its isophotal width, but the FWHM remains constant at
roughly 1\arcsec \citep[this was first noted by][]{jetII}.

One might hope to determine the true extent of the jet on the images
or radio maps with the highest available resolution. While this is
possible on the HST images for the entire jet, the resolution of
$\approx 0\farcs1$ for the optical and $\approx 0\farcs2$ for the
infrared is not reached by the VLA with sufficient signal-to-noise for
most part of the jet. This is a consequence of the large dynamic range
of over 30,000 between the fainter parts of the jet and the radio
core.  Therefore, the current radio data do not permit a comparison of
the jet width at the resolution reached by HST.  The dynamic range of
these radio images actually fell short of expectations for reasons we
are currently trying to understand. 

Since the optical jet has fairly sharp boundaries on the WFPC2
$R$-band image at original resolution, we use the average width of the
$3\sigma$ isophote of 0\farcs7 as the width of the jet channel
\citep[this is identical to the value reported by][]{Baheta95}.  Thus,
the jet channel is described as a cylinder of constant radius
$0\farcs7/2 = 0\farcs35$ for regions A1--D2/H3
($r=12\arcsec$--$21\arcsec$). This width agrees with the hot spot
diameter given by \citet{hs_II}, so to lowest order, we can extend the
cylindrical model to cover the entire jet out to the tip of the jet
H1.

As the cocoon emission (\cf Fig.\,\ref{f:res.im.morph.cuts}) is much
fainter than the jet's, its morphology is much more difficult to
establish. To assess any likely contribution of the cocoon emission to
the flux observed from the jet, the radio emission surrounding the
optical jet channel is described as a cylindrical shell with inner
radius 0\farcs35 (enclosing the optical jet channel without a gap) and
outer radius 0\farcs8, close to the isophotal width of the radio jet
there and roughly twice that of the optical jet.

The volume belonging to each photometry aperture (or pixel in
Fig.\,\ref{f:res.ima.images}), \ie the effective jet volume sampled by
each photometry aperture, is calculated by explicitly convolving the
model assumed for the jet (a filled cylinder of radius 0\farcs35) with
the observing beam of 0\farcs3 FWHM at the location of each aperture,
assuming that the symmetry axis of the cylinder lies along the radius
vector at position angle 222\fdg2. The obtained values are tabulated
in Tab.\,\ref{t:res.im.width.jetvol}.  We use them here to assess the
likely relative volume emissivity of the jet and the cocoon. The jet
volume sampled by each aperture will be used in the calculation of the
minimum-energy field in \S\ref{s:ana.equi}, where we also consider the
effect of the inclination of the jet to the line of sight.
\begin{table}
\caption{\label{t:res.im.width.jetvol} Effective volume of the jet
sampled by photometry apertures at distance $\delta_\mathrm{y}$ from
the symmetry axis of the model. \emph{Jet volume}, contribution from
the jet channel visible in the optical, assumed as filled cylinder
extending from $\delta_\mathrm{y}=0\arcsec$ to
$\delta_\mathrm{y}=0\farcs35$.  \emph{Cocoon volume}, contribution
from the cocoon, modelled as hollow cylinder wrapped extending from
$\delta_\mathrm{y}=0\farcs35$ to $\delta_\mathrm{y}=1\arcsec$ around
the jet channel. These values assume a fully side-on view.}
\centering
\begin{tabular}{ccc}
\hline\hline $\delta_\mathrm{y}$ & Jet volume & Cocoon
volume\\
\arcsec & $(h_{70}^{-1}\kpc)^3$ & $(h_{70}^{-1}\kpc)^3$\\
\hline 
0.0 &  1.3       & 1.8 \\
0.1 &  1.2      & 1.9 \\
0.2 &  .98      & 2.0 \\
0.3 &  .62      & 2.3 \\
0.4 &  .28      & 2.4 \\
0.5 &  .084     & 2.3 \\
0.6 &  .015     & 1.9 \\
0.7 & .0016     & 1.3 \\
0.8 & .0001     & .71 \\
0.9 & .00003    & .26 \\
1.0 & .0000007  & .06 \\
\hline\hline
\end{tabular}
\end{table}

Using Tab.\,\ref{t:res.im.width.jetvol}, we can now also estimate the
contribution of the cocoon or backflow material along the line of
sight to the central part of the jet. The effective volume contributed
by the cocoon volume is much larger than that of the jet
channel. Therefore, the cocoon's volume emissivity can be no more
than about 1\% of the jet's volume emissivity (the exact ratio depends
on the azimuthal extent of the cocoon). Otherwise, the cocoon
emission would completely dominate the jet emission and the profile
would not appear centrally peaked, or fall off more slowly than
observed. In the central part of the jet, the cocoon will then also
contribute only about 1\% of the jet emission.  The same constraint
from the observed brightness profile implies that the contribution of
the cocoon to the central jet flux cannot be appreciable even if the
true width of the cocoon is different from the assumed 0\farcs8.

\subsubsection{Morphology summary} \label{s:res.im.morphsum}

In summary, the overall morphology of the jet is similar at all
observed wavelengths from 3.6\,cm to 300\,nm.  The exceptions to this
are the radio-quiet ``extensions'' \citep[In1, In2, S;
Fig.\ref{f:obs.nir.IRmap} and][]{Jes01} to the optical jet, and region
B1, in which the SED of the southern strand of emission peaks in the
radio, but that of the northern strand peaks at optical wavelengths.
All of the emission blueward of 1.6\micron\ and the bulk of the radio
emission at the wavelengths of 3.6\cm{} and shorter considered here
are emission from the jet itself, while the steep-spectrum radio
cocoon (or ``backflow'') south of the jet makes a negligible
contribution along the line of sight to the jet channel
\citep[\cf][]{jetII,jetIII}.  We therefore use the full data set from
3.6\cm{} to 300\nm{} to analyse the spectrum of the jet emission here.

\subsection{Spectral indices}\label{s:res.alpha}

\begin{figure*}
\parbox[c]{0.1\hsize}{\hspace*{0.1\hsize}}\parbox[c]{0.9\hsize}{\includegraphics[width=0.9\hsize,clip]{all-knot-labels.eps}}
\parbox[c]{.1\hsize}{\hspace*{.1\hsize}}\parbox[c]{0.9\hsize}{\includegraphics[angle=270,width=0.9\hsize,clip]{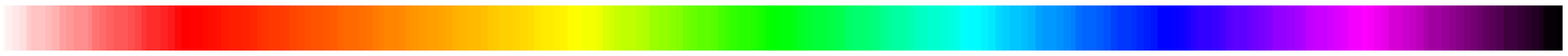}}
\parbox[c]{.1\hsize}{a)\\ $-2$\ldots0\\\\\aopt\\\\ $\frac{300\,\mathrm{nm}}{620\,\mathrm{nm}}$}\parbox[c]{0.9\hsize}{\includegraphics[width=0.9\hsize,clip]{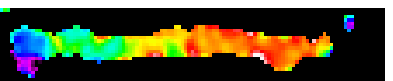}}\\
\parbox[c]{.1\hsize}{\hspace*{.1\hsize}}\parbox[c]{0.9\hsize}{\includegraphics[angle=270,width=0.9\hsize,clip]{colorbar-rain2.eps}}
\parbox[c]{.1\hsize}{b)\\ $-2.5$\ldots\\\ldots$-0.4$\\\\\aIRo\\\\$\frac{620\,\mathrm{nm}}{1.6\,\mu\mathrm{m}}$}\parbox[c]{0.9\hsize}{\includegraphics[width=0.9\hsize,clip]{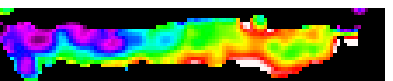}}\\
\parbox[c]{.1\hsize}{\hspace*{.1\hsize}}\parbox[c]{0.9\hsize}{\includegraphics[angle=270,width=0.9\hsize,clip]{colorbar-rain2.eps}}
\parbox[c]{.1\hsize}{c)\\ $-1.35$\ldots\\\ldots$-0.8$\\\\$\alpha^\mathrm{IR}_\mathrm{1.3}$\\\\ $\frac{1.6\,\mu\mathrm{m}}{1.3\,\mathrm{cm}}$}\parbox[c]{0.9\hsize}{\includegraphics[width=0.9\hsize,clip]{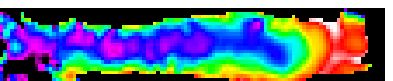}}\\
\parbox[c]{.1\hsize}{\hspace*{.1\hsize}}\parbox[c]{0.9\hsize}{\includegraphics[angle=270,width=0.9\hsize,clip]{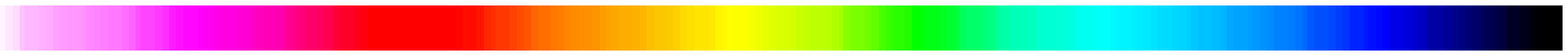}}
\parbox[c]{.1\hsize}{d)\\ $-2$\ldots0\\\\$\alpha^\mathrm{1.3}_\mathrm{2.0}$\\\\ $\frac{1.3\,\mathrm{cm}}{2.0\,\mathrm{cm}}$}\parbox[c]{0.9\hsize}{\includegraphics[width=0.9\hsize,clip]{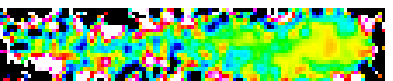}}\\
\parbox[c]{.1\hsize}{\hspace*{.1\hsize}}\parbox[c]{0.9\hsize}{\includegraphics[angle=270,width=0.9\hsize,clip]{colorbar-rain3.eps}}
\parbox[c]{.1\hsize}{e)\\ $-2$\ldots0\\\\$\alpha^\mathrm{2.0}_\mathrm{3.6}$\\\\ $\frac{2.0\,\mathrm{cm}}{3.6\,\mathrm{cm}}$}\parbox[c]{0.9\hsize}{\includegraphics[width=0.9\hsize,clip]{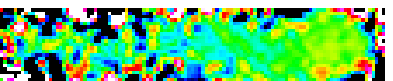}}
\parbox[c]{.1\hsize}{radius/\arcsec}\parbox[c]{0.9\hsize}{\includegraphics[width=0.9\hsize,clip]{skala.eps}} 
\caption{\label{f:res.alpha.alpha} Spectral index maps at 0\farcs3
resolution generated from the photometry data in
Fig.\,\ref{f:res.ima.images}.  Only pixels with a signal-to-noise
ratio of at least 5 per beam are shown. Images are combined pairwise
in order of increasing wavelength. Linear colour scales (shown above
the respective images) have been chosen to stress variations within
one map.\newline a, optical spectral index (range: $-2$\ldots$0$); b,
optical-infrared ($-2.5$\ldots$-0.4$); c, infrared-radio
($-1.35$\ldots$-0.8$); d, radio $\lambda 1.3\,$cm-$\lambda 2.0\,$cm
($-2$\ldots$0$); e, radio $\lambda 2.0\,$cm-$\lambda 3.6\,$cm
($-2$\ldots$0$)\newline The variations of both radio spectral indices
in the inner part of the jet are mainly due to low signal-to-noise and
the associated imaging uncertainties.  Compare with
Fig.~\ref{f:res.alpha.run} to gauge the relative magnitude of
variations of the different spectral indices.}
\end{figure*}
\begin{figure*}
% turn sidecaption off for astro-ph 
% \sidecaption
\begin{center}
 \includegraphics[origin=rB,angle=270,width=12cm,clip]{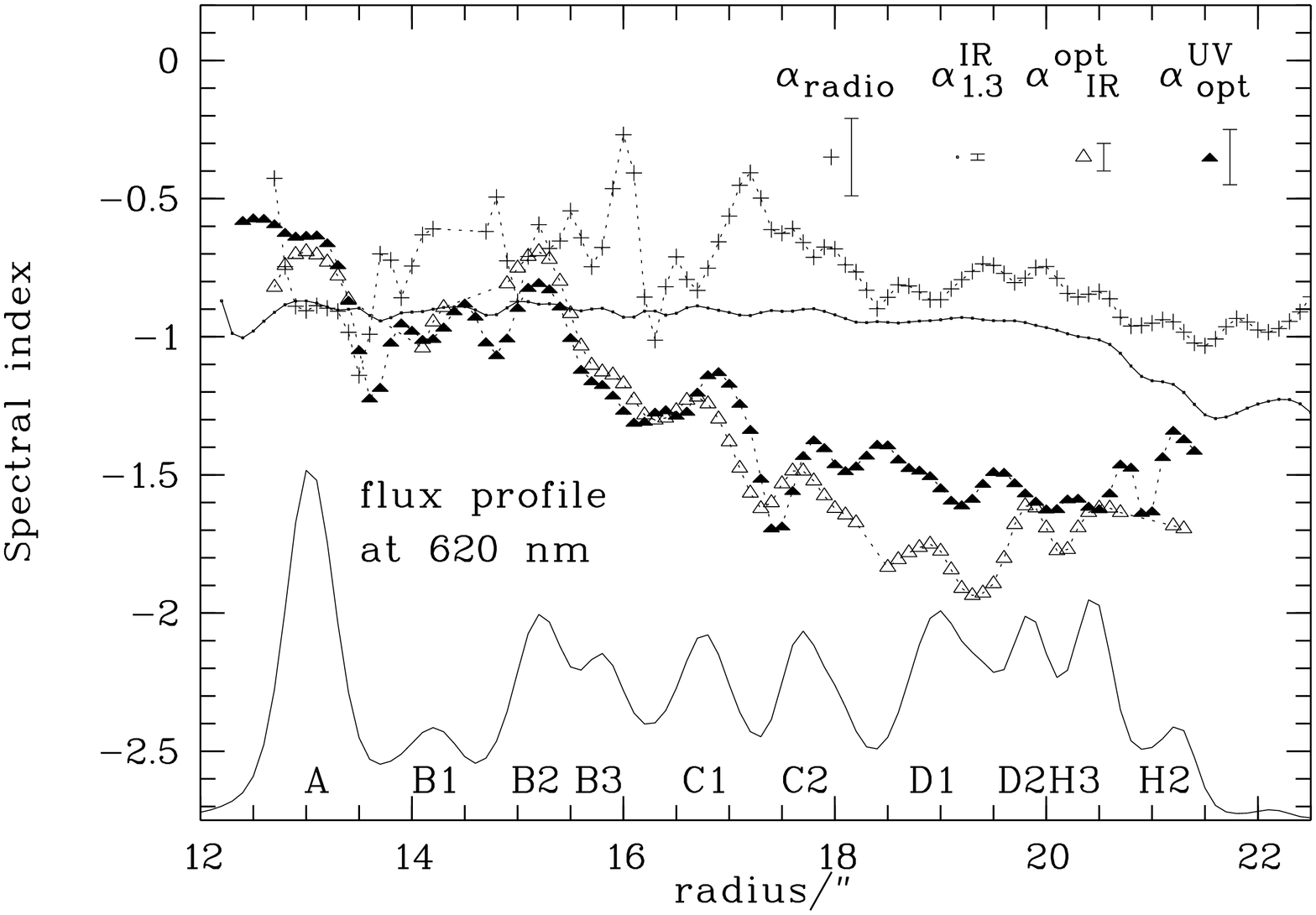}
\end{center}
 \caption{\label{f:res.alpha.run}Run of the spectral indices along the
 jet at 0\farcs3 resolution, sampled in 0\farcs1 intervals (cut along
 radius vector at position angle 222\fdg2).  For sake of clarity, only
 typical $2\sigma$ error bars are shown for the random error.
 Systematic flux calibration uncertainties are of the same order and
 would shift an entire curve.  The radio spectral index
 $\alpha_\mathrm{radio}$ is obtained by a fit to the radio data at
 3.6\cm, 2.0\cm, and 1.3\cm.  The other spectral indices are derived
 from the jet photometry at the given wavelengths
 ($\alpha_{1.3}^{\mathrm{IR}}$: 1.3\cm\ and 1.6\micron, \aIRo:
 1.6\micron\ and 620\nm, \aopt: 620\nm\ and 300\nm).  The optical flux
 profile is shown for reference.  Reprinted from \protect\citet{Jes02} for
 reference.}
\end{figure*}

From the photometry results shown in Fig.\,\ref{f:res.ima.images}, we
compute pairwise spectral indices $\alpha$. The spectral index maps
are shown in Fig.\,\ref{f:res.alpha.alpha}.  Their run along the jet's
centre line is shown in Fig.\,\ref{f:res.alpha.run} (Fig.\,1 in
\citealt{Jes02}). The error bars in Fig.\,\ref{f:res.alpha.run} have
been calculated from the total random error.  There is an additional
systematic error from flux calibration uncertainties (typically 2\%,
\S\ref{s:photo}). These change all flux measurements through one
filter by the same factor.  Its effect on the spectral index
determination is to offset a given spectral index by a constant amount
for the entire jet. The magnitude of the effect is smaller than any of
features which we detect at high statistical significance. Our
conclusions are, therefore, unaffected by this residual systematic
uncertainty.

The two radio spectral indices (between 3.6\cm\ and 2\cm, and between
2\cm\ and 1.3\cm) behave erratically out to a radius of about
19\arcsec. These variations are not significant: the radio jet is
detected at low signal-to-noise ratio at the inner end, and the
deconvolution involved in the reconstruction of the brightness
distribution on the sky from the observed interferometric data is only
accurate (in the sense of achieving an image representing the true
brightness distribution) for high signal-to-noise. We therefore show
an overall radio spectral index in Fig.\,\ref{f:res.alpha.run} which
has been determined by a least-squares straight-line fit to the three
radio data points.  For the outer part of the jet, the radio spectral
index shows a steepening of the hot spot (H2) spectrum compared to the
remainder of the jet. The run of the spectral index between 6\cm\ and
3.6\cm, for which only lower-resolution data at 0\farcs5 are available
agrees with the spectral index run determined at the wavelengths
considered here \citep[\emph{cf.} ][]{jetII}.

The infrared-radio spectral index (Fig.\,\ref{f:res.alpha.alpha}c) is
nearly constant at $\arIR \approx -0.9$ along the centre of the
optical jet, with some flattening in optically bright regions and a
pronounced steepening in the transition from D2
($r\approx20\arcsec$) to the radio hot spot H2. It steepens markedly
to $\approx -1.2$ away from the centre line. These features are
identical to those identified by \citet{NMR97} on a spectral index map
at 1\farcs3 resolution generated from observations at 73\cm\ and
2.1\micron. There is no spectral index feature uniquely corresponding
to the hot spot H2, as is the case on the radio spectral index map. On
the other hand, the tip of the jet H1 is identifiable as region with
radio-infrared spectral index slightly flatter than H2.

We use the term ``high-frequency'' to refer to the infrared-optical
and optical-ultraviolet spectral indices.  As already noted by
considering the spectral index map derived from the optical and
near-ultraviolet imaging at 0\farcs2 resolution \citep{Jes01}, there
are none but smooth changes in the optical-UV spectral index along the
jet.  The same is true for the infrared-optical spectral index. Both
these spectral indices decline globally along the jet.  At the onset
of the optical jet at A and at B1, both have a value of $-0.7$, \ie
the optical-UV spectrum there is flatter than the radio and
radio-infrared.  Both high-frequency indices decrease to about $-1.7$
at C2.  The optical-ultraviolet \aopt\ remains near this value for the
reminder of the jet, while the infrared-optical \aIRo\ steepens
further, reaching a minimum near $-2$ between D1 and D2, and
flattening back to $-1.7$ at D2.

Thus, as already reported in \citet{Jes02}, the spectrum does not
steepen everywhere towards higher frequencies, but flattens between the
near-infrared and optical in nearly all parts of the jet. We will
consider the implications of this finding in \S\S\ref{s:ana} and
\ref{s:disc.pop}.

\begin{figure}
\resizebox{\hsize}{!}{\includegraphics{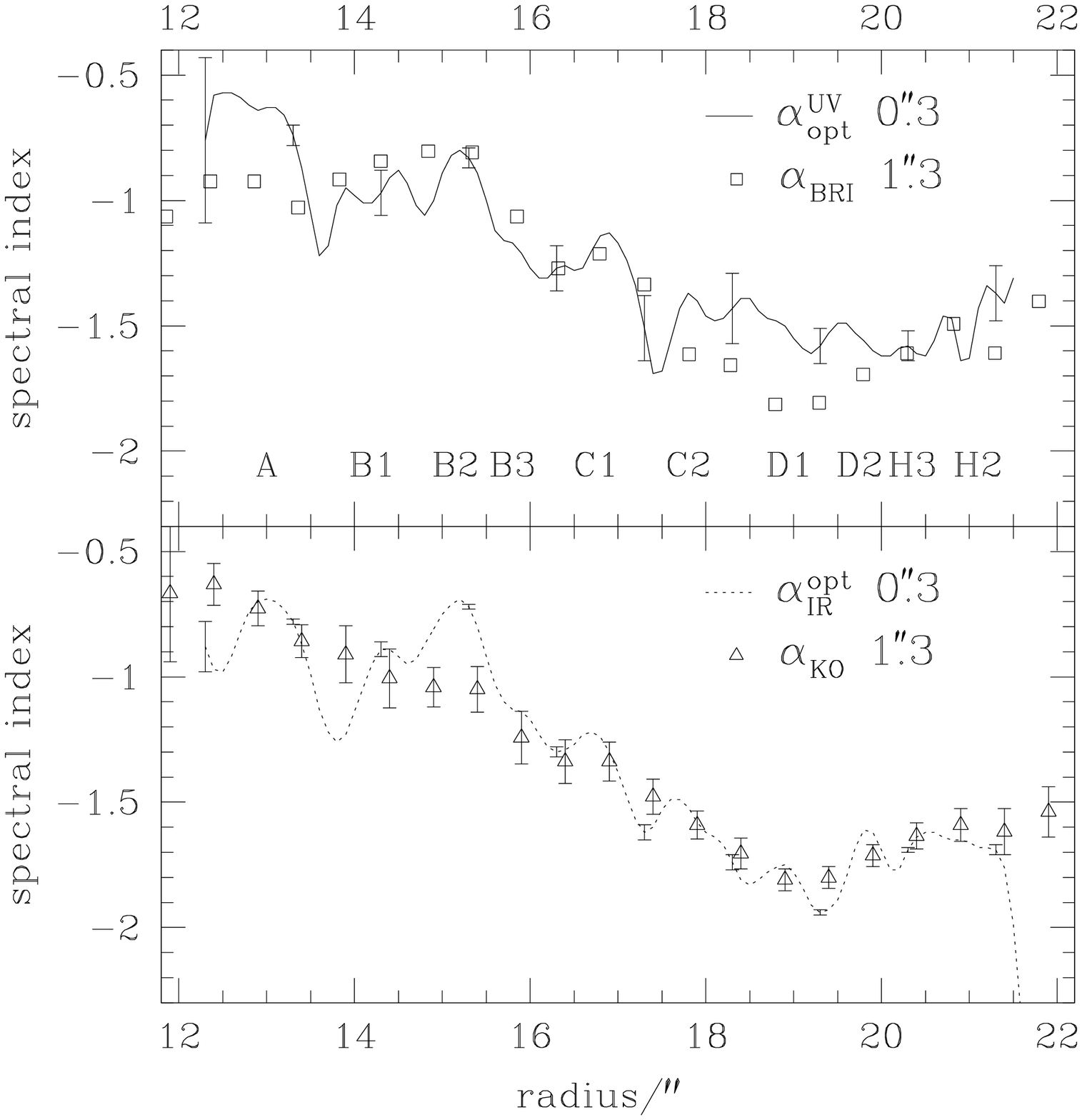}}
\caption{\label{f:res.alpha.compare}Comparison of high-frequency
spectral indices with previous observations at 1\farcs3
resolution. Top, comparison of optical spectral index $\alpha_{BRI}$
at 1\farcs3 resolution from \protect\citet{RM91} with optical-UV spectral
index \aopt\ at 0\farcs3 from this work (error bars are similar for
\aBRI\ and \aopt). Below, comparison of infrared-optical spectral
index $\alpha_\mathrm{KO}$ from \protect\citet{NMR97} and \aIRo\ from this
work (only every tenth error bar shown).  The overall agreement
between the spectral indices at 0\farcs3 and 1\farcs3 is surprisingly
good.  Differences between \aopt\ and \aBRI\ occur in A, C2, D1, and
D2, those regions in which the observed spectrum
(Fig.\ref{f:res.alpha.run}) does not show the expected steepening
towards higher frequencies.}
\end{figure}
The general outward steepening of the high-frequency spectral indices
is in agreement with previous determinations of the knots' synchrotron
spectrum which showed a decrease of the cutoff frequency outward
\citep{MNR96,Roe00}.  There is an excellent correspondence of the run
of the optical-infrared spectral index \aIRo\ along the jet with the
spectral index $\alpha_\mathrm{KO}$ between $2.1\,\mu$m and the
optical as determined by \citet{NMR97} at 1\farcs3 resolution
(Fig.\,\ref{f:res.alpha.compare}).  As has been noted in
\citet{Jes01}, the overall run of the optical-ultraviolet spectral
index \aopt\ at 0\farcs3 agrees overall with the run of the optical
spectral index \aBRI\ at 1\farcs3, with discrepancies in a few
regiosn: a comparison of Figs.\,\ref{f:res.alpha.run} and
\ref{f:res.alpha.compare} shows that these discrepancies arise
precisely in those regions in which the spectrum flattens at high
frequencies. The discrepancies can thus be ascribed to the different
wavelength of the bluest HST and ground-based images (300\,nm compared
to 400\,nm) and the fact that the flattening occurs just in this
wavelength region (between 600 to 300 nm).

Optical spectral index variations are not strongly correlated with
brightness variations.  Some local peaks of \aopt\ coincide with
brightness maxima, while others coincide with minima.  In contrast,
there is a correlation between the optical-infrared spectral index
\aIRo\ and the jet's surface brightness, in the sense that brighter
regions show a flatter spectrum (smaller $|\aIRo|$, see
Figs.\,\ref{f:res.alpha.alpha} b and \ref{f:res.alpha.run}). This
correlation is most clearly visible on the spectral index map for the
outer end of the jet. The bright regions D2/H3 together with the hot
spot H2 appear as an island of $\aIRo \approx -1.6$ surrounded by
regions with steeper spectrum.  The correlation between local maxima
in surface brightness and \aIRo\ is also present in the inner part of
the jet, although the spectral index maxima are displaced sideways
from the brightness peaks due to a transverse spectral index gradient.

This spectral index gradient is suggestive of a residual misalignment between
the optical and infrared images, \eg, a rotation between the two about
a point close to D2/H3. It could also have been caused by an
overestimation of the diffraction spike signal which has been modelled
and subtracted (see \S\ref{s:obs.nir.IRmap}).  Since \citet{NMR97} did
not detect a significant change of the infrared-optical spectral index
transversely to the jet at 1\farcs3, and although the alignment
procedures described above (\S\ref{s:photo}) should have ensured that
such an error should not have occurred, we reconsidered this
possibility to avoid the introduction of spurious gradients.  After a
detailed investigation \citep[details are contained in][]{JesterDiss},
we concluded that the misalignment necessary to produce such a
gradient was far greater than compatible with the alignment precision
established previously. Neither can the gradient firmly be linked to
the diffraction spike subtraction or any obviously detectable
misalignment.  In the given situation, we rely on the data with the
offsets established to the best of our knowledge. The clarification of
this matter has to await new observational data.

In summary, there are two surprising findings regarding the spectral
indices. Firstly, the knots, \ie the local brightness peaks occurring
at nearly the same position at all wavelengths, have a flatter
infrared-optical spectrum than the regions separating them.  Thus,
there is a local positive correlation between the jet brightness at
any wavelength and the infrared-optical spectral index. This contrasts
with the global anti-correlation between energy output and spectral
index: the radio surface brightness increases while the high-frequency
spectrum steepens considerably.  Secondly, the spectrum flattens in
the optical-UV wavelength region. In fact, the optical-ultraviolet
\aopt\ spectral index is \emph{nowhere} significantly below the
infrared-optical \aIRo, in contrast to the expectation of a
synchrotron spectrum which steepens at higher frequencies.  We will
consider the implications in \S\ref{s:disc.corr} and first turn to the
determination of the maximum particle energy from the observations.

\section{Analysis}\label{s:ana}

The present observations have been obtained to study the behaviour of
the maximum particle energy along the jet in 3C\,273, with the aim of
identifying acceleration and/or loss sites within the jet.  In order
to determine the maximum particle energy from the observed spectral
energy distributions, we need to determine both the cutoff frequency
\nuc\, which is the characteristic synchrotron frequency corresponding
to the highest particle energy, and the magnetic field in the jet,
assumed to be uniform along lines of sight.  We first consider the
determination of the cutoff frequency and then estimate the magnetic
field strength using the minimum-energy argument in
\S\ref{s:ana.equi}.

\subsection{Spectral fits: determination of \nuc}\label{s:ana.fits}

\begin{figure}
\resizebox{\hsize}{!}{\includegraphics{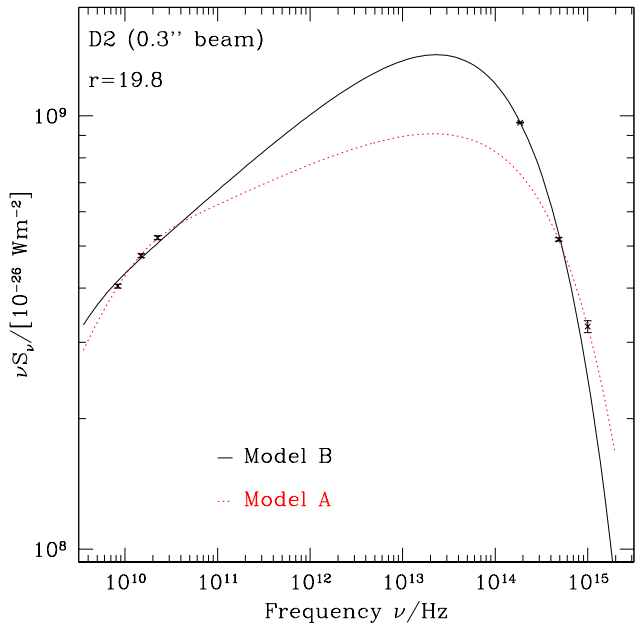}}
\caption{\label{f:ana.fits.fitab}Illustration of the two different
spectral fits performed. Model A assumes infrared emission in excess
of the cutoff given by the optical-ultraviolet spectral index, while
Model B assumes an ultraviolet excess above the infrared-optical
cutoff.}
\end{figure}

The cutoff frequency is to be determined from the observed spectral
energy distributions by fitting them with model synchrotron spectra.
All previous studies have used single-population models to describe
the jet's synchrotron spectrum \citep{Roe00,MNR96,NMR97,hs_II}.  We
noted above that there is a flattening of the observed spectrum
towards the ultraviolet (Fig.\,\ref{f:res.alpha.run} in
\S\ref{s:res.alpha}).  This means that a description using a single
electron population is, in fact, inadequate: any non-idealised
power-law electron population (\emph{i.e.}, with finite maximum
particle energy and rapid pitch-angle scattering) always gives rise to
a synchrotron spectrum with steepening slope in the $\log
B_\nu$--$\log \nu$ plane.  The observed high-frequency flattening
implies that a second high-frequency emission component must be
present, which contributes either predominantly to the jet's
near-infrared or near-ultraviolet flux to produce the observed
high-frequency flattening.  The discrepancy can be explained by
assuming that one of the two spectral indices reflects the true
synchrotron spectrum, while the other is contaminated by flux not due
to the same population as the remainder of the jet.  To assess the
likely reason for this discrepancy between the observations and the
expectations from synchrotron theory, we perform two separate fits
which differ in the determination of the cutoff frequency
(Fig.\,\ref{f:ana.fits.fitab}): either the cutoff is described by the
optical-ultraviolet spectral index leading to an infrared excess
(Model A), or conversely, the true cutoff is described by the
infrared-optical spectrum and there is additional flux in the
ultraviolet \citep[Model B; see][]{Jes02}.  This allows us to perform
the fits using a single-population model.

Following previous studies, we use the method of computing synchrotron
spectra with a smooth cutoff from \citet{HM87} to determine the cutoff
frequency from the observed spectral energy distribution.  They model
the synchrotron source as a region of constant magnetic field into
which a power-law distribution of electrons extending up to a maximum
electron Lorentz factor \gmax\ is continuously injected.  The
resulting spectrum has a low-frequency power law part with spectral
index $\alpha_\mathrm{low}$, which steepens by one-half power at a
break frequency \nub\ and cuts off exponentially above the cutoff
frequency $\nuc>\nub$.  The break is produced by adding up the
contributions from the electron population observed at increasing times
since acceleration, \ie with different cutoff frequencies.  The
magnitude of the break of $1/2$ is fixed by the cooling mechanism.
Although the model was originally devised to describe the spectra of
hot spots, the use of such a continuous-injection model is justified
by the fact that optically emitting electrons must be accelerated
within the jet: our calculation in \citet{Jes01} showed that
relativistic beaming and/or sub-equipartition magnetic fields cannot
remove the discrepancy between light-travel time along 3C\,273's jet
and the lifetime of electrons emitting optical synchrotron radiation.

The model spectrum has four free parameters: the low-frequency
spectral index $\alpha_\mathrm{low}$, the ratio of cutoff energy to
break energy of the emitting electron population = $\sqrt{\nuc/\nub}$,
the observed cutoff frequency \nuc, and a flux normalisation.  Since
we only observe the spectrum at six frequencies, we introduce
additional constraints to obtain meaningful fits. First, we restrict
$\sqrt{\nuc/\nub}$ so that the break is in the range
$10^{9}$\,Hz--$10^{12}$\,Hz, \ie within the range of the observed
radio data. Secondly, we artificially fix $\alpha_\mathrm{low} \approx
-0.4$, so that the observed radio-infrared spectral index of about
$-0.9$ (Fig.\,\ref{f:res.alpha.run}) corresponds to
$\alpha_\mathrm{low} -0.5$.  In effect, only the cutoff frequency
\nuc\ is determined by this fitting procedure. The fit is performed
using a $\chi^2$ minimisation technique; the detailed steps can be
found in \citet[ \S4.2 and references therein]{JesterDiss}.  We
discuss the implications of these constraints on the interpretation of
the fit results in \S\ref{s:ana.assume} below.

\subsubsection{Model A and Model B}\label{s:ana.fits.modelAB}

Model A assigns a low weight to the near-infrared flux point in the
data set. This is motivated by the indication that the radio cocoon
(possibly a ``backflow'') around the jet \citep{jetIV} may also be
detectable at 2.1\micron\ \citep{NMR97}, suggesting that the flux from
the jet at 1.6\micron\ may be contaminated by emission from the
cocoon as well.  Hence, the cutoff in Model A is determined by the
optical and near-ultraviolet points at 620\nm\ and 300\nm,
respectively.  Conversely, in Model B, the location of the cutoff is
dominated by the infrared and optical points at 1.6\micron\ and
620\nm, respectively.

In contrast to the remainder of the jet, the hot spot shows an offset
between optical and radio hot spot position
(Fig.\,\ref{f:res.im.morph.HS}) and a change of the radio-infrared
spectral index (\S\ref{s:res.alpha}). Therefore, the spectra from the
hot spot regions H2 and H1 at radii beyond $r=21\arcsec$ are fitted
differently by allowing $\alpha_\mathrm{low}$ to vary between $-0.8$
and $-0.3$. This model is referred to as Model HS.

In the bright peaks of A, B1 and B2, there is no cutoff within the
observed frequency range (the high-frequency spectral indices are
flatter than $-1$ there (Fig.\,\ref{f:res.alpha.run}), so that the
spectral energy density continues to rise towards higher frequencies).
Where this is the case, an artificial high-frequency data point is
introduced to allow the fit to proceed, which assumes the presence of
a local maximum in the data set. The artificial point is chosen so
that it has a spectral index relative to the observed UV point of
$-1.2$ and is introduced at a frequency $10^{18}$\,Hz, 1000 times
higher than the frequency of UV emission. It is assigned a large error
so that it does not influence the goodness-of-fit. The results
obtained for \nuc\ for these points are then only lower limits to the
true cutoff frequency as it would be inferred from observations at
higher frequencies.  On the other hand, it is possible that the
optical/UV emission in these regions has a substantial contribution
from the second, high-energy emission component (\S\ref{s:disc.pop}).
In this case, there may be a cutoff to the ``low-energy''
radio-optical synchrotron component at \emph{lower} energies than
obtained here.  Observations in additional wavebands, in particular at
longer infrared wavelengths, or a comparison of the radio and optical
polarisation of these regions at high resolution may shed light on
this issue.  However, given the similarity of radio and optical
polarisation at low resolution, it appears unlikely that the emission
is \emph{dominated} by the high-energy component, although its
contribution may be significant.

\subsubsection{Fit results}\label{s:ana.fits.shape}

\begin{figure}
\resizebox{\hsize}{!}{\includegraphics{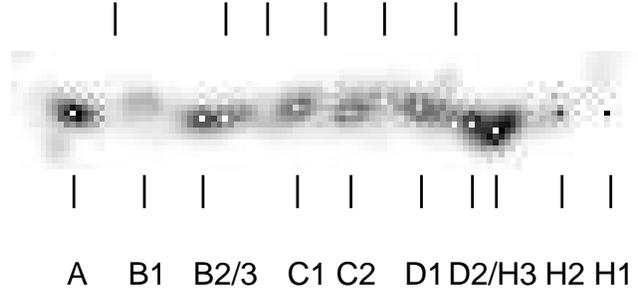}}
\resizebox{\hsize}{!}{\includegraphics{all-knot-labels.eps}}
\caption{\label{f:ana.fits.spec-loc}Location of apertures
for which spectra are shown in Fig.\,\ref{f:ana.fits.obs1}.  Locations
at local peaks (``knots'') are marked by a line below the image, while
locations in inter-knot regions are marked by a line above.}
\end{figure}
\begin{figure*}
\includegraphics[width=.4\hsize]{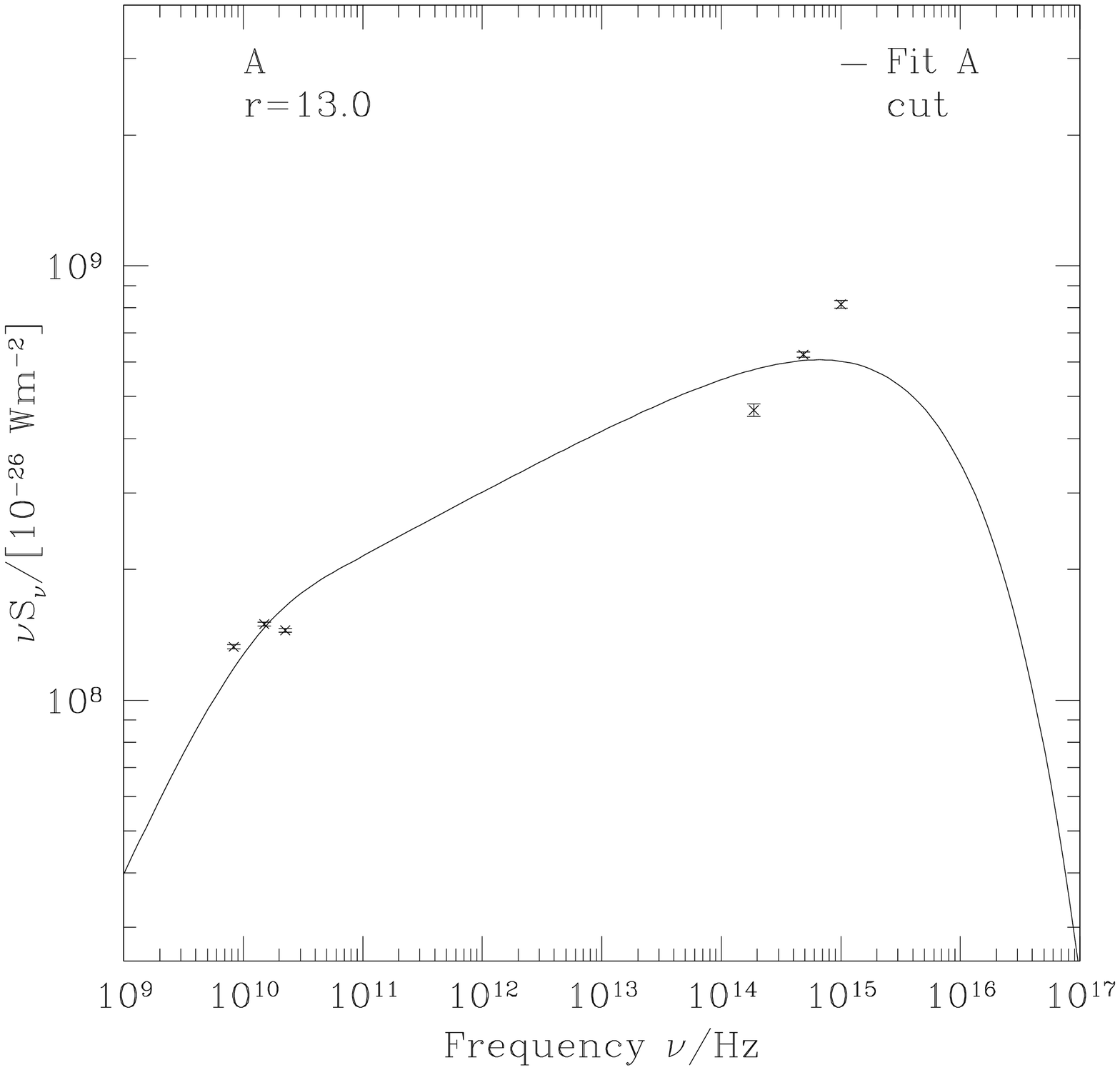}
\includegraphics[width=.4\hsize]{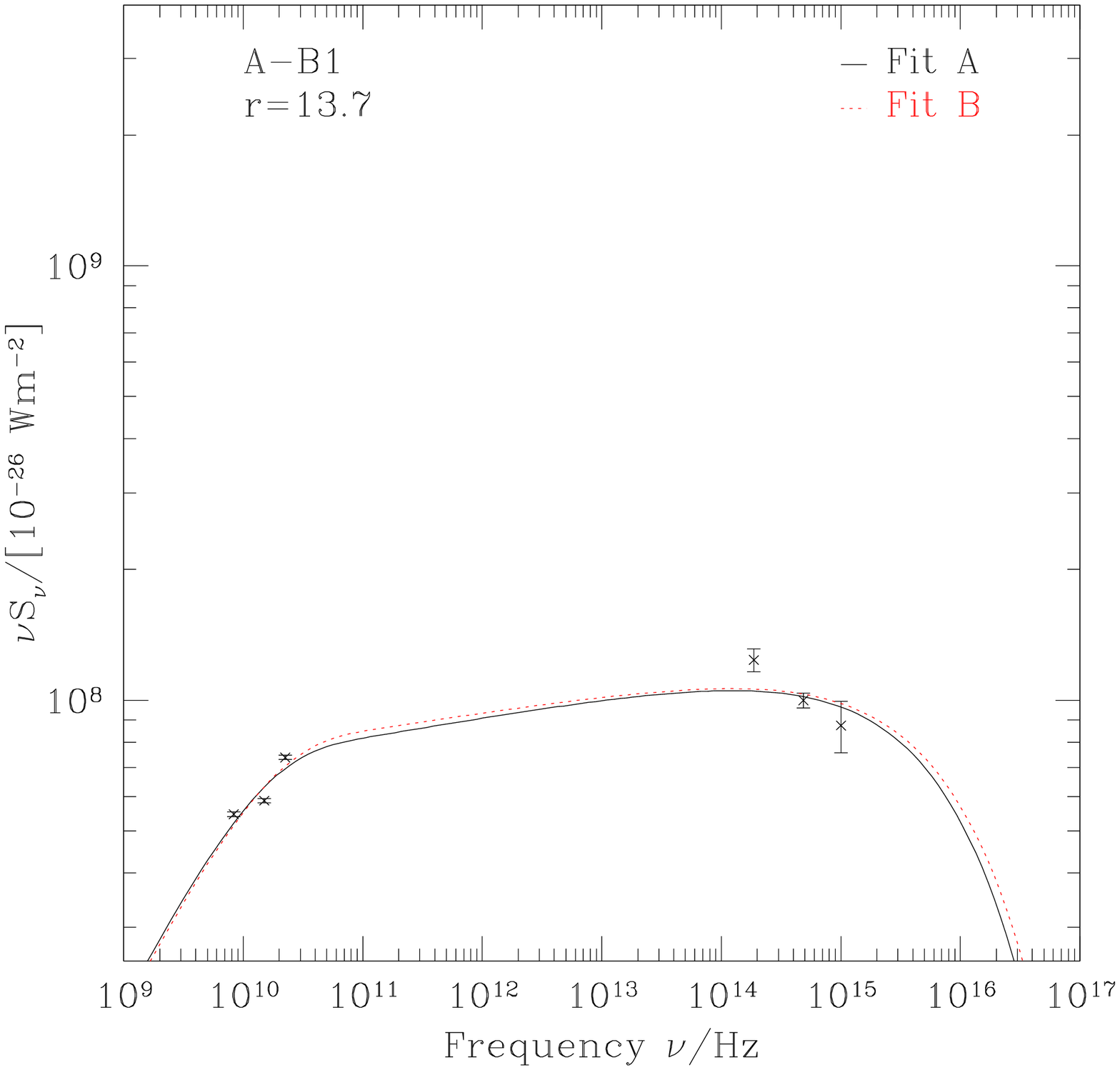}
\includegraphics[width=.4\hsize]{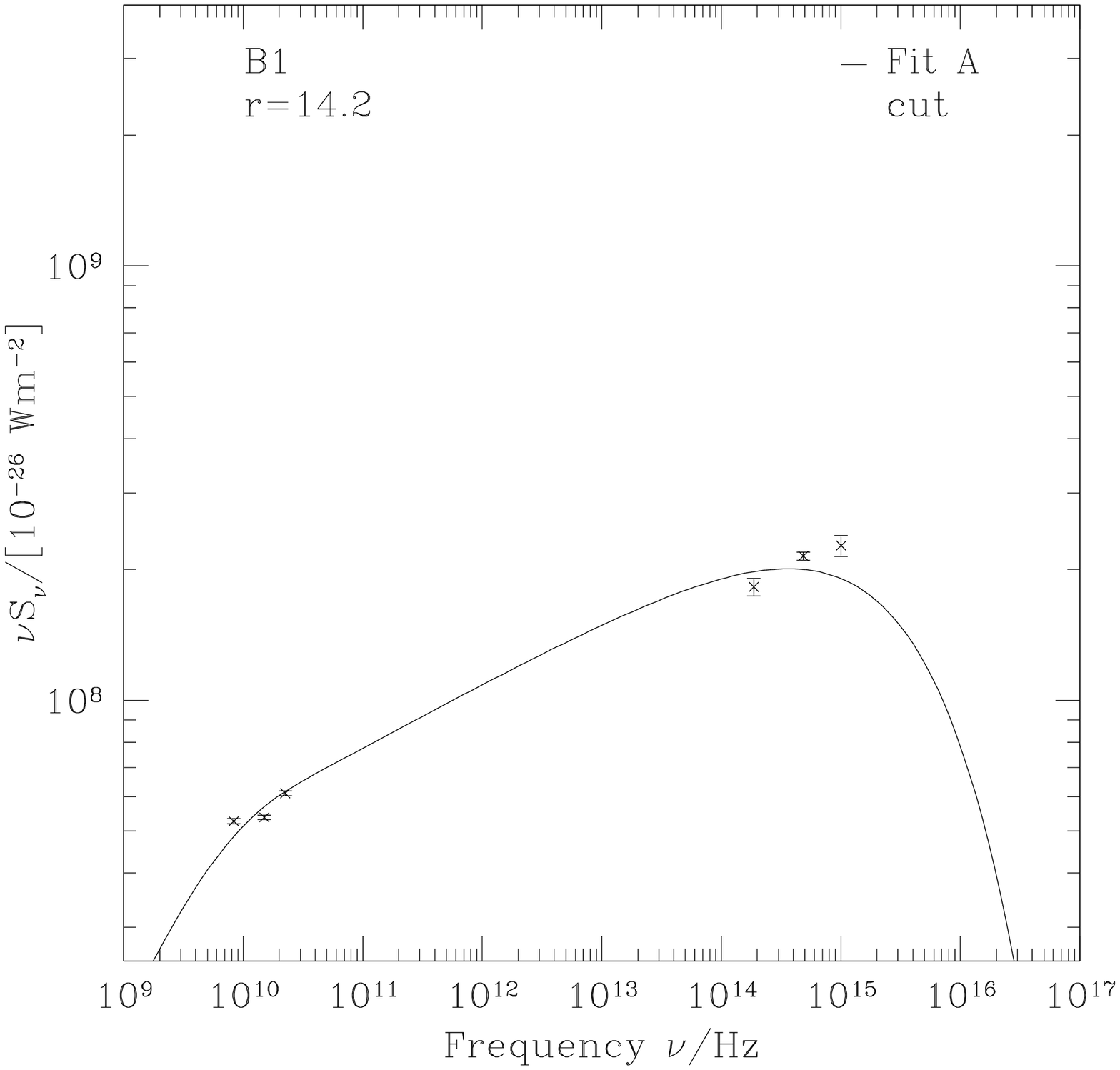}
\includegraphics[width=.4\hsize]{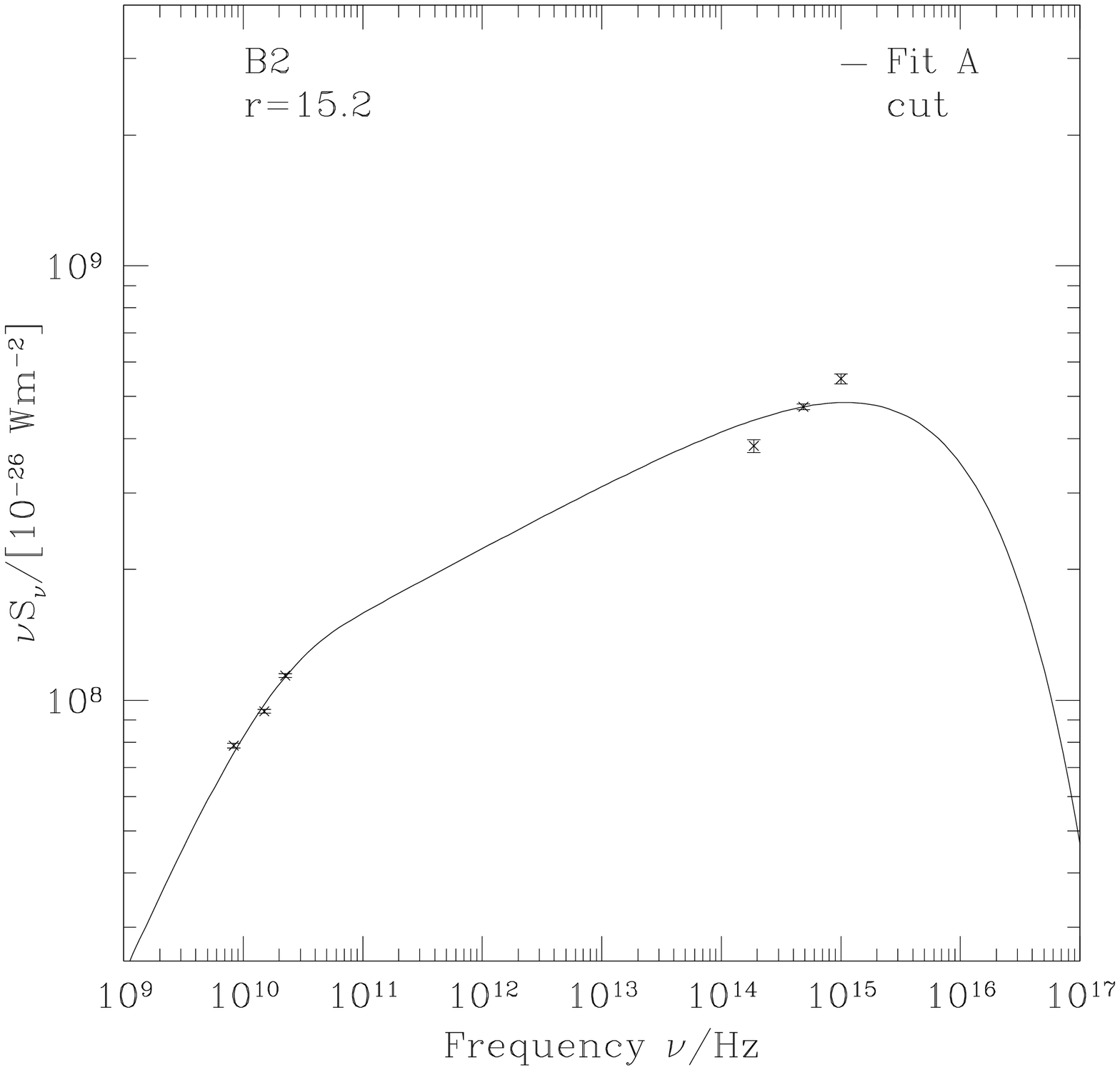}
\includegraphics[width=.4\hsize]{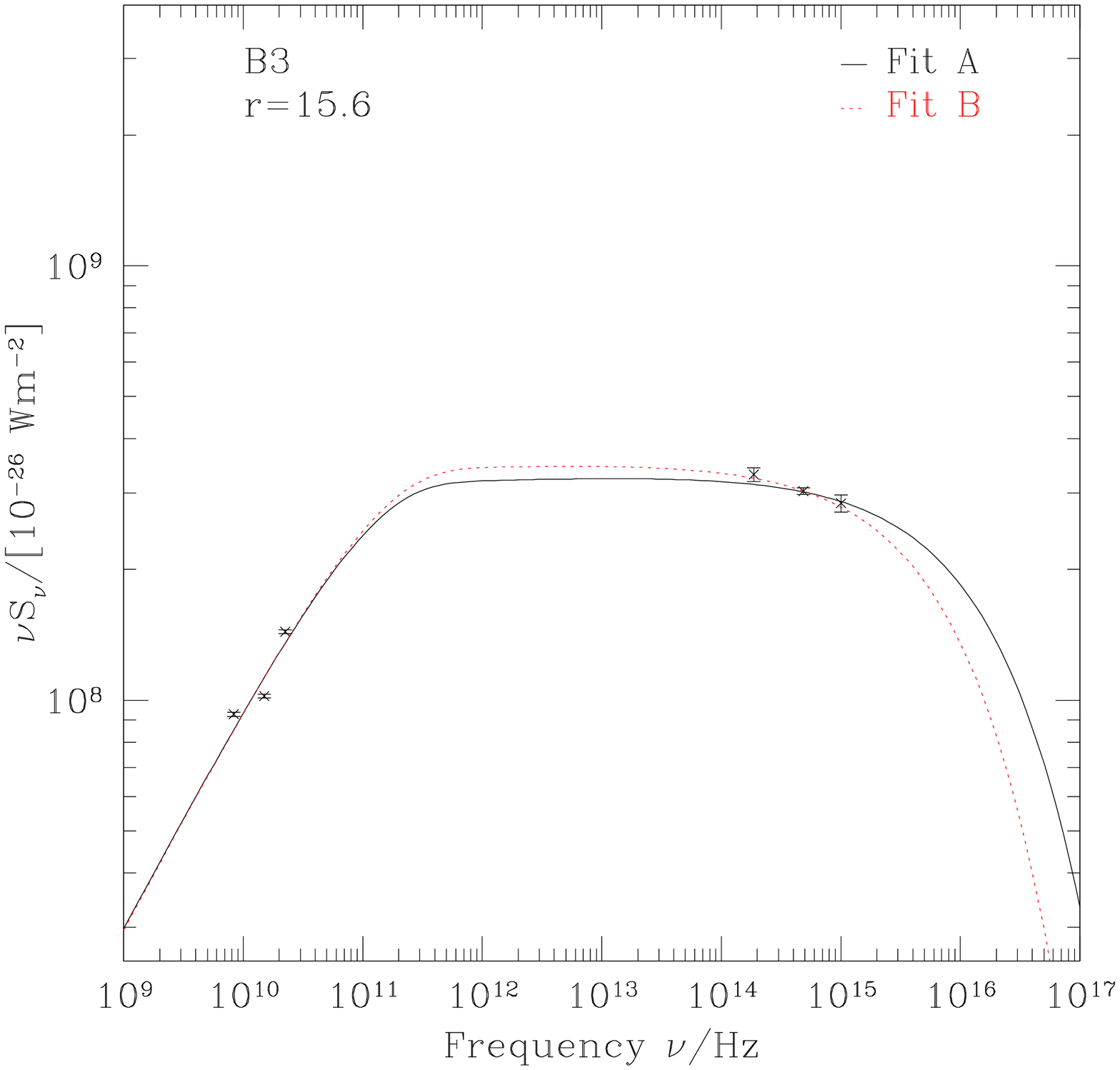}
\includegraphics[width=.4\hsize]{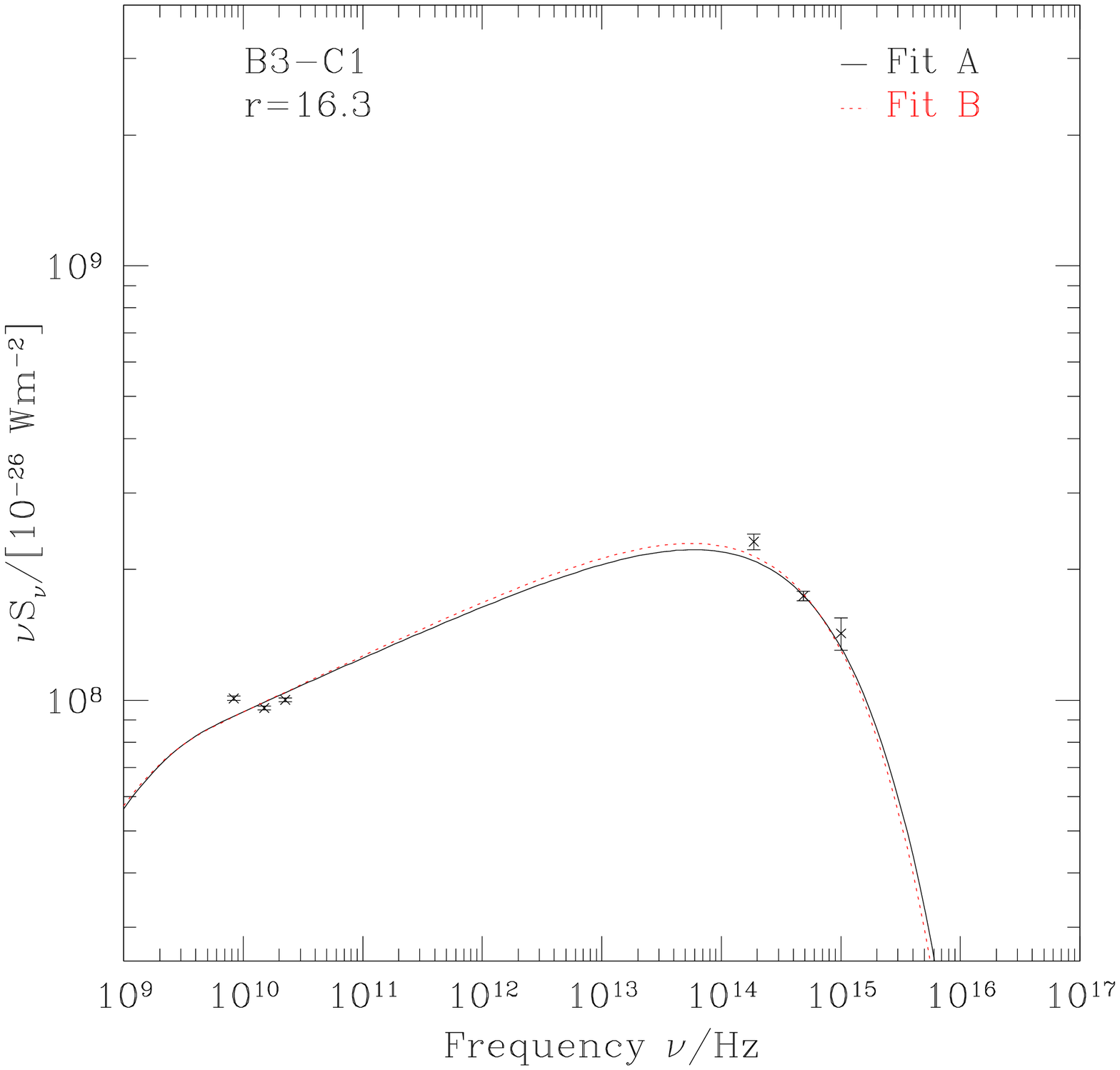} \centering
\caption[Synchrotron spectral fits]{\label{f:ana.fits.obs1}Observed
data points with fitted spectra for the points shown in
Fig.\,\ref{f:ana.fits.spec-loc}. To account for the observed
flattening of the spectrum towards the ultraviolet, Model A assumes a
contamination in the infrared, so that the cutoff is determined by the
optical-ultraviolet spectral index there, while the cutoff in Model B
is determined by the infrared-optical spectral index.  Those spectra
which require an artificial high-frequency data point to obtain a fit
result are labelled ``cut''; for these, Model A and Model B are
identical, and the artificially obtained value for \nuc\ is a lower
limit to the actual value.  The spectrum may extend up to X-rays in
those locations where an artificial cut is necessary
\protect\citep[\emph{cf. }][]{Roe00,Mareta01,Jes02}. \emph{Continues.}}
\end{figure*}

At the chosen resolution of 0\farcs3, the inter-knot regions are
fairly well-resolved from the knots themselves, so that we obtain
meaningful values for the cutoff frequency for the entire jet. For the
following discussion, we have chosen 16 locations corresponding to
distinct features, the inter-knot regions and the brightness peaks.
These locations are indicated in Fig.\,\ref{f:ana.fits.spec-loc}.
Figure \ref{f:ana.fits.obs1} shows the corresponding spectra, table
\ref{t:ana.fits.fitpars} shows the fit parameters describing the shape
of the spectrum (low-frequency spectral index, break frequency, cutoff
frequency).

The fit results for the Models A and B are very similar in most cases.
As described in detail in \citet[][section 3]{Jes02}, we reject the large
near-infrared residuals obtained with Model A as implausible and
prefer Model B, which has a significant excess in the near-UV.  This
implies that the jet emission consists of a ``low-frequency''
component responsible for emission from radio through optical, and at
least one ``high-energy'' component, responsible for the near-UV
emission, and possibly the X-rays at well.  We will nevertheless show
all results for both models to illustrate that the quantities derived
from either model do not differ significantly.

\begin{table*}
\caption{\label{t:ana.fits.fitpars} Fit parameters describing the
  shape of the spectra (low-frequency spectral index, break frequency, cutoff
frequency)  shown in Fig.\,\ref{f:ana.fits.obs1}.}
\begin{minipage}[t]{\hsize}
\centering
\renewcommand{\footnoterule}{}
\begin{tabular}{lrr|rll|rll}
\hline\hline
 & & & \multicolumn{3}{c}{Model A} & \multicolumn{3}{c}{Model B/HS} \\
Location & $r$\footnote{Distance along radius vector} &  $\delta_\mathrm{y}$\footnote{Distance from the
 radius vector; negative offsets are to the south.} & $\alpha_\mathrm{low} $  & \nub  &
\nuc\footnote{The $\dagger$ mark indicates locations where there is no cutoff to
 the optical spectrum in the observed range and the value given is a lower limit to the true
 cutoff frequency. Both fits are identical in this case.}  & $\alpha_\mathrm{low} $  & \nub  & \nuc \\
 & \arcsec & \arcsec & & Hz & Hz & & Hz & Hz \\
\hline
A & 13.0 & 0.0 & $-$0.35 & $1.83\times 10^{10}$ & $1.83\times
 10^{16}$ $\dagger$ & \ldots & \ldots & \ldots \\
A-B1 & 13.7 & 0.0 & $-$0.38 & $2.25\times 10^{9}$ & $2.25\times 10^{15}$ & $-$0.38 & $2.15\times 10^{9}$ & $2.15\times 10^{15}$ \\
B1 & 14.2 & 0.1 & $-$0.35 & $7.13\times 10^{9}$ & $7.14\times 10^{15}$
 $\dagger$ &  \ldots & \ldots & \ldots \\
B2 & 15.2 & $-$0.1 & $-$0.44 & $2.78\times 10^{11}$ & $3.81\times
 10^{16}$ $\dagger$ &  \ldots & \ldots & \ldots \\
B3 & 15.6 & $-$0.1 & $-$0.48 & $2.71\times 10^{11}$ & $3.34\times 10^{16}$ & $-$0.46 & $1.98\times 10^{11}$ & $2.38\times 10^{16}$ \\
B3-C1 & 16.3 & 0.0 & $-$0.38 & $3.12\times 10^{9}$ & $3.13\times 10^{15}$ & $-$0.35 & $1.36\times 10^{9}$ & $1.36\times 10^{15}$ \\
C1 & 16.8 & 0.1 & $-$0.39 & $1.65\times 10^{10}$ & $5.24\times 10^{15}$ & $-$0.38 & $1.59\times 10^{10}$ & $3.74\times 10^{15}$ \\
C1-C2 & 17.3 & 0.0 & $-$0.37 & $1.45\times 10^{11}$ & $4.42\times 10^{14}$ & $-$0.35 & $1.35\times 10^{11}$ & $3.67\times 10^{14}$ \\
C2 & 17.7 & 0.0 & $-$0.36 & $2.86\times 10^{10}$ & $9.38\times 10^{14}$ & $-$0.35 & $3.07\times 10^{10}$ & $8.19\times 10^{14}$ \\
C2-D1 & 18.3 & 0.2 & $-$0.49 & $2.39\times 10^{10}$ & $7.88\times 10^{15}$ & $-$0.35 & $8.65\times 10^{9}$ & $5.46\times 10^{14}$ \\
D1 & 18.9 & 0.1 & $-$0.48 & $1.06\times 10^{10}$ & $4.34\times 10^{15}$ & $-$0.35 & $9.58\times 10^{8}$ & $4.65\times 10^{14}$ \\
D1-D2 & 19.5 & $-$0.2 & $-$0.35 & $4.62\times 10^{9}$ & $3.16\times 10^{14}$ & $-$0.35 & $5.78\times 10^{9}$ & $3.23\times 10^{14}$\\
D2 & 19.8 & $-$0.2 & $-$0.35 & $7.92\times 10^{9}$ & $5.26\times 10^{14}$ & $-$0.35 & $9.27\times 10^{9}$ & $5.15\times 10^{14}$ \\
H3 & 20.2 & $-$0.3 & $-$0.39 & $2.20\times 10^{9}$ & $8.75\times 10^{14}$ & $-$0.35 & $1.08\times 10^{9}$ & $5.10\times 10^{14}$ \\
H2 & 21.3 & 0.0 & \ldots & \ldots & \ldots & $-$0.60 & $1.14\times 10^{10}$ & $7.50\times 10^{14}$ \\
H1 & 22.1 & $-$0.2 & \ldots & \ldots & \ldots & $-$0.44 & $2.28\times 10^{9}$ & $5.79\times 10^{13}$ \\
\hline\hline
\end{tabular}
\end{minipage}
\end{table*}

%% next figure with same number as previous
%\addtocounter{figure}{-1}
\begin{figure*}
\centering
\includegraphics[width=.4\hsize]{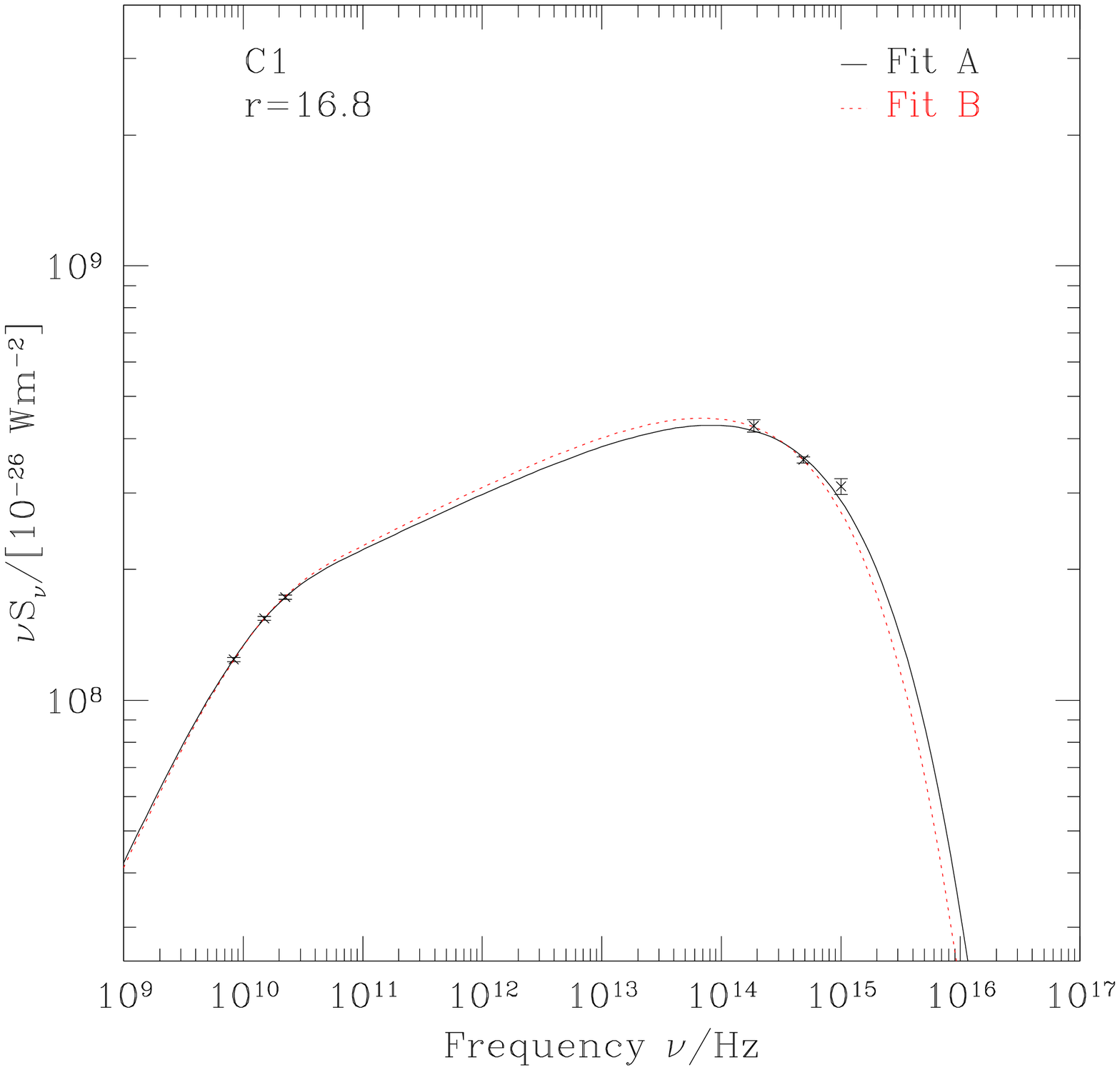}
\includegraphics[width=.4\hsize]{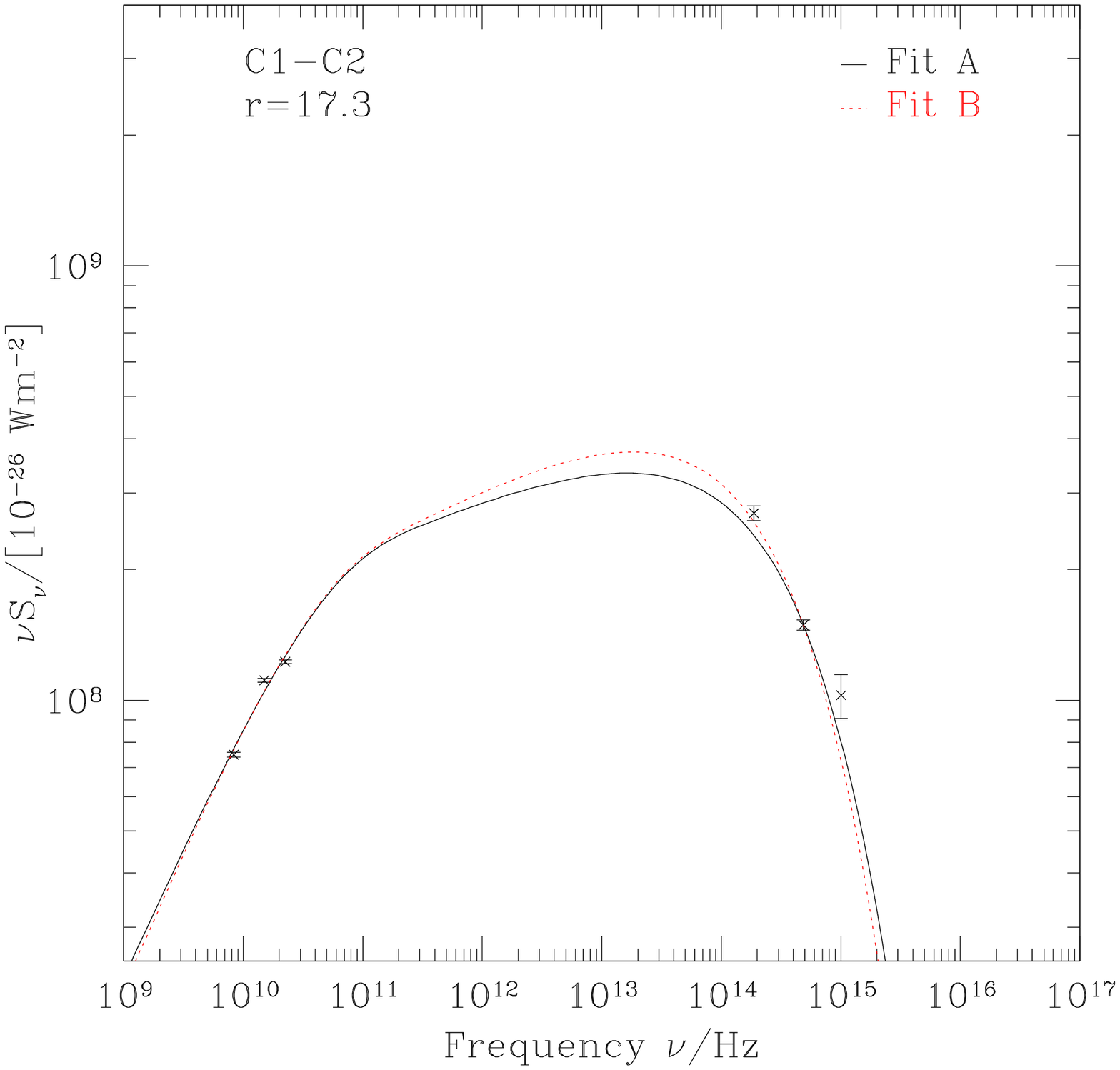}
\includegraphics[width=.4\hsize]{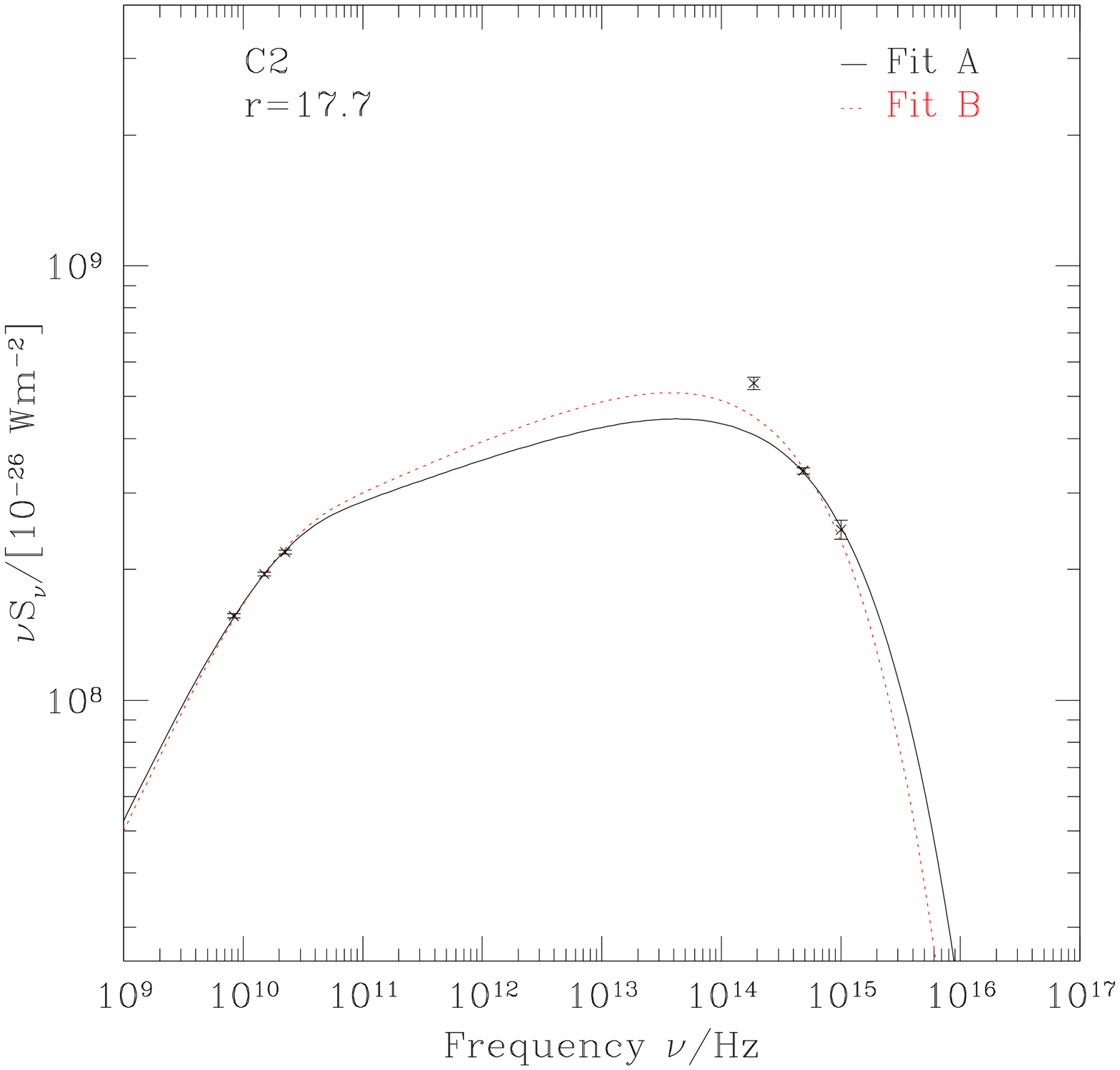}
\includegraphics[width=.4\hsize]{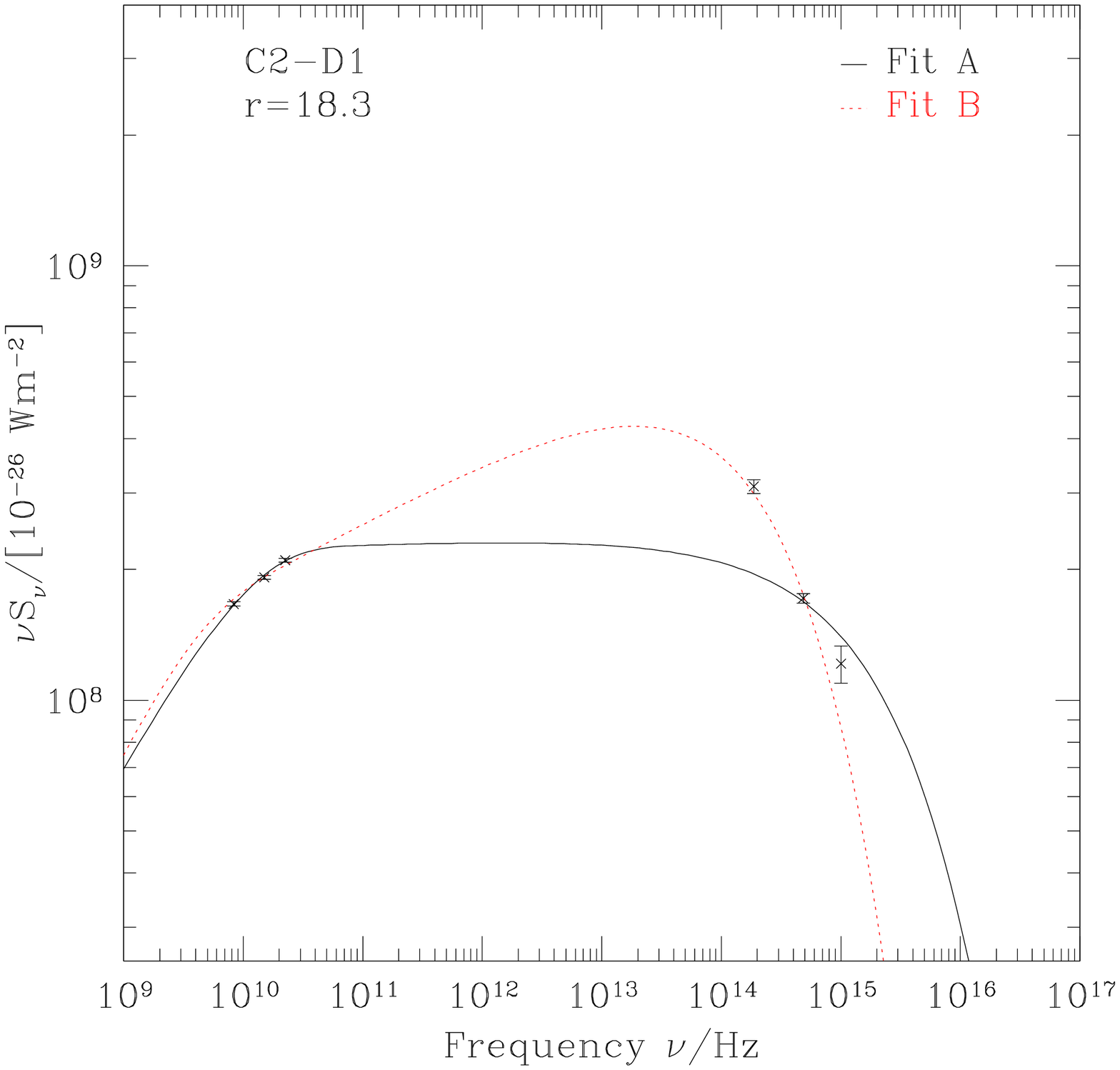}
\includegraphics[width=.4\hsize]{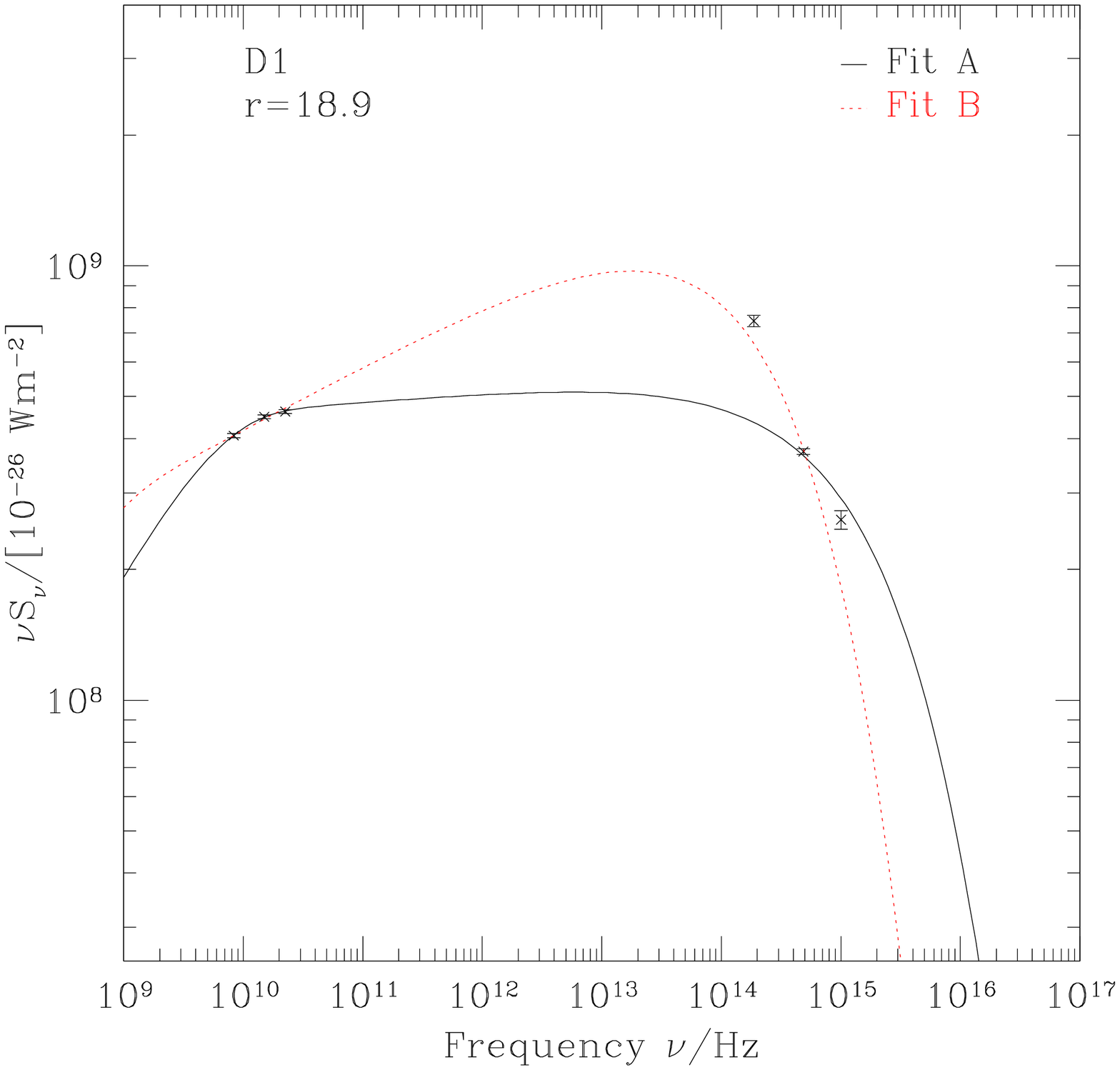}
\includegraphics[width=.4\hsize]{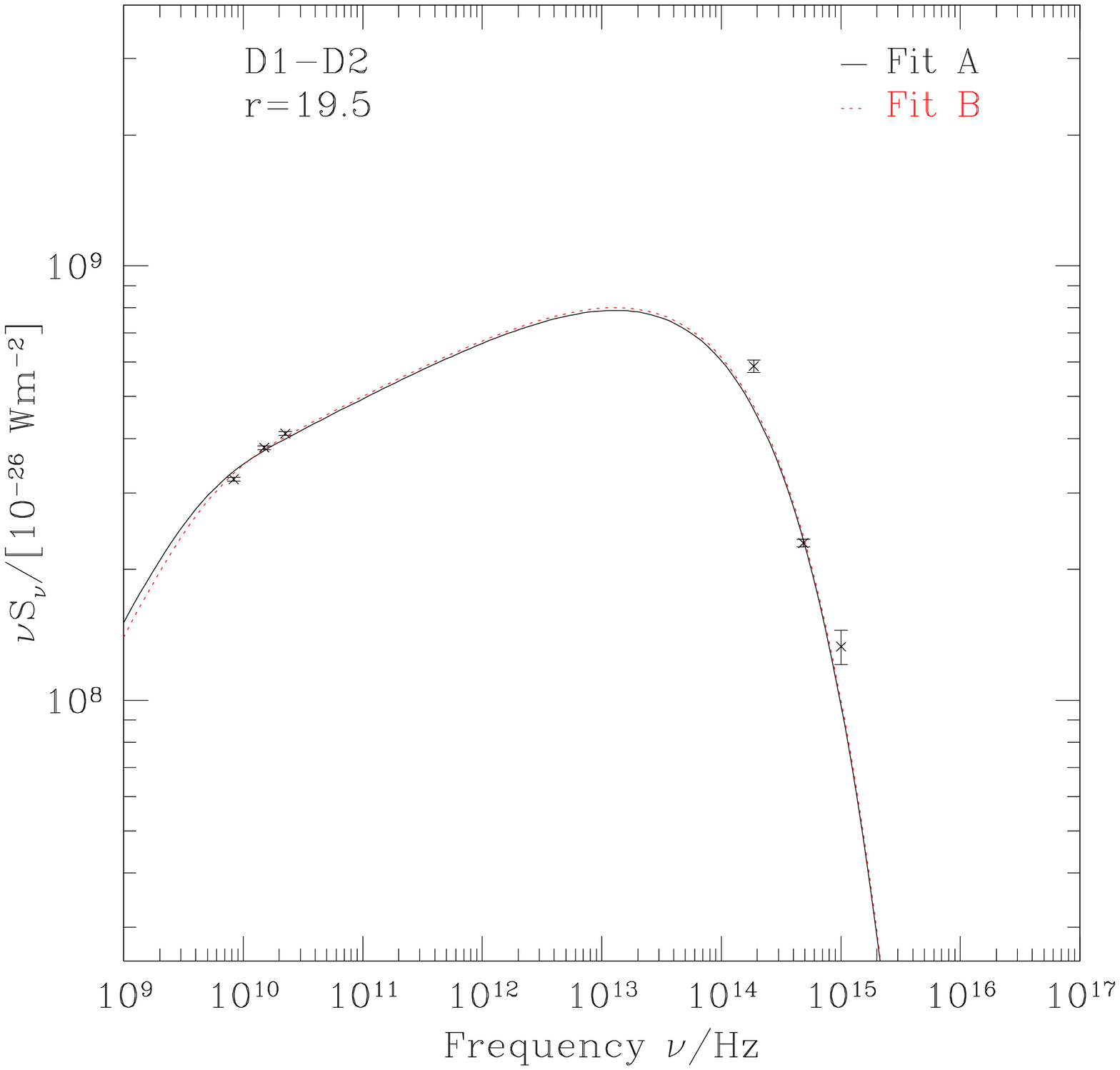}
\caption[Synchrotron spectral fits]{As Fig.\,\ref{f:ana.fits.obs1}, \emph{continued}}
\end{figure*}
%% next figure with same number as previous
%\addtocounter{figure}{-1}
\begin{figure*}[t]
\centering
\includegraphics[width=.4\hsize]{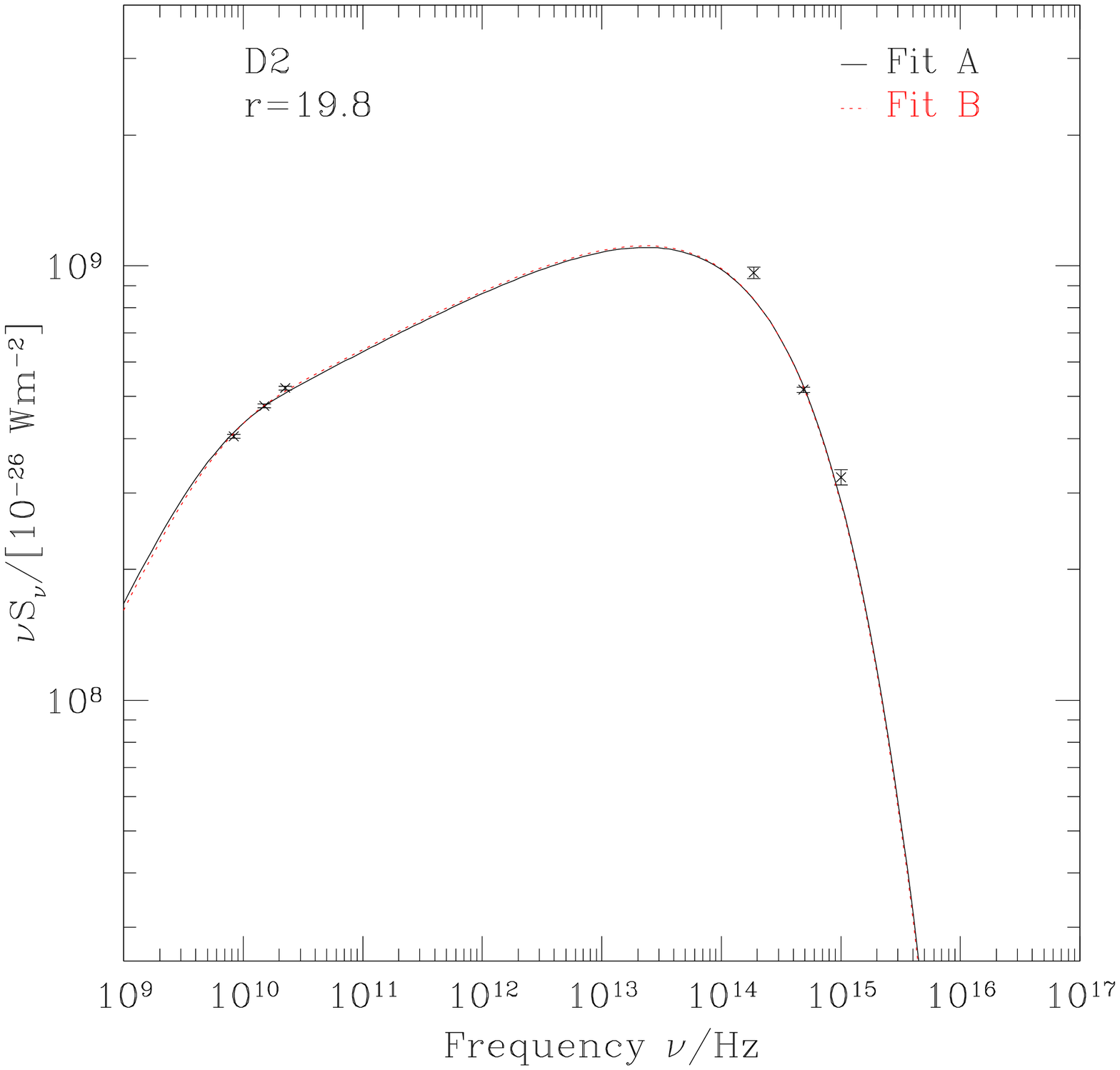}
\includegraphics[width=.4\hsize]{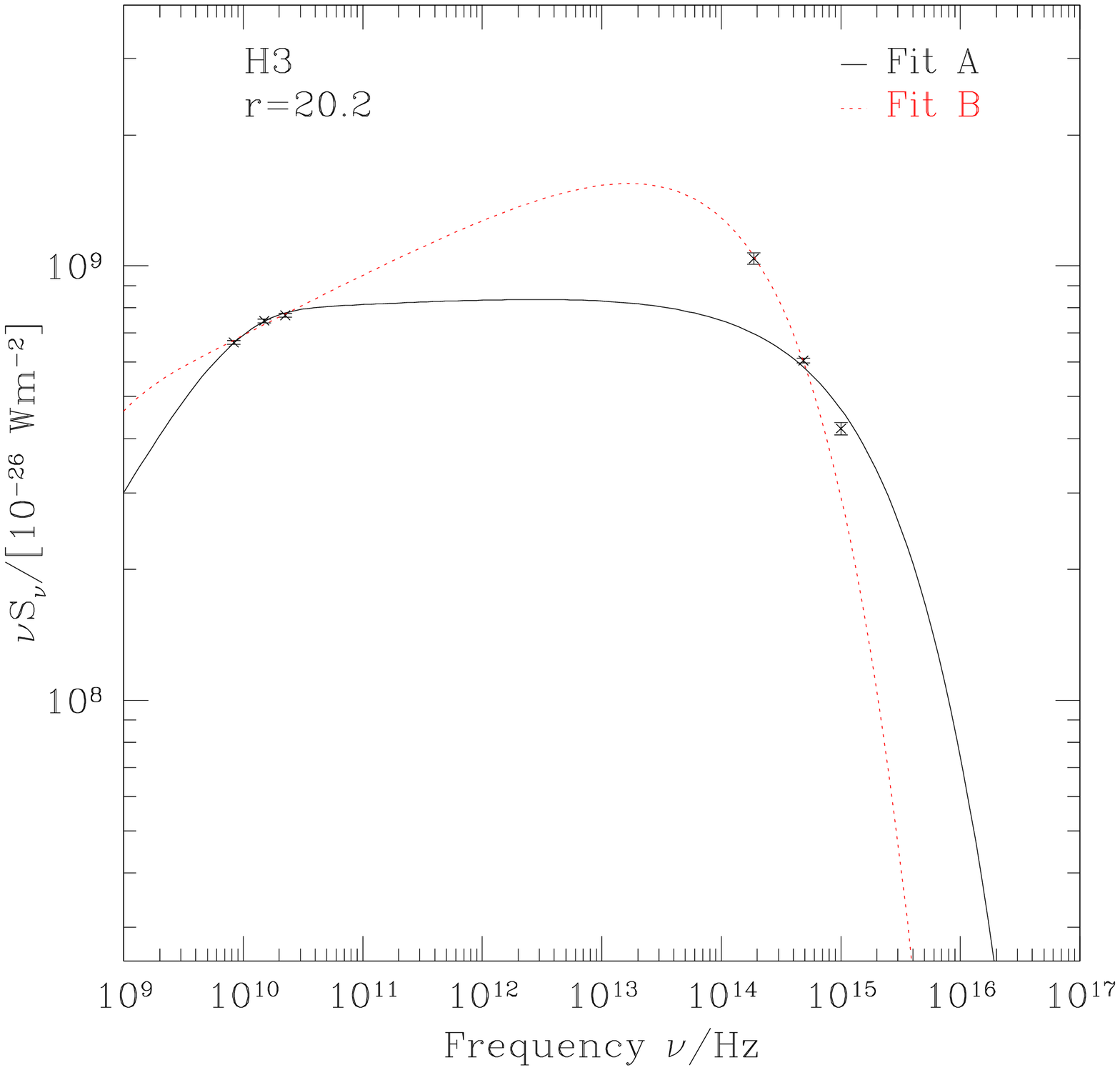}
\includegraphics[width=.4\hsize]{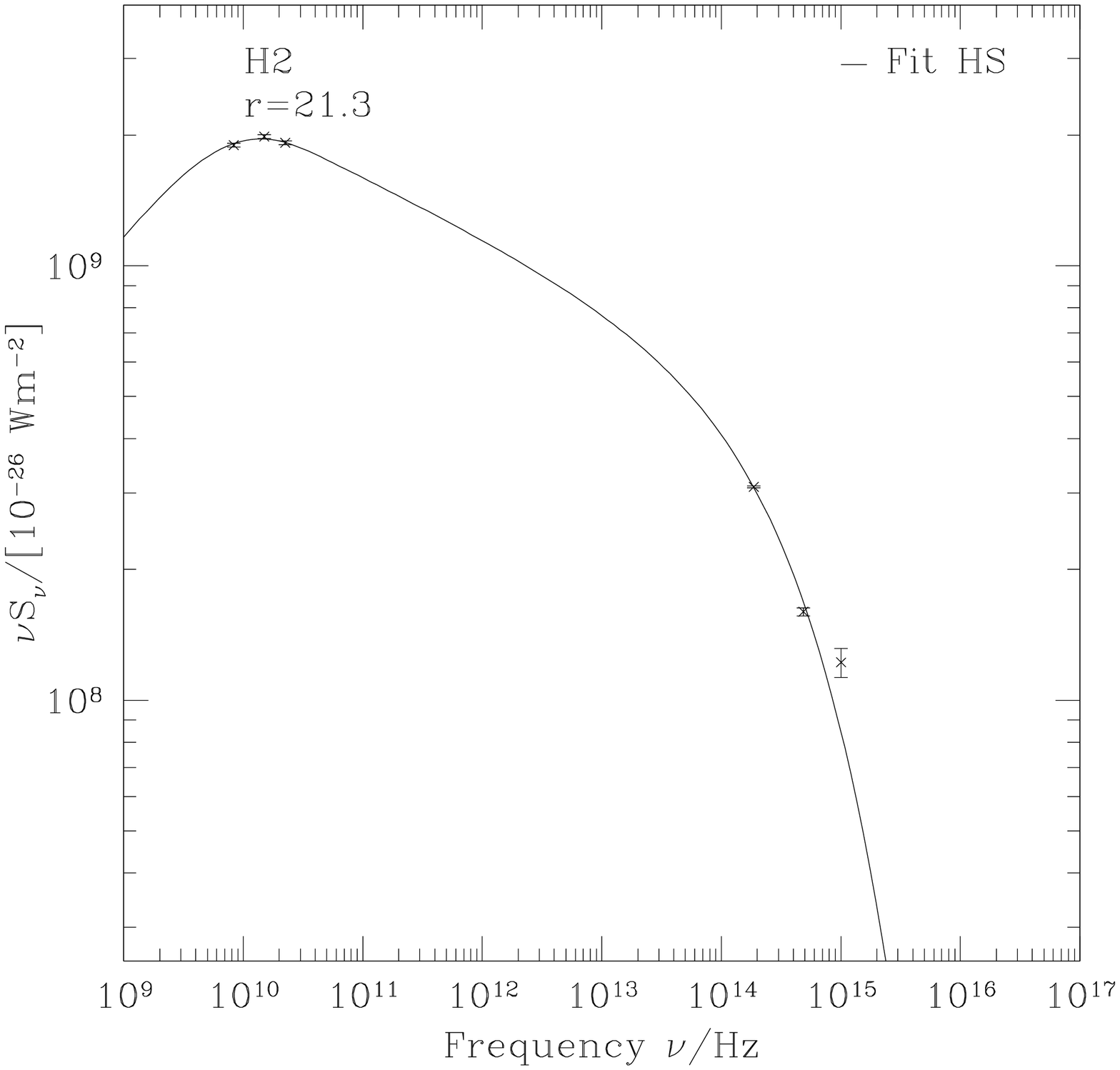}
\includegraphics[width=.4\hsize]{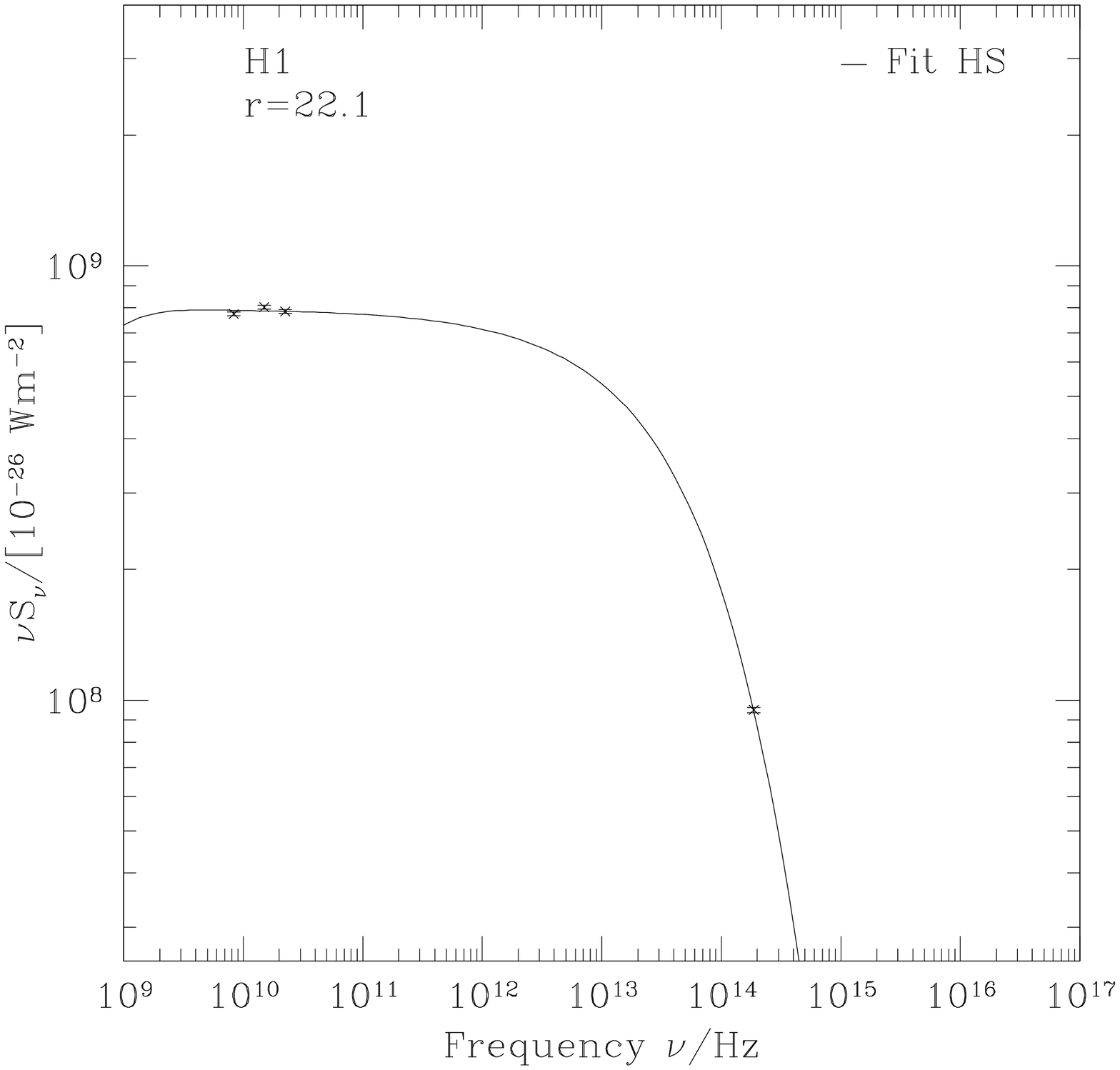}
\caption[Synchrotron spectral fits]{As Fig.\,\ref{f:ana.fits.obs1}, \emph{continued}}
\end{figure*}

Figure \ref{f:ana.fits.obs1} highlights the development of the spectra
along the jet: a global increase in luminosity coupled with a decrease
in cutoff frequency.  As expected from the brightness profiles, the
spectral energy distribution peaks at lower frequencies at larger
radii from the core. Simultaneously, the peak flux density increases.

Compared to earlier studies at lower resolution
\citep{MNR96,NeumDiss,Roe00}, the contributions from individual knots
are now clearly separated from each other.  While previously only knot
A showed a spectrum without a cutoff, it is now seen that there is no
cutoff at the brightness peaks in B1 and B2, either. Not only is there
no cutoff, but the infrared-optical-UV spectrum is harder than the
radio-infrared spectrum in these locations, as already inferred from
the spectral indices (compare Fig.\,\ref{f:res.alpha.run}).  There is,
however, a cutoff in the transition A-B1, between these peaks. The
presence of a cutoff in the southern part of B1 can already be
inferred from the fact that it is optically much fainter than the
northern part, while it is brighter in the radio (\cf
Fig.\,\ref{f:res.ima.images}). The differences between the spectra
fitted at the brightness peaks and the regions connecting them are
nowhere else as drastic.  However, for the first time we see that
there is an increase of the cutoff frequency in the knots, compared to
inter-knot regions, corresponding to the slightly flatter spectrum
there (\cf \S\ref{s:res.alpha}).  This local increase is a modulation
of the global outward decrease of the cutoff frequency.

The synchrotron radiation at the cutoff frequency is emitted by those
particles with the highest energy. The maximum particle energy can
therefore be computed from the cutoff frequency through the
synchrotron \emph{characteristic frequency}, the frequency around
which most of the synchrotron emission of an electron of energy
$\gamma m_\mathrm{e} c^2$ in a magnetic field of flux density $B$ is
emitted: 
\begin{equation} \nu_{\mathrm{char}} = 4.2\times 10^{15}
\left(\frac{\gamma}{10^7}\right)^2 \left(\frac{B \sin\psi}{\rm
nT}\right) {\rm Hz}. \label{eq:nu_c_prop} \end{equation}
The dependence on the pitch angle $\psi$ requires one to make an
assumption about the pitch angle distribution, which we take to be
isotropic.  This calculation requires the knowledge of the magnetic
field in the source.  In order to discuss the behaviour of the cutoff
frequency and the maximum particle energy together, we first consider
how to derive the magnetic field value.

In the absence of any other reliable method, the magnetic field for the jet
can only be estimated by making use of the minimum-energy assumption.
We therefore present the derivation of the value of the minimum-energy
magnetic field for the type of spectra described here.

\subsection{Equipartition field strength}\label{s:ana.equi}

There is a firm minimum value for the total energy density (in
relativistic particles and magnetic field) necessary to generate a
given synchrotron luminosity. This energy density can be deduced from
observations with the help of assumptions about the source geometry
and the range of frequencies over which synchrotron emission is
emitted.  \citet{Pach70} presents a clear derivation of the minimum
energy density of a synchrotron source for the assumption of a
power-law electron energy distribution in the source.

The minimum-energy field is that value of the source's magnetic field
corresponding to the minimum total energy sufficient to give the
observed emission. Due to the lack of other diagnostic tools, this
minimum-energy magnetic field is often used as estimate for the true
magnetic field. Doing so, one introduces a local correlation between
the magnetic field strength and the energy density in particles. This
correlation provides the most efficient way to produce a given
synchrotron luminosity. One may imagine magneto-hydrodynamical
processes causing this correlation, because the particles are both
tied to magnetic field lines and create them by their motion. However,
no detailed microphysical feedback process has been identified which
maintains this correlation on all scales, and it appears dangerous to
us to draw conclusions deriving from this correlation until it is is
established that it is a natural, not an artificial one.
Nevertheless, it seems reasonable to assume that the minimum-energy
magnetic field is a good indicator of the order of magnitude of the
source's magnetic field \citep{magnumopus89}.

With this caveat in mind, we use the minimum-energy magnetic field as
a measure of the jet's magnetic field, calculating it following
\citet[ p.~168\,ff.]{Pach70} but explicitly assuming a broken power
law.  To account for the break in the electron spectrum at Lorentz
factor $\gamma_\mathrm{b}$ leading to a break in the model spectrum at
$\nu_\mathrm{b}$, we write the electron density as
\begin{equation}
n(\gamma) = \left\{ \begin{array}{l@{\mathrm{\ for\ }}r}
n(\gamma_0) \left(\frac{\gamma}{\gamma_0}\right)^{-p_\mathrm{low}}
& \gamma < \gamma_\mathrm{b}\\
n(\gamma_0) \left(\frac{\gamma}{\gamma_0}\right)^{-p_\mathrm{high}}
& \gamma \geq \gamma_\mathrm{b}
\end{array}\right.
\label{eq:ana.minE.broken.pl}
\end{equation}
Similarly, the observed spectrum is approximated by
\begin{equation}
S(\nu) = \left\{ \begin{array}{l@{\mathrm{\ for\ }}r}
S(\nu_0) \left(\frac{\nu}{\nu_0}\right)^{\alpha_\mathrm{low}}
& \nu < \nu_\mathrm{b}\\
S(\nu_0) \left(\frac{\nu}{\nu_0}\right)^{\alpha_\mathrm{high}}
& \nu \geq \nu_\mathrm{b}
\end{array}\right.,
\label{eq:ana.minE.broken.S} \end{equation} with $\alpha=-(p-1)/2$.  
The spectra in \S\ref{s:ana.fits} have been constrained to break near
$\nu_\mathrm{b} \approx 10^{10}$\,Hz (see discussion below), from
$\alpha_\mathrm{low}\approx -0.5$ to $\alpha_\mathrm{high} \approx
-1$, corresponding to $p_\mathrm{low}\approx 2$ and
$p_\mathrm{high}\approx 3$, respectively.  Even the separate hot spot
fit, in which the spectral index is allowed to vary, does not have a
significantly steeper best-fit spectral index. All integrals can
therefore be approximated by setting $\alpha_\mathrm{low} = -0.5$ and
correspondingly $\alpha_\mathrm{high}= -1$. It is useful to write the
electron energy spectrum in terms of $\gamma_0 = \gamma_\mathrm{b}$,
and correspondingly the observed spectrum in terms of $\nu_0 =
\nu_\mathrm{b}$.

The calculation of the minimum energy density involves integrating
both the electron energy over the electron distribution, and the
observed flux density over the corresponding frequency range.  We use
a fixed low-frequency limit of $\nu_\mathrm{min} = 10^7$\,Hz.  While a
choice of a fixed lower electron energy limit would have been more
physical, it turns out that the value of $\nu_\mathrm{min}$ has little
influence on the result for $\alpha_\mathrm{low} = -0.5$ because terms
in $\nu_\mathrm{min}/\nu_\mathrm{b}$ enter logarithmically or can be
neglected altogether. Similarly, for the observed cutoff in the optical
range, a term in $\nu_\mathrm{b}/\nu_\mathrm{c}$ can be neglected.
With these approximations, we obtain the following expressions for the
total energy in electrons, the luminosity following from the electron
energy distribution, and the observed luminosity, respectively:
\begin{eqnarray}
\label{eq:ana.minE.broken.U_el} U_\mathrm{el} &=& m_\mathrm{e}c^2
n(\gamma_\mathrm{b})
K \nu_\mathrm{b} B^{-1} \left(1+\frac{1}{2} \ln \frac{\nu_\mathrm{b}}{\nu_\mathrm{min}}\right)\\
\label{eq:ana.minE.broken.L} L(n_\mathrm{e},B) & = & n(\gamma_\mathrm{b}) \phi V
\frac{4}{3} \sigma_{\rm T} c U_{\rm mag} \nonumber \\
& & \times K^{\frac{3}{2}} \nu_\mathrm{b}^{\frac{3}{2}}  B^{-\frac{3}{2}}
\left(1+\frac{1}{2}\ln\frac{\nu_\mathrm{c}}{\nu_\mathrm{b}}\right) \\
\label{eq:ana.minE.broken.L_obs} L_\mathrm{obs} & = & 8\pi
d_\mathrm{L}^2 S(\nu_\mathrm{b}) \nu_\mathrm{b}
\left(1+\frac{1}{2}\ln\frac{\nu_\mathrm{c}}{\nu_\mathrm{b}}\right)
\end{eqnarray}
$K^{-1} = 4.2\times10^{10} \mathrm{T}^{-1}$\,Hz is the numerical
constant from Eqn.\,\ref{eq:nu_c_prop}, $V$ is the source volume, a
fraction $\phi$ of which is filled by radiating particles,
$d_\mathrm{L}$ is the luminosity distance to the source, and the
remaining symbols have their conventional meaning.  We solve for the
minimum-energy magnetic field in the usual way by using
Eqns.~\ref{eq:ana.minE.broken.L} and \ref{eq:ana.minE.broken.L_obs} to
solve for the electron number density as function of the magnetic
field and substituting into Eqn.~\,\ref{eq:ana.minE.broken.U_el}, then
minimising the total energy density in the source
\begin{equation}
U_\mathrm{tot}(B) = (1+k) U_\mathrm{el}(B) + \frac{B^2}{2\mu_0}
\label{eq:Utot}
\end{equation}
 with respect to $B$ to obtain the minimum-energy (or
``equipartition'') field $B_\mathrm{min}$. Here, $k$ is the ratio of
energy in other relativistic particles to energy in relativistic
electrons. If the jet is an electron-positron jet, $k=1$ since
positrons are accelerated in the same way as electrons. If the
charge-balancing particles are protons or even heavier ions, the value
of $k$ depends on details of the injection and acceleration process. A
typical number found in cosmic-ray particles is $k\approx 100$. We
choose $k=1$ to obtain a lower limit to $B_\mathrm{min}$, which scales
as $(1+k)^{2/7}$. The final expression for $B_\mathrm{min}$ is of
the form
\begin{equation}
B_\mathrm{min} \propto \left(S(\nub) \sqrt{\nub}\,
g(\nu_\mathrm{min},\nub,\nuc)\right)^{\frac{2}{7}}
\label{eq:Bmin}
\end{equation}
where $g(\nu_\mathrm{min},\nub,\nuc)=1+1/2 \ln
(\nub/\nu_\mathrm{min})$ in our case. In the more general case of a
broken power law with arbitrary low-frequency spectral index \alo,
$g(\nu_\mathrm{min},\nub,\nuc) = c(\alo)\,
h(\nu_\mathrm{min},\nub,\nuc)$, where both $c$ and $h$ are a
slowly-varying functions, and $h$ contains the integration limits from
the $U_\mathrm{el}$ term analogous to Equation
\ref{eq:ana.minE.broken.U_el}.  The integration limits for the
observed luminosity and the luminosity following from the electron
energy distribution drop out of the final expression for
$B_\mathrm{min}$, leaving only the dependence on the spectral index in
$c(\alo)$.  This important property of the minimum-energy field is not
obvious from the form of the equations as given by \citet{Pach70}, but
becomes evident when writing the observed bolometric luminosity in
terms of an integral over a power-law flux density distribution
\citep[\cf][Equation 2]{Miley80}.

We next discuss the impact of the jet orientation the assumptions we
have made about the shape of the spectrum before discussing the
behaviour of the minimum-energy field and hence the maximum particle
energy along the jet.

\subsection{Impact of our assumptions}
\label{s:ana.assume}

\subsubsection{Impact of line-of-sight angle and Doppler factor}
\label{s:ana.assume.los}

If the jet does not lie in the plane of the sky but is inclined to the
line of sight by an angle $i$, all lines of sight passing through the
jet, and hence the total jet volume, are longer by a factor $1/\sin i$
compared to the side view (ignoring edge effects at the end of the
jet).  The minimum-energy field is influenced by the variation in
volume $V$ as $V^{-\frac{2}{7}}$. A line-of-sight angle $i \approx
45\degr$ has been inferred for the flow into the hot spot from
independent considerations of the jet's polarisation change there and
the hot spot's morphology \citep{jetIII,hs_II}. If the jet is at the
same angle, the values in Tab.\,\ref{t:res.im.width.jetvol} need to be
scaled up by $1/\sin 45\degr \approx 1.4$. Hence, the minimum-energy
field needs to be scaled down by about 10\%, and correspondingly the
maximum energy up by 10\%. Since we do not expect the minimum energy
to be accurate to this level, we simply assume the jet is in the plane
of the sky.  Even if the jet flow is relativistic, the value of \gmax\
inferred from the minimum-energy field and the observed \nuc\ is
nearly independent of the value of the Doppler factor since both \nuc\
and $\Bmin \propto S_\nu^{2/7}$ are modified by the beaming in nearly
the same way \citep{NeumDiss,MNR96}.

\subsubsection{Impact of assumptions about the spectral shape}
\label{s:ana.assume.shape}

Our fits of the observed spectra with theoretical models
(\S\ref{s:ana.fits}) constrained the spectral shape to have a
low-frequency spectral index of $\alpha=-0.4$ or flatter and a break
of $|\Delta \alpha| = 0.5$ in the frequency range of the VLA
observations, leaving the cutoff frequency as the only ``shape
parameter'' to be determined by the observations.  This is an
appropriate model of the observations since all radio observations
show a nearly constant radio spectral index $\alpha \approx -0.8$
everywhere in the jet, while there are variations in the spectral
indices only in the infrared/optical/UV wavelength range
\citep[\S\ref{s:res.alpha} and][]{jetII}.  We are primarily interested
in the value of the cutoff frequency, but also in the behaviour of the
bolometric luminosity of the radio-optical emission component.  We
show now that the impact of the assumptions about the spectral shape
on these quantities only has negligible impact on our conclusions.

To check the impact of our assumptions, we repeated the minimum-energy
argument for a simple power law with spectral index $\alpha = -0.8$
(approximating the radio observations).  For spectral indices $\alpha
< -0.5$ and for $\nuc \gg \nu_\mathrm{min}$, the only spectral shape
parameters that remain in the expression for the minimum-energy field
are the spectral index itself and the assumed minimum frequency.  The
value of the cutoff frequency has negligible impact on the
minimum-energy field.  The the spectral index only appears in a
slowly-varying function, so that small changes in the assumed $\alpha$
do not have a large impact on the result.  Changing the break
frequency $\nub$, which could in principle lie in the unobserved gap
between the VLA and HST data, by as much as a factor of $10^4$ only
changes the derived minimum-energy field by a factor of about two.
Similar changes will be produced by much more modest variations in the
assumed values of the filling factor $\phi$ and the ratio $k$ of
energy in other relativistic particles to energy in relativistic
electrons.  Thus, our lack of knowledge about $\phi$ and $k$
constitutes the dominant systematic uncertainty.  Even these
uncertainties do not have a strong impact on our results because the
inferred maximum particle energy scales as $\gmax \propto
\sqrt{\nuc/B_\mathrm{min}}$ so that $\gmax$ depends at most on the
7$^{\mathrm{th}}$ root of any quantity of interest. The results
presented in the following section are therefore very robust with
respect to changing any of the underlying assumptions, at least for
the low-energy emission component responsible for the radio-optical
synchrotron emission (\cf \S\ref{s:disc.pop}).

\subsection{Run of \nuc, \Bmin, and \gmax\ along the jet}
\label{s:ana.gmax}

\begin{figure}
\parbox[c]{.15\hsize}{Mod.\ A\\and
HS}\parbox[c]{.85\hsize}{\resizebox{\hsize}{!}{\includegraphics[clip]{log_nuc_pa.eps}}}\\
\parbox[c]{.15\hsize}{Mod.\ B\\and HS}\parbox[c]{.85\hsize}{\resizebox{\hsize}{!}{\includegraphics[clip]{log_nuc_pb.eps}}}\\
\parbox[c]{.15\hsize}{\hspace*{.15\hsize}}\parbox[c]{.85\hsize}{\resizebox{\hsize}{!}{\includegraphics[clip]{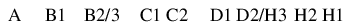}}}\\
\parbox[c]{.15\hsize}{radius/\arcsec}\parbox[c]{.85\hsize}{\resizebox{\hsize}{!}{\includegraphics[clip]{skala.eps}}}\\
\parbox[c]{0.15\hsize}{\hspace*{0.15\hsize}}\parbox[c]{.85\hsize}{\includegraphics[angle=0,width=.9\hsize,clip]{bw-colorbar-log.eps}}
\caption[Map of the cutoff frequency in the jet of
3C\,273]{\label{f:ana.res.nuc}Maps of the cutoff frequency. Grey
levels run from $10^{13}$\,Hz (white) to $10^{17}$\,Hz (black)
with a pseudo-logarithmic stretch as indicated by the greyscale
bar. The values fitted in A and B2 are lower limits. As expected,
Model A bears a closer resemblance to the optical-ultraviolet spectral
index map, while Model B is dominated by the infrared-optical spectral
index map (\cf Fig.\,\ref{f:res.alpha.alpha}). Regions H2 and H1 are
fitted with a different model HS.}
\end{figure}
For each photometry aperture (or pixel in
Fig.\,\ref{f:res.ima.images}), the values for $\nu_\mathrm{b},
S(\nu_\mathrm{b})$ and \nuc\ from the fitted spectra together with the
appropriate volume from Tab.\,\ref{t:res.im.width.jetvol} and the
assumed value of $\nu_\mathrm{min}$ are used to calculate
$B_\mathrm{min}$.  With this knowledge, the maximum particle energy
\gmax\ can be inferred from the fitted value of \nuc by Eqn.\,\ref{eq:nu_c_prop}.

We consider the determined values of the cutoff frequency,
minimum-energy field and hence maximum particle energy and their
relation to the jet morphology. In particular, we will be
interested whether it is possible to identify localised
acceleration regions in the jet.

\subsubsection{The cutoff frequency \nuc}\label{s:ana.gmax.nuc}

\begin{figure}
\resizebox{\hsize}{!}{\includegraphics[angle=0]{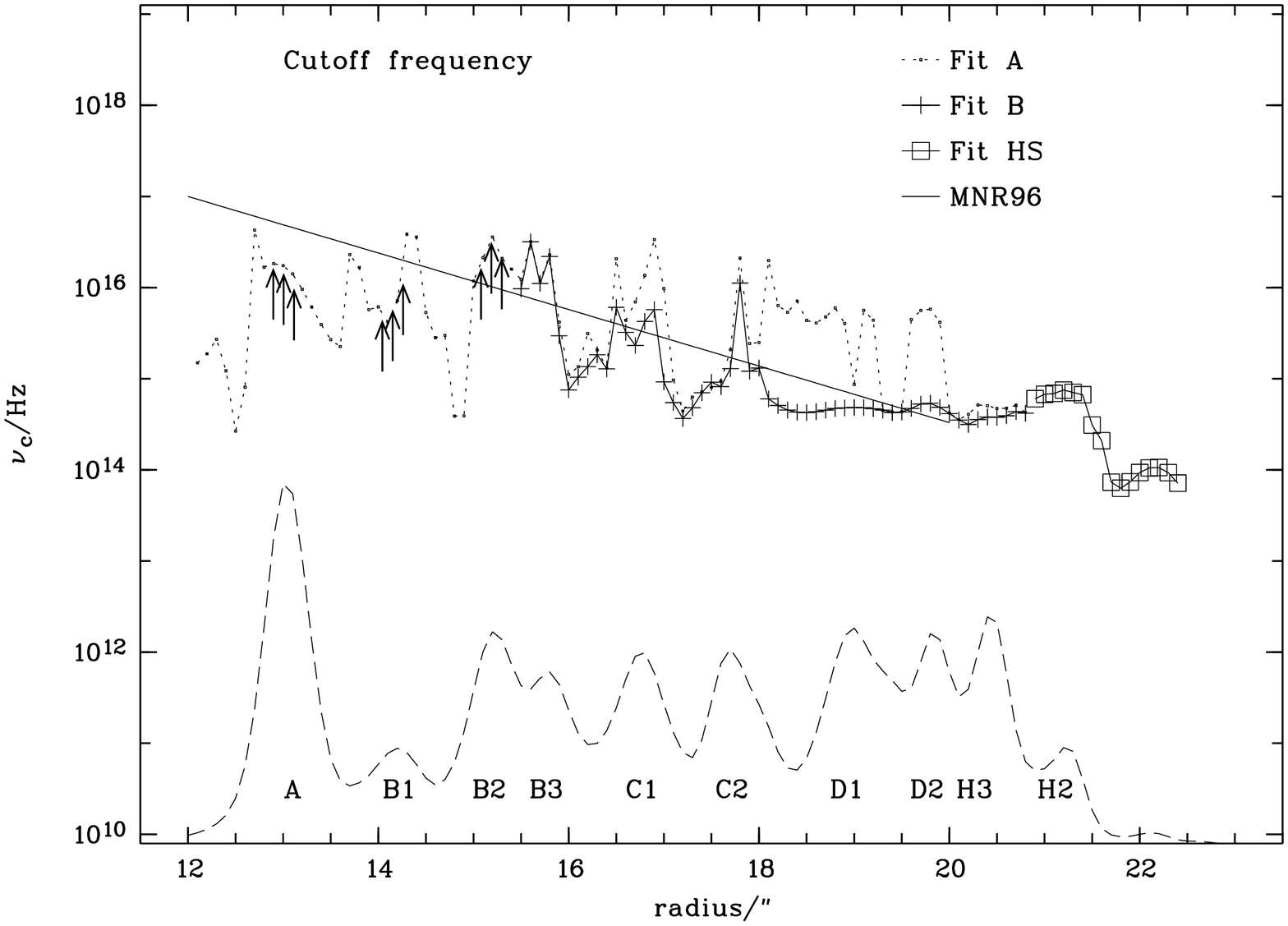}}
  \caption[Run of cutoff frequency]{Run of the fitted cutoff frequency
  \nuc\ along the jet.  The thin line represents the run of the cutoff
  frequency at 1\farcs3 resolution, parameterised as $\nuc =
  10^{17}\,\mathrm{Hz} \exp(-(r-12\arcsec)/1\farcs4)$ by
  \protect\citet{MNR96}.  Model A has a slightly higher \nuc\ than
  Model B. Lower limits to the true cutoff frequency are indicated by
  arrows. Error bars are not shown because the errors on the fitted
  cutoff frequency are correlated in a complicated manner with the
  observational flux errors and the assumed spectral shape.
  Variations in the cutoff frequency are expected to be significant
  where the variations in the high-frequency spectral indices are
  significant.} \label{f:ana.res.run-nuc}
\end{figure}
We present maps of the fitted cutoff frequency \nuc\ for the three
different fits in Fig.\,\ref{f:ana.res.nuc}, and its run along the
radius vector at position angle 222\fdg2 in
Fig.\,\ref{f:ana.res.run-nuc}. No cutoff is observed in the regions at
$r=13\arcsec$ (A1) and at 15\arcsec\ (B2) (\cf \S\ref{s:res.alpha}),
so the fitted values there are only lower limits to the true cutoff
frequency. The overall trend is a decrease in \nuc\ with increasing
distance from the core.  The lowest value of \nuc\ is reached at the
hot spot, which is characterised by a sharp drop in \nuc\ from
$10^{15}$\,Hz at 21\farcs2 (optical hot spot position) to
$10^{14}$\,Hz at $r=21\farcs6$.  As expected from the overall
similarity of the spectral indices determined at 0\farcs3 resolution
and in earlier work at 1\farcs3 resolution
(Fig.\,\ref{f:res.alpha.compare}), the cutoff frequency determined
here agrees well with the run of the cutoff frequency $\nuc =
10^{17}\,\mathrm{Hz} \exp(-(r-12\arcsec)/1\farcs4)$ determined by
\citet{MNR96}.

All variations are rather smooth. Any sharp jumps are due to the
uncertainties introduced into the fitting by the fact that there are
only three high-frequency data points which cannot constrain the
cutoff frequency and the UV excess well simultaneously.  Therefore,
only those local peaks are significant which correspond to significant
peaks in the high-frequency spectral indices
(Fig.\,\ref{f:res.alpha.run}). In fact, local peaks in the cutoff
frequency are not as pronounced as expected from the peaks in the
infrared-optical spectral index, e.g. at C2 and D2.  Considering the
fits in these locations (Fig.\,\ref{f:ana.fits.obs1}), it is apparent
that the cutoff frequency is likely overestimated in those locations.
In any case, there is a tendency for the cutoff frequency to be
slightly higher in the brighter regions than in the immediate
surroundings.  However, the variation in cutoff frequency is less
pronounced than the corresponding variation in the jet's surface
brightness at all wavelengths.  Thus, there is a local positive
correlation between brightness and cutoff frequency, while globally,
the cutoff frequency decreases along the jet as the (radio) surface
brightness increases.

Only small discrepancies arise in the value of the cutoff frequency
between Model A and Model B for most of the jet (for a discussion, see
\S\ref{s:disc.pop} below). Between $r=17.5$ and $r=19.5$, where the
spectral hardening is most pronounced, the value of \nuc\ in {Model
A}, in which the cutoff is determined mainly by the optical-UV
spectral index, is a factor of 3--10 larger than in the preferred
Model B, in which the steeper (by $\Delta\alpha \approx 0.2$)
infrared-optical spectral index determines the cutoff frequency.
Before we discuss the variation of the particles' maximum energy along
the jet, we consider the derived minimum-energy field.

\subsubsection{The equipartition magnetic field}\label{s:ana.gmax.bmin}

\begin{figure}
\resizebox{\hsize}{!}{\includegraphics[angle=270]{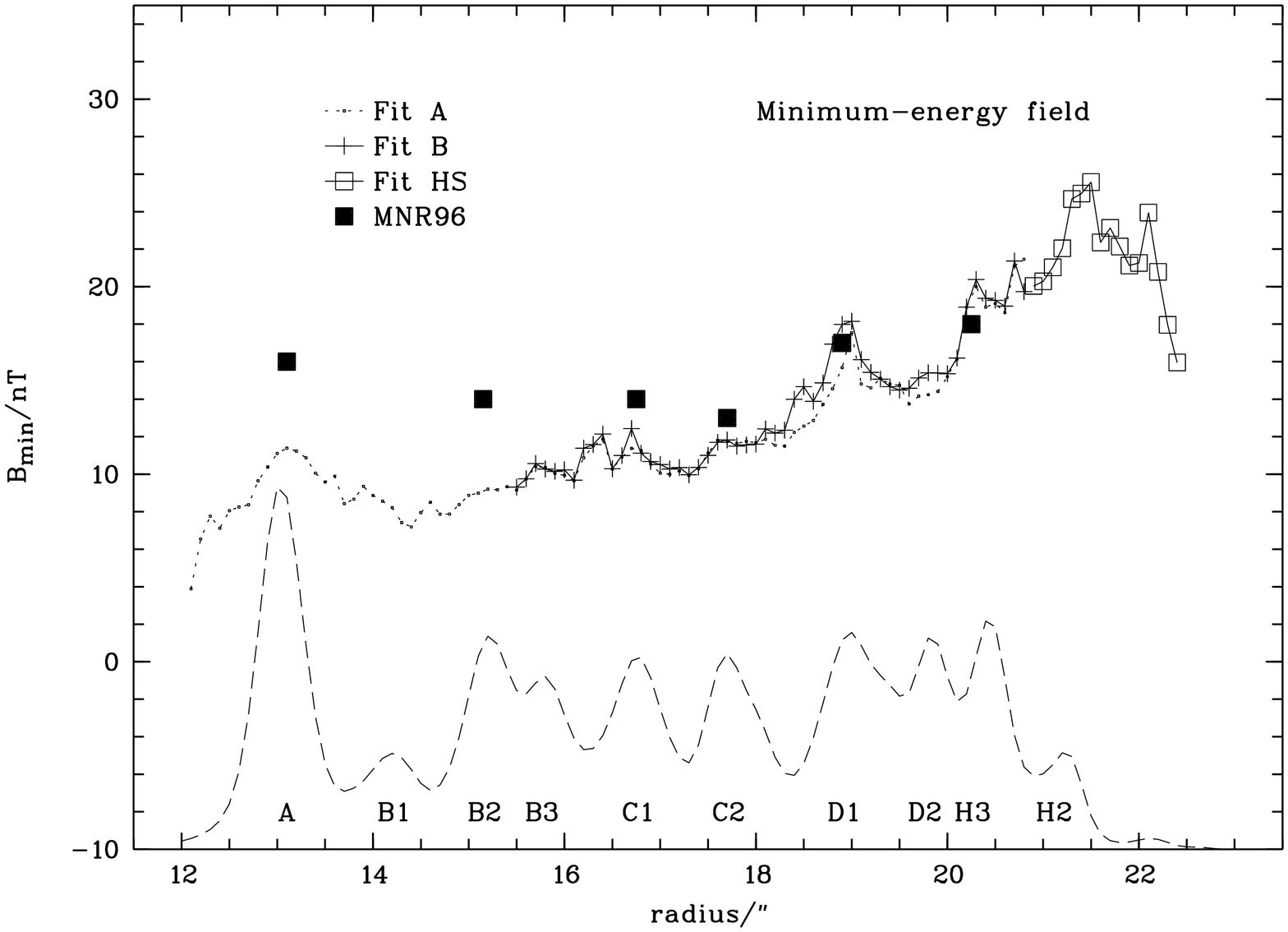}}
  \caption[Run of minimum-energy field]{Run of the minimum-energy
  field \Bmin\ along the jet.  The solid points show the values
  determined for individual regions by \protect\citet{MNR96}. The
  overall run corresponds to that of the jet luminosity. The most
  recent value for the magnetic field determined for the hot spot is
  $(39^{+24}_{-10})$\,nT.\label{f:ana.res.run-Bmin}; correcting for a
  different value of $k$ in Eqn.\,\ref{eq:Utot} brings this into
  agreement with the present determination.  Other discrepancies are
  due to the different volumes assumed for the emission regions,
  which were modeled as individual blobs by \protect\citet{MNR96}.}
\end{figure}
\begin{figure}
\resizebox{\hsize}{!}{\includegraphics[width=.85\hsize]{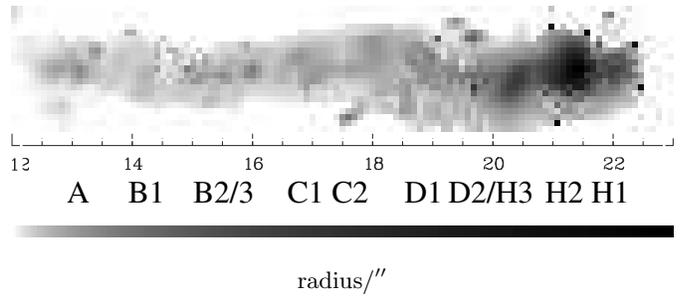}}
\resizebox{\hsize}{!}{\includegraphics[width=.85\hsize]{skala.eps}}
\resizebox{\hsize}{!}{\includegraphics{knot-labels-times.eps}}
\resizebox{\hsize}{!}{\includegraphics{bw-colorbar-log.eps}}
\begin{center}radius/\arcsec\end{center}
\caption[Map of bolometric surface brightness]{ \label{f:ana.res.Lobs}
Map of the bolometric surface brightness assuming isotropic
emissivity.  Values range from $1.5\times 10^{34}$\,W/pixel at the
peak of knot A to $1.9\times10^{35}$\,W/pixel in the hot spot with a
pseudo-logarithmic stretch as indicated by the greyscale bar.  The hot
spot has a luminosity of approximately $1.31\times 10^{37}$\,W (sum of
H2 and H1), which is just a little more than the total luminosity of
the jet at $r\leq 20\farcs7$ of $1.26\times10^{37}$\,W. The luminosity
of regions A and B2 could be much larger than indicated here if the
synchrotron spectrum extends up to X-rays.}
\end{figure}
The variation of the derived minimum-energy field perpendicular to the
jet axis is dominated by the assumed geometry. We therefore present
only its run along the assumed jet axis in
Fig.\,\ref{f:ana.res.run-Bmin}. It starts at just below 10\,nT at the
onset of the optical jet, increasing to 20--25\,nT in the hot
spot. The corresponding value of the electron number density ranges
from about 1-5\,m$^{-3}$. The evolution of the minimum-energy field
along the jet corresponds qualitatively to the run of the bolometric
surface brightness (Fig.\,\ref{f:ana.res.Lobs}). This is expected
because for the constant spectral shape assumed here (justified by the
constant radio spectral index) and for a constant emitting volume, the
minimum-energy field scales with the surface brightness $S(\nu_0)$ (at
any frequency $\nu_0$ significantly below the cutoff frequency) as
$S^{2/7}$. As already noted in \S\S\ref{s:ana.equi} and
\ref{s:ana.assume.shape} above, the cutoff frequency has negligible
influence on the minimum-energy field for spectra with $\alpha <
-0.5$. Therefore, the possibility that the synchrotron spectrum extends
up to X-rays in regions A, B1 and B2 has no influence on the
minimum-energy estimate.

The minimum-energy field of $\approx 25$\,nT determined for the hot
spot seems to be low compared to the magnetic field value determined
by \citet{hs_II}, both from spectral fits ($(39^{+24}_{-10})$\,nT) and
from the minimum-energy argument ($(35^{+8}_{-4})$\,nT; \cf
Fig.\,\ref{f:ana.res.run-Bmin}). This difference is entirely due to
the different values assumed for the ratio $k$ in Eqn.\,\ref{eq:Utot},
chosen as $k=1$ here but $k=10$ in \citet{hs_II}.  Thus, our
determination of the minimum-energy field agrees with previous
spectral fits at lower resolution.

As discussed below, only the order of magnitude of the minimum-energy
field matters in the considerations here. We therefore do not discuss
its behaviour in detail.

\subsubsection{The maximum particle energy \gmax}\label{s:ana.gmax.gmax}

\begin{figure}
\resizebox{\hsize}{!}{\includegraphics[angle=0]{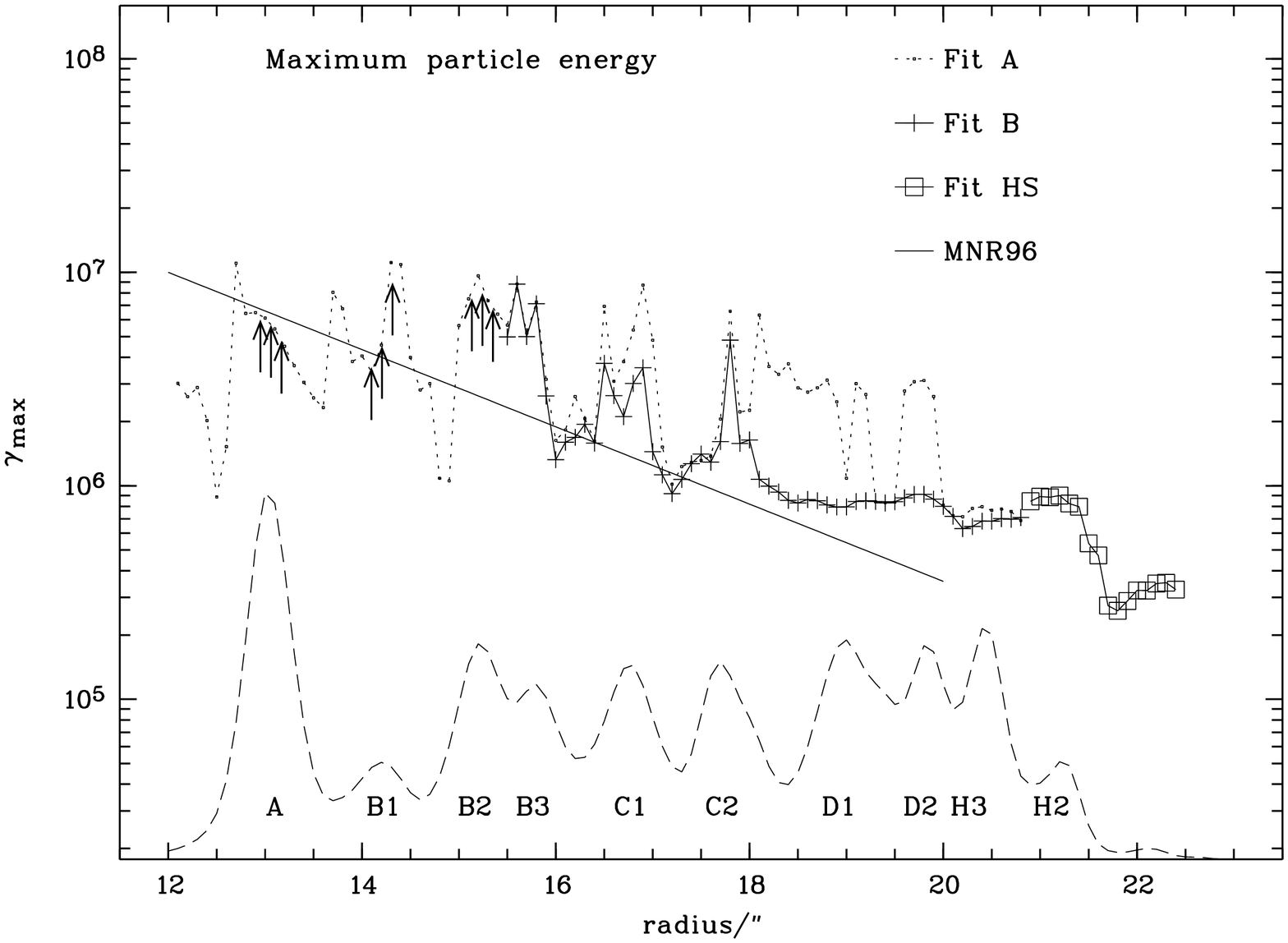}}
  \caption[Run of maximum particle energy]{Run of the maximum particle
  Lorentz factor \gmax\ along the jet.  The thin line represents the
  run parameterised as $\gmax = 10^{7} \exp(-(r-12\arcsec)/2\farcs4)$
  by \protect\citet{MNR96}.  Their points lie above our new 
The overall run of \gmax\ is identical to
  that of \nuc.}\label{f:ana.res.run-gmax}
\end{figure}
\begin{figure}
\parbox[c]{.15\hsize}{image\\620\nm}\parbox[c]{.85\hsize}{\resizebox{\hsize}{!}{\includegraphics{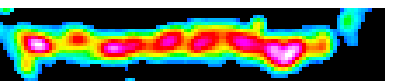}}}\\
\parbox[c]{.15\hsize}{$\log \gmax$\\Model A}\parbox[c]{.85\hsize}{\resizebox{\hsize}{!}{\includegraphics[width=.9\hsize,clip]{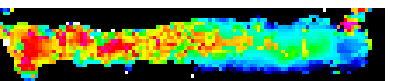}}}\\
\parbox[c]{.15\hsize}{$\log \gmax$\\Model B}\parbox[c]{.85\hsize}{\resizebox{\hsize}{!}{\includegraphics[clip,width=.9\hsize]{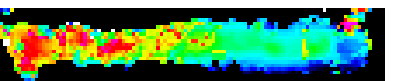}}}\\
\caption[Map of the maximum particle Lorentz
factor]{\label{f:ana.res.gmax}Map of the maximum particle Lorentz
factor \gmax. The results for Model HS have been inserted in both.
Only in regions A and B2 does \gmax\ show a strong correlation to
features of the radio-optical jet morphology. The only differences
between Model A and Model B occur in regions C--H3, where the
discrepancies between the infrared-optical and optical-ultraviolet
spectral indices are strongest.}
\end{figure}

The run of the maximum particle energy inferred from the
minimum-energy field and the cutoff frequency is shown in
Fig.\,\ref{f:ana.res.run-gmax}. This run is very similar to the run of
the cutoff frequency in Fig.\,\ref{f:ana.res.run-nuc}; this arises
from the relation used to derive the maximum particle energy
(Eqn.\,\ref{eq:nu_c_prop}):
\begin{displaymath}
 \gmax = 10^7\times \left(\frac{\nuc}{4.2\times10^{15}{\rm
 Hz}}\right)^\frac{1}{2} \left(\frac{B}{\rm nT}\right)^{-\frac{1}{2}} .
\end{displaymath}
The maximum particle energy and cutoff frequency are both plotted
logarithmically, that is, we are comparing their order of magnitude.
Hence, the run of the two can only differ if the order of magnitude of
the magnetic field changes significantly along the jet. This is not
the case, as expected from the fact that the minimum-energy field
scales as $S_\nu^{2/7}$ (\cf \S\ref{s:ana.gmax.bmin} and
Fig.\,\ref{f:ana.res.run-Bmin}). This implies that although the
variations of \Bmin\ perpendicular to the jet axis are dominated by
the assumed geometry, the variations in \gmax\ are not affected as
greatly.  We therefore show a map of the maximum particle energy in
Fig.\,\ref{f:ana.res.gmax} in addition to the development of \gmax\
along the jet axis presented in Fig.\,\ref{f:ana.res.run-gmax}.

As expected from the run of the spectral indices from previous
studies, the global trend is a decrease of \gmax\ from $>10^7$
down to $10^{5}$--$10^{6}$. Simultaneously, there is a strong increase
of the jet's luminosity, \ie luminosity and maximum energy are
anti-correlated.  We consider the local variations next. As we noted
above, local extrema of the infrared-optical spectral indices
correspond to extrema of \nuc\ and hence of \gmax\ because
the minimum-energy field does not change much along the jet.
Significant local extrema of \gmax\ are found in the following
regions:
\begin{itemize}
 \item A and B2 (13\arcsec and 15\farcs5), in which both
  high-frequency spectral indices peak simultaneously at a value
  significantly flatter than the radio-infrared spectral index
 \item C1 and C2 (16\farcs75 and 17\farcs75), where both
  high-frequency spectral indices show a local peak; here, the
  infrared-optical spectral index peaks at the brightness maximum,
  while the optical-UV maxima are slightly offset radially outward
\item The radio hot spot H2, at which all high-frequency spectral
  indices and the radio-infrared index drop significantly. \gmax\
  starts to drop from a value $\gmax \approx 10^6$ at the optical
  counterpart to H2 and reaches a plateau beyond the radio hot spot at
  one-third of the pre-hot spot value. This drop corresponds to the
  absence of optical and ultraviolet emission beyond the radio hot
  spot in the tip of the jet H1. The detailed run of the cutoff frequency
  in the hot spot is determined not only by the spectral evolution
  downstream of the Mach disk, which is located at the
  highest-frequency emission peak, but also by the effect of telescope
  resolution and integrating along the line of sight through the
  cylindrical emission region oriented at 45\degr to the line of sight
  \citep{HM87}.
\end{itemize}
Thus, like for the cutoff frequency, the \emph{local} variations of
\gmax\ show the exact opposite behaviour of the global variations: in
regions A and B2, there is a strong positive correlation between
energy output and cutoff energy.  A much weaker, but still positive
correlation is observed for the remainder of the jet's
knots. Globally, however, the cutoff frequency decreases as the radio
jet becomes brighter.  We consider possible explanations in
\S\ref{s:disc.corr} below.

\section{Discussion}\label{s:disc}

\subsection{The hot spot}\label{s:disc.HS}
\citet{MH86} presented a model for particle acceleration in the hot
spot of 3C\,273. They showed that the hot spot spectrum can be
understood by modelling the hot spot as a planar shock at which
first-order Fermi acceleration of particles takes place (details of
the spectral modelling are contained in \citet{HM87}, while further
observations of this and other hot spots have been presented by
\citet{magnumopus89,hs_II}). This model predicts an offset between the
peaks of the radio and optical emission of 0\farcs2, arising from the
vastly different synchrotron loss scales of electrons being advected
downstream from the shock itself: ``radio'' electrons lose a
negligible fraction of their total energy, while ``optical'' electrons
quickly lose their entire energy as they move away from the
acceleration region.  Recall that the strong compression at the shock
leads to an increased magnetic field compared to the jet flow, and
hence to stronger losses.

Figure\,\ref{f:res.im.morph.HS} clearly shows the predicted offset in
the position of the hot spot (H2) between radio and optical
wavelengths. This strengthens the identification of H2 with a planar
shock with strongly localised particle acceleration.

We note that again the minimum-energy magnetic field determined for
the hot spot agrees with the magnetic field determined from the hot
spot spectrum by \citet[][ the only difference arises due to different
values of the parameter $k$ in the minimum-energy field
determination]{hs_II}.

Thus, at 0\farcs3 resolution, the \citet{HM87} hot spot model is
consistent with all observations. Moreover, our new observations have
confirmed the predicted offset in the hot spot's position.  However,
it is remarkable that even the hot spot shows a UV excess, indicating
that the hot spot, too, has a more complicated internal structure than
previously assumed.  In addition, the presence of emission from the
tip of the jet H1 \citep[which has been called a ``precursor'' in the
literature; see][e.g.]{FD85} and its relation to the hot spot must be
explained.  The next tests of this model will involve a detailed
analysis of the hot spot's radio morphology at the highest available
resolution, which we defer to a future paper.

\subsection{Flatter spectra at higher frequencies: the need for a
  second emission component}\label{s:disc.pop}

When describing synchrotron spectra as power laws, the underlying
assumption is a power-law electron energy distribution.  As noted
above, a non-idealised spectrum with finite maximum particle energy
and rapid pitch-angle scattering will furthermore exhibit a
quasi-exponential cutoff at some frequency \nuc, the synchrotron
frequency corresponding to the highest electron energies.  This cutoff
implies that the spectrum has a convex shape, so that the spectrum
steepens progressively towards higher frequencies. Any high-frequency
flux must therefore lie below a power-law extrapolation from lower
frequencies.

In employing \emph{broken} power laws, the only physical break is a
steepening of the spectrum from $|\alpha_\mathrm{low}|$ to
$|\alpha_\mathrm{high}| = |\alpha_\mathrm{low}| + 1/2$ near some break
frequency $\nub < \nuc$.  This break arises in a model of the
synchrotron source as a loss region into which a power-law
distribution of electrons with approximately constant number density
and maximum energy is continuously injected. The break is produced by
adding up the contributions from the electron population observed at
different times since acceleration, \ie with different cutoff
frequencies.  The magnitude of the break of $1/2$ is fixed by the
cooling mechanism\footnote{Slightly larger breaks may be possible by
assuming certain source conditions, systematically varying magnetic
fields, for example \citep{Wil75}.}. Describing synchrotron
spectra with arbitrary breaks is therefore inconsistent with any
physically motivated model; in violation of our reasoning above, one
could then even invent breaks to flatter spectra.

% clearly not motivated by
% any physical model; for example, there would be no reason to exclude
% breaks to flatter spectral indices at higher frequencies.  

We used the detailed shape of cutoff spectra as computed by
\citet{HM87}. The spectral shape is the result of observing a mixture
of electron populations, and the appearance of the entire source can
be understood in terms of the temporal evolution of the initially
accelerated population.  As long as the spectral shape is determined
only by radiation losses, not even homogeneous source conditions are
required.  The resulting spectrum rarely will follow a power law over many
decades in frequency, but will be curved and in any case exhibit a
convex shape.

In contrast, a more complex electron distribution could be produced if
different parts of a source accelerate electrons to different maximum
energies, and contain different numbers of electrons.  Such a source
could be a jet with an inhomogeneous distribution of particles and
magnetic fields. This source must be described by more than one
electron population (or equivalently, as the sum of the emission from
more than one source).  This corresponds precisely to the situation
encountered in 3C\,273's jet: the observed flattening of the spectrum
implies that a description using a single electron population is
inadequate.  This implies unambiguously that there are at least two
components to the observed spectrum: one is the ``low-frequency''
synchrotron spectrum from radio through optical wavelengths. In
addition, there is a ``high-frequency'' component responsible for the
UV excess, and hence the observed flattening. We have suggested that
this second component is the same as that responsible for the jet's
X-rays \citep{Jes02}, which could be a second synchrotron component
\citep{Roe00,Mareta01} or beamed inverse Compton emission
\citep[\cf][]{Sam01} . In any case, our spectra prove that 3C\,273's
jet is not a homogeneous synchrotron source.

It is notable that besides 3C\,273, the spectral energy distributions
(SEDs) of M87's jet shows the same upward curvature: \citet{Mar02}
have presented SEDs for that jet from radio through X-rays (their
Fig. 3).  They fitted two different single-population models, a
Jaffe-Perola model assuming a randomly tangled magnetic field
structure and pitch-angle scattering, and a Kardashev-Pacholczyk
model, also assuming a tangled field but no pitch angle scattering.
Only the latter provides an acceptable fit to the observed SEDs, but
it is unclear how pitch-angle scattering could be avoided in a tangled
field geometry.  Moreover, the spectra for knots HST-1 and D clearly
show a steeper downturn near $10^{15}$\,Hz \citep[already hinted at by
the ground-based spectra obtained by][ their Figs. 7 and 8]{MRS96}
than the Kardashev-Pacholczyk spectrum, which is therefore inadequate.
Instead, again a second spectral component must be invoked to account
for the X-ray flux lying above the cutoff observed in the
near-infrared--optical region.  \citet{Mar02} viewed this as evidence
in favour of the spatially stratified model for M87's jet presented by
\citet{Pereta99}.

\subsection{Slow decline and smooth changes in \gmax: the need for an
  extended acceleration mechanism}\label{s:disc.acc}

Our determination of the maximum energy of synchrotron-emitting
electrons shows a lower limit for the Lorentz factors of $\gmax>10^7$
at the onset of the optical jet, 12\arcsec\ from the quasar core
(Fig.\,\ref{f:ana.res.run-gmax}).  The maximum Lorentz factor first
drops outward along the jet, and then stays nearly constant at a
Lorentz factor of a few times $10^6$ at radii $>17\arcsec$. It drops
again by factor of 3 within the hot spot. This drop most likely is caused by the
stronger magnetic field there.  In any case, the Lorentz factor of the
particles emitting optical synchrotron radiation cannot be much lower
than $10^5$ anywhere in the jet. Electrons with such high energies can
only be observed close to their acceleration sites: as noted in all
previous studies, synchrotron cooling timescales for these particles
are of the order of a few hundred or thousand years \citep[see][
\emph{e.g.}]{GN75,RM91,MNR96}.  The corresponding loss scales are a
few hundred parsec for the case of electrons freely streaming at
the speed of light, which is unrealistic since they will have to follow
tangled field lines.

The development of the maximum particle energy along the jet is
therefore consistent with the absence of any synchrotron cooling. The
same conclusion can already be drawn from the smooth changes of just
the optical spectral index, which should most strongly reflect the
synchrotron losses the particles undergo, and from the mere presence
of optical synchrotron radiation along the entire jet \citep{Jes01}.
The apparent absence of cooling in the presence of synchrotron
radiation can only be explained by a distributed acceleration mechanism.

The smooth changes of the spectral features along the jet indicate a
correspondingly smooth variation of the physical conditions along the
jet.  The present detailed high-resolution study of the jet in 3C\,273
set out with the aim to map those regions in which particles are
preferentially accelerated. Since the observed cooling is much less
drastic than expected, the conclusion is that the \emph{entire} jet is
the particle acceleration region.  The fact that the jet has a much
higher emissivity than the radio cocoon around it (see
\S\ref{s:res.im.width}) also implies a difference in physical
conditions between the jet and the surrounding material.  Furthermore,
the jet's X-ray emission is a further sink of energy (whatever the
emission process) which needs to be filled by the re-acceleration
mechanism. Note that this does not preclude the possibility that the
knots in the jet are due to shocks, at which particle acceleration can
also take place.  But even in this case, particles must be accelerated
between the shocks as well.

This corresponds to the ``jet-like'' acceleration mechanism proposed
by \citet{hs_II}.  The detailed physics of this distributed
acceleration mechanisms are unknown, but several theoretical
suggestions have been made.  Magnetic reconnection is one way to tap
the energy stored in the jet fluid and convert it continuously into
relativistic particles \citep{Lit99}.  Another possible energy source
is the velocity shear between the surface of the jet and the
surrounding medium \citep{StaOst02}.  This latter mechanism would be
limited to particle acceleration within a thin surface layer. Since it
appears that the jet emission is not edge-brightened, as would be
expected for emission from a surface layer (\S\ref{s:res.im.vol}),
surface velocity shear is unlikely to be the acceleration mechanism
providing optical electrons along the entire jet.  It may, however, be
appropriate to explain the second higher-energy emission component
identified above (\S\ref{s:disc.pop}), provided this turns out to be
edge-brightened.  A similar mechanism has recently been proposed by
\citet{RieMan02}: these authors suggest that velocity shear and
centrifugal forces in a rotating jet flow may provide efficient
particle acceleration.

The global energy budget of the jet is a related question. If the
entire jet operates at the minimum energy condition, and if there is
no strong, variable beaming, the increase in the minimum-energy field
along the jet (linked to the increase in synchrotron luminosity)
corresponds to an increase of the energy density stored in magnetic
fields and relativistic particles by about one order of magnitude.
This raises interesting questions about the energy budget of the jet:
If the total energy flux in magnetic fields and relativistic particles
is to be conserved, the jet would need to slow down
\citep[\cf][]{GK03}. The necessity to account for the kinetic energy
and momentum flux imposes further constraints.  On the other hand, it
may be possible to convert jet kinetic energy into magnetic and
particle energy, as in some of the acceleration mechanisms just
mentioned, leading to a different mutual dependence of jet speed and
magnetic energy density than in the pure energy conservation case.  A
detailed examination of the jet's energy budget addressing these
issues is certainly a worthy exercise, but beyond the scope of the
present publication.

\subsection{Correlations between brightness and maximum particle energy}\label{s:disc.corr}

As described in \S\ref{s:ana.gmax.gmax}, there is a local positive
correlation between surface brightness and maximum particle energy,
although these quantities are \emph{anti-}correlated globally.  The
opposite correlations must be caused by two different physical
mechanisms.  Although the cutoff frequency is the direct observable,
the underlying physical parameters are the maximum particle energy and
the jet's magnetic field.  In 3C\,273, the global run of the maximum
particle energy is practically identical to that of the cutoff
frequency, and the local variations in cutoff frequency are much
smaller than those in surface brightness (see
Figs.\,\ref{f:ana.res.run-nuc} and \ref{f:ana.res.run-gmax}).

We have established above (\S\ref{s:disc.acc}) that particles must be
accelerated continuously within the jet.  In this case, the value of
the maximum particle energy is obtained by equating the acceleration
timescale $\tau_{\mathrm{acc}}$, which is the time during which a
particle's energy is increased by a factor of two, with the
synchrotron loss timescale $\tau_\mathrm{syn}$
\citep[\eg,][]{Longair_eloss}:
\begin{eqnarray}
\tau_{\mathrm{acc}} &=& \tau_\mathrm{syn}(\gamma_\mathrm{max}) \propto
\frac{1}{B^2 \gamma_\mathrm{max}}\nonumber\\
\Leftrightarrow \gamma_\mathrm{max} & \propto &  \frac{1}{B^2
  \tau_{\mathrm{acc}}}. \label{eq:gmax_tacc}
\end{eqnarray}
The acceleration timescale is energy-independent both for diffusive
shock acceleration \citep[\eg,][]{Bell78a} and for acceleration by
reconnection \citep[\eg,][]{Lit99}. Thus, the behaviour of the
maximum particle energy as a function of magnetic field depends
critically on the nature of the acceleration mechanism, in particular
the scaling of the acceleration time scale with the magnetic field
strength.

We have taken the jet Doppler factor to be $\mathcal{D}=1$ in all
calculations up to here, which is justified because the value of the
maximum particle energy as inferred from the jet's surface brightness
and minimum-energy magnetic field only depends very weakly on the true
value of the Doppler factor \citep[cf. \S\ref{s:ana.equi}
and][]{NeumDiss,MNR96}.  The observed surface brightness $S$ and
cutoff frequency \nuc\ have the following scalings with $B$ and
$\mathcal{D}$ for a continuous jet \citep[see][ e.g.]{BBR84}:
\begin{eqnarray}
S & \propto & B^{1-\alpha}\, \mathcal{D}^{2-\alpha} \label{eq:S_BD}\\
\nuc & \propto & B\, \gamma_\mathrm{max}^{2}\, \mathcal{D}. \label{eq:nuc_BD}
\end{eqnarray}
In the following discussion, we will \emph{not} use the minimum-energy
magnetic field, because its computation has introduced correlations
between the magnetic field strength, the electron number density, and
the surface brightness by construction.

\subsubsection{Local correlation between brightness and maximum particle energy}

The only other jet which has been studied similarly well is that in
M87. There are key differences between the behaviour of the maximum
particle energy between these two jets. \citet{MRS96} showed that M87
shows a strong correlation between surface brightness and cutoff
frequency, while the maximum particle energy remains practically
unchanged.  This correlation can be understood purely in terms of
variations of the magnetic field strength, while the constancy of the
maximum particle energy is explained by invoking acceleration in
shocks \citep[\eg][]{SBM96}, by an extended acceleration mechanism
\citep{MRS96}, or by assuming a sub-equipartition magnetic field and
relativistic time dilation which both lower the loss times
sufficiently even without particle acceleration \citep{HB97}.
\citet{MRS96} conclude that re-acceleration is responsible, and that
the mechanism in M87's jet obeys the scaling $\tau_{\mathrm{acc}}
\propto B^{-2}$ to achieve a constant maximum particle energy
(cf. Eqn.\,\ref{eq:gmax_tacc}).  It is useful to consider whether this
acceleration mechanism proposed for M87 could be identical to the one
at work in 3C\,273, \ie whether it is universal.

Since the maximum particle energy changes significantly along
3C\,273's jet, the acceleration mechanism invoked by \citet{MRS96} for
M87 must be different from the one acting in 3C\,273's jet.  To
produce a local correlation between brightness and maximum particle
energy via magnetic field variations, the acceleration time scale in
3C\,273 would need to follow a scaling $\tau_{\mathrm{acc}} \propto
B^{s}$ with $s>2$ so that both \gmax\ and $S$ correlate with $B$
(Eqns.\,\ref{eq:gmax_tacc} and \ref{eq:S_BD}).  In that case, local
magnetic field variations could explain the observed
correlations. However, since the brightness changes are larger than
the changes in the maximum particle energy and cutoff frequency, it
seems more likely that small-scale variations in the Doppler factor
$\mathcal{D}$ are responsible because $S$ varies more strongly with
$\mathcal{D}$ than with $B$.  But variations in $\mathcal{D}$ by
themselves are insufficient for 3C\,273 because they do not produce a
correlation between brightness and maximum particle energy.  To
clarify the origin of this correlation, it will be necessary to
disentangle the UV excess from the remainder of the jet emission to
remove any uncertainty about the true value of the cutoff frequency,
and ideally to obtain a measurement of the jet's magnetic field other
than the minimum-energy estimate.

\subsubsection{Global anti-correlation of brightness and maximum
  particle energy}

The global anti-correlation of brightness and maximum energy appears
intuitive for synchrotron radiation: stronger synchrotron emission
implies larger energy losses, and hence a lower maximum energy.
However, as noted above, the actual behaviour of the maximum particle
energy in the presence of particle acceleration depends on the details
of the acceleration mechanism.  If the local correlation of brightness
and \gmax\ can be explained by the nature of the acceleration
mechanism, the global anticorrelation must be due to an entirely
different effect which reverses the local correlation.  It may be
possible that the local correlations are produced by a ``shock-like''
acceleration mechanism acting only in the brighter regions (\eg if the
bright regions \emph{are} shocks), while electrons elsewhere are
accelerated by the ``jet-like'' mechanism \citep{hs_II}.  However,
such a scenario is somewhat unappealing because there would need to be
a certain fine-tuning to explain the overall smoothness of the
spectral index changes.  Furthermore, although this may explain the
different global and local behaviour within 3C\,273 as well as the
presence of two distinct emission components (\S\ref{s:disc.pop}), the
difference to M87 remains, in which the ``jet-like'' mechanism is
assumed to produce the positive correlation.  We therefore need to
appeal to further differences.  \citet{Mei96} suggested that the
observations of M87 and 3C\,273 could be unified if there was a link
between the maximum particle energy and the velocity gradient along
the jet, and the jet velocity was relativistic, so that changes in the
velocity would lead both to changes in the maximum particle energy and
in the Doppler factor and hence the apparent brightness.  We will
discuss this idea elsewhere.

\section{Summary and future work}\label{s:summary}

One of the main unsolved questions in the study of extragalactic jets
is posed by the observations of optical synchrotron radiation over
scales much larger than typical synchrotron loss scales and far from
the ``working surfaces'' of the standard model \citep{BlaRee74}.
Another puzzle has been posed by the recent \emph{Chandra} observations
of X-rays from extragalactic jets, whose emission mechanism remains
debated and which form a further sink of energy that has to be filled
within the jet. The study of the physical conditions giving rise to the
observed emission by keeping up the particle energy against strong
synchrotron losses at optical wavelength and supplying the energy
observed in X-rays can only be performed via a study of the synchrotron
continuum, at the highest possible spatial resolution and covering the
largest possible wavelength range.

\subsection{Observations}

We present new HST and VLA images of the jet at $\lambda\lambda$
3.6\cm, 2.0\cm, 1.3\cm, 1.6\micron, 620\nm\ and 300\nm\ matched to a
common resolution of 0\farcs3 (Fig.\,\ref{f:res.ima.images}).  We
combine the data to obtain spectral energy distributions which we fit
with theoretical synchrotron spectra (Fig.\,\ref{f:ana.fits.obs1}).
Contrary to expectations for synchrotron emission, the observed
spectra show a significant flattening in the infrared-ultraviolet
wavelength range, implying that the emission cannot be modelled as
synchrotron emission due to a single electron population, as has been
assumed in previous studies of this jet's emission
\citep{MNR96}. Instead, additional emission must be present which can
lead to the observed flattening. The most likely explanation is an
additional flat-spectrum component in the ultraviolet. The same
component may simultaneously be responsible for the jet's X-ray
emission \citep[for details, see][]{Jes02}.

The optical-ultraviolet spectral index map generated at 0\farcs3
resolution (Fig.\,\ref{f:res.alpha.alpha}a) shows no strong
correlation between local spectral index and surface brightness
variations. There is, however, such a correlation on the
radio-infrared and infrared-optical spectral index maps. Even these
correlations are less pronounced than those found in the jet of M87
\citep{Per01,PMB01}, where they are taken as evidence of localised
particle acceleration in the knots of this jet (but see
\citealp{Mei96} for an alternative explanation).  As confirmed by our
spectral fits (see below), the spectral index variations reflect
variations of the maximum particle energy along the jet.

We find an offset between the radio and optical images of the hotspot
that is consistent with the prediction by \citet{HM87}. This adds
further to the confidence that the assumed hot spot model is correct
\citep{hs_II}.  However, the role of the tip of the jet H1 and its
relation to the hot spot remains to be clarified, perhaps in a model
fully describing the hot spot flow in three dimensions, instead of
just one.  If the jet flow is still highly relativistic just upstream
of the hot spot, the details of the flow's deceleration in the hot
spot may also change its appearance \citep{GK03}.

\subsection{Synchrotron spectral fits}

We fitted the observed spectra with model synchrotron spectra to
extract physical information, in particular the maximum particle
energy and its variation along the jet.

Model synchrotron spectra according to \citet{HM87} have been used to
determine the cutoff frequency \nuc.  It is mainly determined by the
infrared-optical spectral index.  We infer the maximum particle energy
from the fitted cutoff frequency by assuming an equipartition magnetic
field $B_\mathrm{min}$.  It is a general result that the value of
$B_\mathrm{min}$ is very robust with respect to assumptions about the
details of the spectral shape, as long as the spectrum has a power law
index that is steeper than $\alpha=-0.5$.

The cutoff frequency decreases from above $5\times10^6$ in region A at
$r\approx 13\arcsec$ and settles to a plateau of order $10^6$ at $r\ga
18\arcsec$. A further decrease to $\gmax \approx 3\times10^5$ is only
observed at the position of the radio hot spot, beyond which there is
no optical, but still infrared emission.  This global run corresponds
to the findings of previous studies at 1\farcs3 resolution, which
showed that the maximum particle energy decreases as the radio jet
brightness increases.  As expected from the spectral index variations,
there is an additional weak correlation between local variations of
maximum particle energy and surface brightness.

An earlier examination of the effect of relativistic beaming on the
inferred synchrotron lifetime had shown that the overall slow decrease
of \gmax\ cannot be explained by an enhancement of the true
synchrotron lifetime by beaming effects \citep{Jes01}. We concluded
that particle acceleration must take place distributed across the
entire jet. This conclusion is strengthened by the X-ray observations,
whatever the X-ray emission mechanism: The radiative lifetimes of
X-ray synchrotron electrons in an equipartition field of 20\,nT are a
few tens of years, those of UV-emitting electrons are a few hundreds
of years.  In the inverse Compton scenario, the combined effects of
relativistic beaming and lower magnetic fields lead to lifetimes of a
few times $10^4$ years for UV-emitting electrons.  In both cases,
particles must therefore be accelerated continuously within the entire
jet \citep[``jet-like'' acceleration,][]{hs_II}.  The same conclusion
is drawn from the fact that the physical conditions within the jet of
3C\,273 vary very smoothly down to the spatial scales of
$0.9h_{70}^{-1}\kpc$ resolved here.  This does not preclude the
possibility that the enhanced-brightness regions are shocks at which
particles are accelerated -- but even if they are, re-acceleration
between them is necessary to explain the observed spectral features.

The details of the acceleration mechanism determine the scaling of the
maximum particle energy with changes in the magnetic field.  The
interplay between the ``jet-like'' and ``shock-like'' acceleration
mechanisms might explain the conflicting global and local behaviour of
the maximum particle energy in 3C\,273.  However, both mechanisms must
be different from the one invoked by \citet{MRS96} for M87's jet
(\S\ref{s:disc.corr}).

We stress again the observed flattening of the high-frequency
spectrum, which implies that the jet's emission cannot be assumed to
arise from a single electron population, but requires the presence of
an additional emission component. In particular, the spectrum of knot
A cannot be considered as a single power law from radio through
X-rays, as concluded from lower-resolution data
\citep{Roe00,Mareta01}.  Instead of considering the jet as a
homogeneous synchrotron source filled with a simple electron
population, we must realise that jets have an internal energetic
structure \citep[\emph{cf.}][]{Pereta99,Per01}.  Future work to
investigate the acceleration mechanism at work in this and other jets
must be based on these results.

\subsection{Future work}

With the present radio-optical data, it has only been possible to
detect the high-frequency hardening of the jet emission, and thus the
presence of a second emission component. No statements can be made
about the spectral shape of this ``high-energy'' component beyond the
approximate match of the X-ray extrapolation and the UV excess.  We
have been awarded HST time to extend the wavelength coverage into the
far-ultraviolet (150\,nm). Using these data, we will be able to
constrain the run of the optical synchrotron cutoff more accurately,
and to characterise the spectrum of the UV excess. This will achieve a
separation of the two spectral components from each other.  Particular
insight can be expected by considering the high-frequency spectra in
those regions of the jet showing the strongest difference between
radio, optical and X-ray morphology (\S\ref{s:res.im.morph}).

In order to test theories for extended particle acceleration, we
require detailed predictions both from theoretical work and from
magneto-hydrodynamical simulations of the jet flow, in particular
regarding the spatial distribution of relativistic particles.  The
most sensitive observational tool is multi-wavelength polarimetry
resolving the jet width.  This allows not only a study the particles'
energy distribution, but also an assessment whether particles at
different energies probe the same underlying magnetic field, that is,
whether they actually occupy the same emitting volume. Thus,
multi-wavelength polarimetry is the natural followup programme for our
detection of multiple emission components. We will use our guaranteed
time for the adaptive optics camera CONICA/NAOS at the VLT to obtain
near-infrared polarimetry at a resolution comparable to the data
presented here. However, optical polarimetry of 3C\,273's jet making
full use of the HST's capabilities is still missing.

Exciting new facts about the jet in 3C\,273 have been revealed by each
advance in optical and radio observing technology. From the surprising
observational facts encountered here, more surprises are yet to be
expected \emph{en route} to an understanding of the physics of this
object. Current advances in numerical work are also expected to shed
further light on the structure and dynamics of jets. As an extreme
object of its kind, 3C\,273 will remain high on the agenda of
astrophysicists studying jet phenomena.

\acknowledgements SJ was supported in part by the U.S. Department of
Energy under contract No.\ DE-AC02-76CH03000. This research has made
use of NASA's Astrophysics Data System. We are grateful to the
referees, Herman Marshal, Eric Perlman and Markos Georganopoulos, for
their thorough review and constructive comments.

% \bibliography{aamnem99,agn,jets,hst,radio,xray}
% \bibliographystyle{aa}

\end{document}